\documentclass[aps,pra,twocolumn,superscriptaddress]{revtex4}
\usepackage{amssymb}
\usepackage{graphicx}
\usepackage{amsmath}
\usepackage{amsfonts}
\usepackage{color}
\usepackage{bm}

\begin{document}

\title{Spin-orbit coupling and topological phases for ultracold atoms}

\author{Long Zhang}
\affiliation{International Center for Quantum Materials, School of Physics, Peking University, Beijing 100871, China}
\affiliation{Collaborative Innovation Center of Quantum Matter, Beijing 100871, China}
\author{Xiong-Jun Liu
\footnote{Correspondence addressed to: xiongjunliu@pku.edu.cn}}
\affiliation{International Center for Quantum Materials, School of Physics, Peking University, Beijing 100871, China}
\affiliation{Collaborative Innovation Center of Quantum Matter, Beijing 100871, China}

\begin{abstract}
Cold atoms with laser-induced spin-orbit (SO) interactions provide promising platforms to explore novel quantum physics, in particular the
exotic topological phases, beyond natural conditions of solids. The past several years have witnessed important progresses in both
theory and experiment in the study of SO coupling and novel quantum states for ultracold atoms. Here we review the physics of the SO
coupled quantum gases, focusing on the latest theoretical and experimental progresses of realizing SO couplings beyond one-dimension (1D),
and the further investigation of novel topological quantum phases in such systems, including the topological insulating phases and topological superfluids.
A pedagogical introduction to the SO coupling for ultracold atoms and topological quantum phases is presented.
We show that the so-called optical Raman lattice schemes, which combine the creation of the conventional optical lattice and Raman lattice with topological
stability, can provide minimal methods with high experimental feasibility to realize 1D to 3D SO couplings.
The optical Raman lattices exhibit novel intrinsic symmetries, which enable the natural realization of topological phases belonging to different symmetry classes,
with the topology being detectable through minimal measurement strategies.
We introduce how the non-Abelian Majorana modes emerge in the SO coupled superfluid phases which can be topologically nontrivial or trivial,
for which a few fundamental theorems are presented and discussed. The experimental schemes for achieving non-Abelian superfluid phases are given.
Finally, we point out the future important issues in this rapidly growing research field.
\end{abstract}

\maketitle

\section{Introduction}

\subsection{Why study spin-orbit coupling?}

Spin-orbit (SO) coupling is a relativistic quantum mechanics effect which characterizes the interaction between the spin and
orbital degrees of freedom of electrons when moving in an external electric field. Due to the special relativity, the electron
experiences a magnetic
field in the rest frame, which is proportional to the electron velocity and
couples to its spin by the magnetic dipole interaction,
rendering the SO coupling with the following form
\begin{eqnarray}\label{PauliSOC}
H_{\rm so}\propto{\bm\sigma}\cdot{\bm B}_{\rm eff}\propto{\bm\sigma}\cdot\left({\nabla}V\times{\bm p}\right)
=\lambda_{\rm so}{\nabla}V\cdot\left({\bm p}\times{\bm\sigma}\right),
\end{eqnarray}
where ${\bm\sigma}$ is the spin, $V({\bm r})$ is the external electric potential experienced by the electron, ${\bm p}$ is
the electron's momentum, and $\lambda_{\rm so}$ denotes the SO coefficient. In atomic physics the SO coupling is responsible
for the fine structure splitting of the optical spectroscopy. In solid state physics the SO interaction of Bloch electrons
exhibits several effective forms by taking into account the crystal symmetries and local orbitals around Fermi energy, and
can strongly affect the band structure of the system. The typical types of the SO coupling includes the Rashba and Dresselhaus
terms, which are due to the structure inversion asymmetry and bulk inversion asymmetry of the materials, respectively, and
Luttinger term which describes the SO coupling for the valence hole bands~\cite{Winkler,Sinova2008}. The study of SO coupling for electrons has
generated very important research fields in the recent years, including spintronics~\cite{AHE2010}, topological insulator~\cite{TIreview1,TIreview2}, and topological
superconductors (SCs)~\cite{TSCreview}, etc. These studies bring about completely new understanding of the effects of the SO coupling in the condensed
matter physics and material science.

{\it Spintronics.}--In a system with SO coupling one can manipulate the electron spins indirectly by controlling the orbital
degree of freedom. This lies in the heart of the study of semiconductor spintronics, with the spin degree of freedom of the
electron being exploited for improved functionality. From the basic form of the SO interaction, one can see that the components
of the spin, momentum, and electric field which couple together are perpendicular to each other: $\bold\sigma\perp\bm p\perp{\nabla}V$.
This implies that applying an external electric field may dynamically drive the electron at opposite spin states to move oppositely
in the real space. In particular, consider a Rashba SO coupled system with the presence of an external electric field, described
by the potential $V_{\rm ex}(\bold r)$, whose Hamiltonian reads $H=\bm p^2/2m_e+\lambda_{\rm so}(k_x\sigma_y-k_y\sigma_x)+V_{\rm ex}(\bold r)$,
where $m_e$ is the effective mass of Bloch electron. This Rashba SO coupling can be equivalently written in terms of a
non-Abelian SU(2) gauge potential and then $H=(\bm p-e\vec A_{\sigma})^2/2m_e+V_{\rm ex}(\bold r)$, where a trivial constant
is neglected and $\vec A_{\sigma}=(m_e\lambda_{\rm so}/e\hbar)(-\sigma_y\hat e_x+\sigma_x\hat e_y)$. Note that the field of a
non-Abelian gauge is given by $F_{\mu\nu}=\partial_\mu A_\nu-\partial_\nu A_\mu+(ie/\hbar c)[A_\mu,A_\nu]$, which is associated
a spin-dependent magnetic field, given by
\begin{eqnarray}
\vec B=\frac{m_e^2\lambda_{\rm so}^2}{\hbar^3}\frac{e}{c}\sigma_z\hat e_z.\label{SOfield}
\end{eqnarray}
When electrons are accelerated by the electric field, their spins are tilted to out-of-plane ($\hat e_z$) direction. The
above formula implies that electron having nonzero spin polarization experiences an effective magnetic field along $+z$ or
$-z$ direction depending on its polarization direction. Thus spin-up and spin-down electrons are deflected to opposite
sides (Fig.1a), leading to a pure spin current and spin accumulation in the edges of the transverse direction. This gives
a spin Hall effect (SHE)~\cite{SHE1,SHE2}. Moreover, in a magnetic semiconductor, the spin of electrons is polarized due to the existence
of a Zeeman coupling term $M_z\sigma_z$, which leads to a population imbalance in the spin-up and spin-down states. In
this case, while the electrons with opposite spin polarization along $z$ axis are deflected oppositely, a nonzero transverse
charge current is also resulted, giving rise to an anomalous Hall effect (AHE)~\cite{AHE2010}. In the anomalous Hall effect both the spin
and charge accumulations are obtained. The SHE and AHE may have important applications designing spintronic devices such as
SHE transistors and spin injectors.

\begin{figure}[h]
\centerline{\includegraphics[width=\columnwidth]{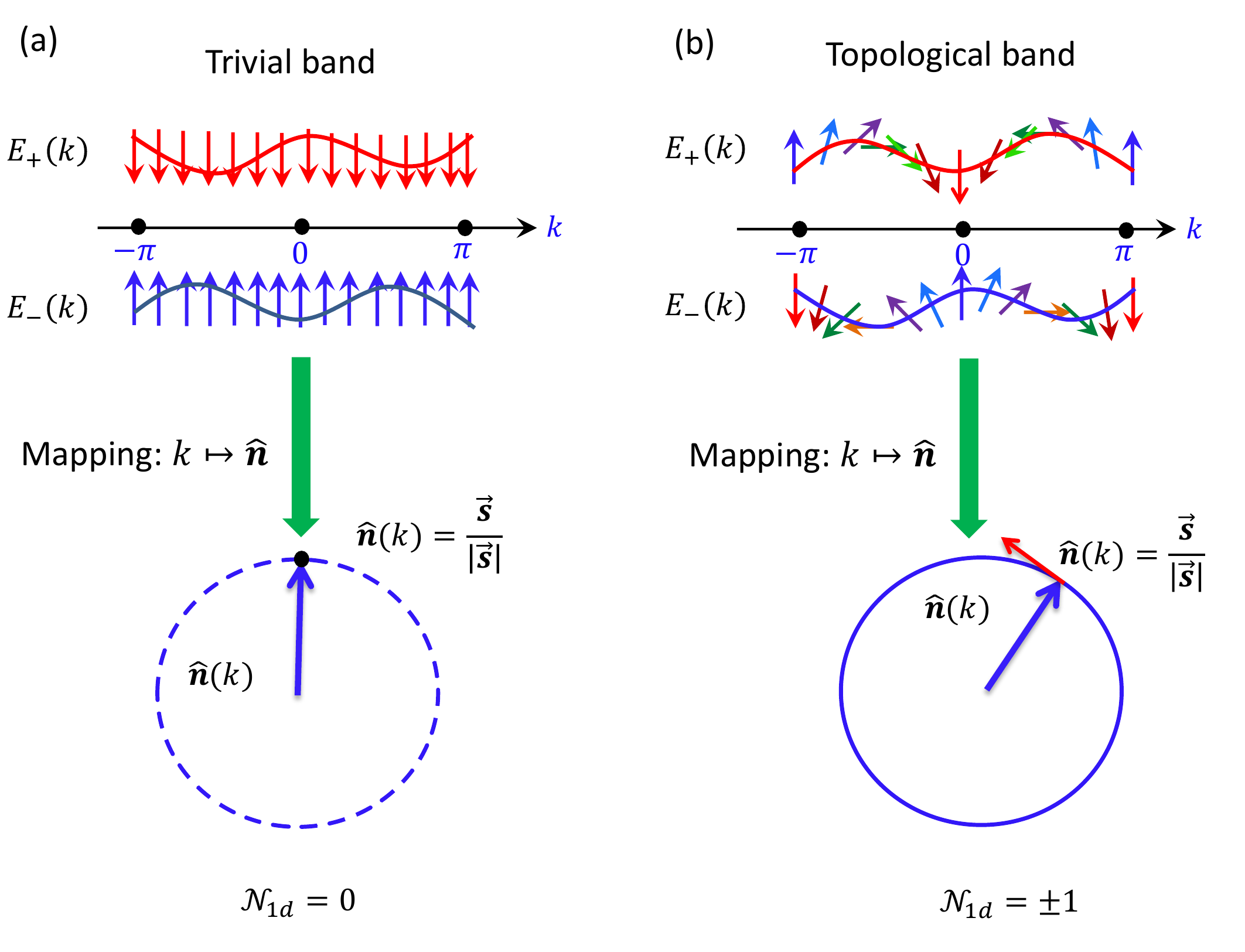}}
\caption{(a) Sketch for a topologically trivial band without SO coupling. In this case the spin states of the whole Brillouin zone does not wind over a circle,
so they map to a single point of the unit ring $S^1$, giving zero winding number. (b) Sketch for a topologically nontrivial band with SO coupling.
The spin states of the whole Brillouin zone winds over a circle. Thus they map to the whole unit ring $S^1$, giving a nonzero winding number.}
\label{Winding1}
\end{figure}

{\it Topological insulators.}--Taking into account the spin degree of freedom can bring rich nontrivial structure of the
Bloch bands for electrons~\cite{TIreview1,TIreview2}. In the presence of SO coupling the spin and momentum of an electron can be entangled, which
causes spin texture in the momentum space. For an insulator, all the momentum states of a single band in the $d$-dimension
form the first Brillouin zone (FBZ), which is a closed manifold known as torus $T^d$. The spin is attached to each Bloch
momentum governed by the SO interaction. A simple illustration of the SO effect on topology is sketched in Fig.~\ref{Winding1}. Fig.~\ref{Winding1} (a) gives a fully spin-polarized configuration, in which case the spin does not wind in the momentum space, thus mapping to a single point of the unit circle. This gives zero winding number (${\cal N}_{1d}=0$) [Fig.~\ref{Winding1} (a)]. In contrast, for case of Fig.~\ref{Winding1} (b), the spin winds over all direction of a circle in the real space when the momentum runs over the FBZ, giving rise of a nonzero winding number
of the Bloch band (${\cal N}_{1d}=1$). The latter case is referred to as a topological phase, while the former is a
trivial one. The transition between a topological phase and a trivial has to experience the band gap closing. As a
consequence, in the interface between the topological and trivial regions, e.g the boundary of the material which is
in a topological phase, the localized in-gap states emerge, mimicking the gap closing at the interface.

More specifically, the terminology {\it topological insulator} typically refers to the topological insulating phases in
2D or 3D with time-reversal symmetry~\cite{TIreview1,TIreview2}, which are a new class of insulating topological phases different from the quantum Hall effect discovered in 1980s~\cite{QHE1,QHE2}. In such insulators the boundary exhibit helical edge or surface states protected
by time-reversal symmetry. Since the concept was introduced in 2005, tremendous proposals and experimental discoveries
for the 2D and 3D topological insulators have been performed in the past years~\cite{TIreview1,TIreview2}. Physically, the SO interaction leads to
the band inversion of the orbital bands with opposite parity around Fermi energy, accounting for the underlying mechanism
for the topological phase transition. The inverted regime corresponds to the topological phase, while the normal un-inverted
regime corresponds to the trivial phase~\cite{TIbook,Basil2016}. The concept of topological insulators with time-reversal symmetry has also been
generalized to the phases protected crystal symmetries, dubbed topological crystalline insulators~\cite{Fu2011,Chiu2016}, and also to the topological
semimetal or metals~\cite{semimetal1,semimetal2,semimetal3,semimetal4}.

{\it Topological superconductors.}--Topological SC is a topological phase similar as the topological insulators,
having a bulk superconducting gap and gapless or midgap excitations in the boundary~\cite{TIreview2,TSCreview}. Unlike the insulating phases, the boundary
excitations in a topological SC are known as Majorana modes which are their own antiparticles~\cite{Franz2015RevModPhys,Wilczek2009,Alicea2012RPP,Franz2013a}. Mathematically, the
annihilation and creation operators of such states are identical $\gamma(x)=\gamma^\dag(x)$. Especially, the Majorana zero
energy modes localized in the point-like topological defects of topological SCs obey non-Abelian statistics~\cite{non-Abelian1,non-Abelian2,non-Abelian3,non-Abelian4,non-Abelian5},
and have potential applications to the topological quantum computation~\cite{TQC1,TQC2,TQC3}, which is an essential motivation for the great efforts
having been driven in both theory and experiment to search for such exotic modes in the recent years. Note that in a SC
the Bogoliubov quasiparticle operators generically takes the form $b_\mu(x)=\alpha_\mu c_\sigma(x)+\beta_\mu c_{\sigma'}^\dag(x)$,
with the spin indices $\sigma,\sigma'=\uparrow,\downarrow$. In an $s$-wave SC whose pairing order is even one has $\sigma\neq\sigma'$ and
the quasiparticle is not a Majorana, while in an spinless (or spin polarized) SC, only a parity-odd ($p$-wave) pairing
phase occurs and then $\sigma=\sigma'$, yielding a Majorana quasiparticle.

An important consequence of the SO interaction is that the parity of superconducting pairings can be manipulated in the SO
coupled materials. In the presence of SO coupling, in general the orbital angular momentum is no longer a good quantum number,
nor the spin. Thus in the superconducting pairing channels the parity-even (like $s$-wave) and parity-odd (like $p$-wave)
pairing states are mixed up. Under proper conditions, e.g. in the presence of an external Zeeman field which can kill the $s$-wave pairing, only parity-odd superconducting states shall survive
and then the Majorana modes can be realized. This idea has been broadly applied to the recent theoretical proposals and experimental
realization of topological SCs based on heterostructures formed by SO coupled materials and conventional $s$-wave
SCs, together with a Zeeman splitting field being applied by magnetic field or ferromagnetic insulators~\cite{Franz2015RevModPhys}. In the
interface the Cooper pairs are forced into $p$-wave type by the SO coupling and Zeeman field, rendering a topological
SC. Nevertheless, while many experimental studies have been performed in observing indirect signatures of
Majorana zero modes, based on $s$-wave SCs and semiconductor nanowires~\cite{Mourik1003,Deng2012,Das2012,Finck2012}, magnetic
chains~\cite{Perge2014,Ruby2015,Meyer2016}, or topological insulators~\cite{JiaJin-Feng2015, JiaJin-Feng2016,HeQL2017}, the rigorous confirmation is yet illusive.

\subsection{Brief history of SO coupling for ultracold atoms}

In cold atoms the spin states refer to the internal electronic hyperfine levels, and the spin of an atom is given by
the summation of the total spin, orbital angular momentum of all electrons of the atom, and the nuclear spin~\cite{Pethick_book}. Compared
with solid state materials, the ultracold atoms are extremely clean systems, with all the parameters being fully controllable
in the experiment~\cite{Bloch_review,Bloch2012}. Such a full controllability enable the ultracold atoms to be ideal platforms to simulate complex
quantum physics with exact models which can be beyond or not precisely achievable in solid state materials.

The ultracold atoms do not have intrinsic SO coupling, while an effective SO interaction can be generated by properly
coupling the atomic spin states to external fields~\cite{Dalibard2011,Goldman2014}. From the case of a Rashba SO coupling we have seen that the SO
interaction is generically equivalent to a non-Abelian gauge potential coupling to the spin degree of freedom of the
system. This picture gives birth to the basic ideas in realizing SO coupling for ultracold atoms through the generation
of synthetic gauge potentials. Before proposing a SO interaction, the realization of an Abelian gauge potential for a
spinless system was first theoretically considered by Jaksch and Zoller~\cite{Jaksch2003} in 2003 for optical lattices, and by Juzeli\=unas
and \"Ohberg~\cite{Slowlight2004} in 2004 for continuum quantum gas. The Abelian gauge potential is associated with an artificial magnetic flux,
similar as that in a rotating Bose-Einstein condensate~\cite{Cooper2008,Fetter2009}. Motivated by the previous study,
in 2004 Liu {\em et al.}~\cite{Liu2004} proposed to realize a spin-dependent gauge potential of the form $\vec A_3(\bold r)\sigma_z$ by
generalizing the previous scheme for magnetic flux
to the spin-dependent regime, yielding an early model for realization of SO coupling. Interestingly, this SO term describes
a coupling between spin and orbital angular momentum, which attracts particular attention very recently for cold atoms~\cite{DeMarco2015,Sun2015,Chen2015,Jiang2016}.
It is noteworthy that the spin-orbital-angular-momentum coupling is very recently realized in experiment by Lin's group~\cite{Chen2018},
which applied a scheme similar to the one proposed in Ref.~\cite{Liu2004}.
Adopting the similar idea of generating spin-dependent gauge potentials, Liu {\em et al.}~\cite{Liu2007} and
Zhu {\em et al.}~\cite{Zhu2006} respectively proposed to observe spin Hall effect for ultracold atoms. In both cases, spin-dependent gauge
potentials are still Abelian with the form $\vec A=\vec A_3(\bold r)\sigma_z$, associated with a spin-dependent magnetic
field $\vec B=B_3\sigma_z\hat e_z$, for which the spin-up and spin-down atoms
couple to artificial magnetic fields along $+z$ and $-z$ direction, respectively, leading to a (quantum) spin Hall effect.
The realization of non-Abelian gauge potentials was first proposed in 2005 by Osterloh {\em et al.}~\cite{Osterloh2005} in optical lattice, and by
Ruseckas {et al.}~\cite{Ruseckas2005} in continuum quantum gas. With the proposed non-Abelian gauge potentials, in principle one can achieve high
dimensional SO couplings.

The early proposals for SO coupling are hard to be realized in real experiments. In 2008, Liu {\em et al.}~\cite{Liu2009} pointed out that a 1D SO
coupling with equal Rashba and Dresselhaus amplitudes can be realized for atoms with a simple $\Lambda$-type configuration
of internal levels. A similar configuration was considered earlier by Higbie and Stamper-Kern~\cite{Higbie2002} to investigate the
properties of a Bose condensate, while not SO coupling. A Lambda type configuration contains two ground hyperfine
levels, mimicking the spin-up and spin-down states. A Raman coupling induced by two light beams drives spin-flip transition
and transfers momentum simultaneously, giving rise to SO coupling. The basic idea has been broadly applied in experiment to
realize the 1D SO coupling for bosons and fermions~\cite{Lin_SOC,Williams2012,Shanxi2012,MIT2012,USTC2012,Qu2013,Olson2014,Fu2014,USTC2014,USTC2015,Hamner2015,Jimenez2015,Burdick2016,
Song2016,Ketterle2016}. The first experimental realization was performed with bosons in Spielman's
group at NIST~\cite{Lin_SOC}. Following this study, Zhang's group at Shanxi University~\cite{Shanxi2012} and Zwerlein's group at MIT~\cite{MIT2012} respectively realized
the 1D SO coupling for $^{40}$K and $^{6}$Li Fermi atoms, respectively. On the other hand, Chen's group at USTC studied the
phonon spectra and observed the roton gap for a SO coupled BEC~\cite{USTC2015}, and further performed a systematic study of the phase diagram
at finite temperature of the SO coupled BEC~\cite{USTC2014}. Engels' group first realized the 1D SO coupling for bosons in an optical lattice.
More recently, the 1D SO coupling has also been realized with lanthanide and alkali earth atoms, including Dy fermions by
Lev's group~\cite{Burdick2016}, Yb atoms by Jo's group at HKUST~\cite{Song2016} and by Fallani's group~\cite{Livi2016}, and Sr atoms by Ye's group at JILA~\cite{Kolkowitz2017}.

Realization of a high-dimensional SO coupling (more than 1D) is more significant. The reason is obvious: a high-dimensional SO
coupling (e.g. the 2D Rashba term) corresponds to a non-Abelian gauge potential and is associated with nonzero Berry's curvature which could have nontrivial geometric or topological effects, while the 1D SO coupling realized in the aforementioned experiments corresponds to an
Abelian potential and does not give a nonzero Berry curvature. The study of broad classes of novel topological states necessitates the realization
of high-dimensional SO couplings, including topological superfluid phases~\cite{Zhang2008,Sato2009}. Many interesting schemes were proposed for realizing
2D~\cite{Osterloh2005,Ruseckas2005,Juzeliunas2010,Campbell2011,Sau2011,Anderson2013a,Xu2013,Liu2014a} and 3D SO couplings~\cite{Anderson2013b}, whereas the experimental realization was not available until very recently. The Rashba type and Dirac type
2D SO couplings are realized by Zhang's group~\cite{Huang2016} and a collaborative team by Liu's group at PKU and Pan-Chen's group at USTC~\cite{Wu2016}, respectively.
The former realization is based on a tripod scheme in a continuum space, and the latter is based on a scheme called optical Raman
lattice with double-$\Lambda$ internal configuration~\cite{Wu2016,Liu2014a}. The achieved 2D Dirac Hamiltonian realizes a minimal
model for quantum anomalous Hall effect driven by SO coupling, which cannot be achieved with solid state materials but has been firstly
obtained for ultracold atoms, and was shown to exhibit novel physics in the bulk and the boundary.

%

\section{Theory for synthetic SO coupling}

A natural way to simulate gauge potentials is to change the dynamical behavior of neutral atoms just as moving charges in electromagnetic fields
by some external forces, like rotation~\cite{Cooper2008,Fetter2009}, atom-light interaction~\cite{Lin2009,Lin_magnetic,Lin_electric,Zhai2015}
or laser-assisted-tunneling~\cite{Aidelsburger2011,Aidelsburger2013,Miyake2013}.
Juzeli\={u}nas {\it et al.}~\cite{Slowlight2004} proposed
to produce an effective magnetic field by Berry's phase~\cite{Berry1984}, which arises from the adiabatic motion of a spatial-dependent
dark state (a light-dressed eigenstate uncoupled with the excited state). Liu {\em et al.}~\cite{Liu2004} proposed to realize a
spin-dependent gauge potential of form $\vec A_3(\bold r)\sigma_z$ by generalizing the previous scheme for magnetic flux
to the spin-dependent regime, yielding an early model for realization of SO coupling. This SO term describes
a coupling between spin and orbital angular momentum, which attracts particular attention very recently~\cite{DeMarco2015,Sun2015,Chen2015}.
A more generic notion of SO coupling,
which corresponds to non-Abelian
gauge potentials, can be generated by employing two or more degenerate dressed states~\cite{Ruseckas2005,Osterloh2005}.
The basic idea of generating adiabatic Abelian
and non-Abelian gauge fields can, in fact, trace back to Wilczek and Zee's seminal work in 1984~\cite{Wilczek_Zee} .

\subsection{Gauge potential for continuum gas}

We start with the generic theory of producing non-Abelian adiabatic gauge potentials~\cite{LiuReview2008}. We consider a $N$-level quantum
system which is coupled to external fields, with
the Hamiltonian $H=-\frac{\hbar^2}{2m}\nabla^2+V({\bm r})+H_{\rm I}({\bm r})$. Here $V({\bm r})$ is the trapping potential,
and $H_{\rm I}({\bm r})$ denotes a spatially varying interacting term between the system and external fields (e.g. laser addressing). One can
diagonalize the coupling Hamiltonian $H_{\rm I}$ through a unitary transformation $U({\bm r})$ via $H^{\rm diag}_{\rm I}=U^\dagger H_{\rm I}U$,
where $H^{\rm diag}_{\rm I}$ is diagonal and is written as
$H^{\rm diag}_{\rm I}=\sum_j^NE_j|\chi_j({\bm r})\rangle\langle\chi_j({\bm r})|$, with
$|\chi_j({\bm r})\rangle$ being the eigenstates of $H_{\rm I}$ and $E_j$ the corresponding energy. With the diagonal bases of $H_I$, the
total Hamiltonian after the transformation can be obtained by
\begin{eqnarray}
H'&=&U^\dagger HU\nonumber\\
&=&\frac{1}{2m}\left[i\hbar\nabla+{\bm A}({\bm r})\right]^2+\widetilde{V}({\bm r})+H^{\rm diag}_{\rm I},
\end{eqnarray}
where $\widetilde{V}= U^\dagger VU$ and the gauge $\bold A$ is a $SU(N)$ Berry's connection, taking the $N\times N$ matrix form, and is introduced by
 \begin{equation}
 {\bm A}({\bm r})=i\hbar U^\dagger({\bm r})\nabla U({\bm r}).
 \end{equation}
 The non-Abelian gauge potential ${\bm A}({\bm r})$ is associated the $SU(N)$ Berry curvature
 \begin{equation}
 F_{\mu\nu}=\partial_\mu{\bm A}_\nu-\partial_\nu{\bm A}_\mu-\frac{i}{\hbar}[{\bm A}_\mu,{\bm A}_\nu],
 \end{equation}
 and the effective magnetic field ${\bm B}$ then given by $B_j=\frac{1}{2}\epsilon_{jkl}F_{kl}$. Note that in the above derivative we have not
 performed adiabatic approximation. In this stage $\bold A$ is a pure gauge given by the $SU(N)$ transformation and the
 Berry curvature is simply zero $F_{\mu\nu}=0$. Further applying the adiabatic condition yields adiabatic gauge potential which may lead to
 nontrivial SO couplings.

Consider that the ground state subspace of $H_I$ has $n$ ($n<N$) degenerate eigenstates. The off-diagonal couplings within the degenerate ground states are not negligible. However, the adiabatic condition can be satisfied and decouple the ground state subspace from excited levels when the ground state energy, $E_g$, is well separated from those of the remaining $N-n$ states, namely, when $|{\bm v}\cdot{\bm A}_{ij}|\ll|E_g-E_j|$ holds for
$i=1,2\cdots,n$ and $j=n+1,n+2,\cdots,N$. In this case, we neglect the couplings between the ground state subspace and excited states and reach the reduced column vector of wave functions only for the degenerate subspace as
$\bar{\Psi}=(\Psi_1,\Psi_2,\cdots,\Psi_n)^{\sf T}$, which satisfies
\begin{equation}
i\hbar\frac{\partial\bar{\Psi}}{\partial t}=\left[\frac{1}{2m}\left(-i\hbar\nabla-{\bm A}^{(n)}\right)^2+V_{\rm eff}\right]\bar{\Psi},
\end{equation}
where ${\bm A}^{(n)}$ is a $n\times n$ matrix reduced from the pure gauge $\bold A$, rendering a $U(n)$ non-Abelian gauge potential if $n>1$.
The effective trapping potential matrix $V_{\rm eff}$ is given by $V_{\rm eff,jl}=E_j\delta_{jl}+\langle\chi_j({\bm r})|V|\chi_l({\bm r})\rangle+\Phi_{jl}$, with the scalar potential $\Phi$ being given by
\begin{equation}
\Phi_{jk}=\frac{1}{2m}\sum_{l>n} {\bm A}_{jl}\cdot{\bm A}_{lk}.
\end{equation}
In the case of $n=1$, i.e. the ground state is non-degenerate, and the reduced adiabatic gauge potential becomes Abelian.

\begin{figure}[btp]
\centering
\includegraphics[width=0.99\columnwidth]{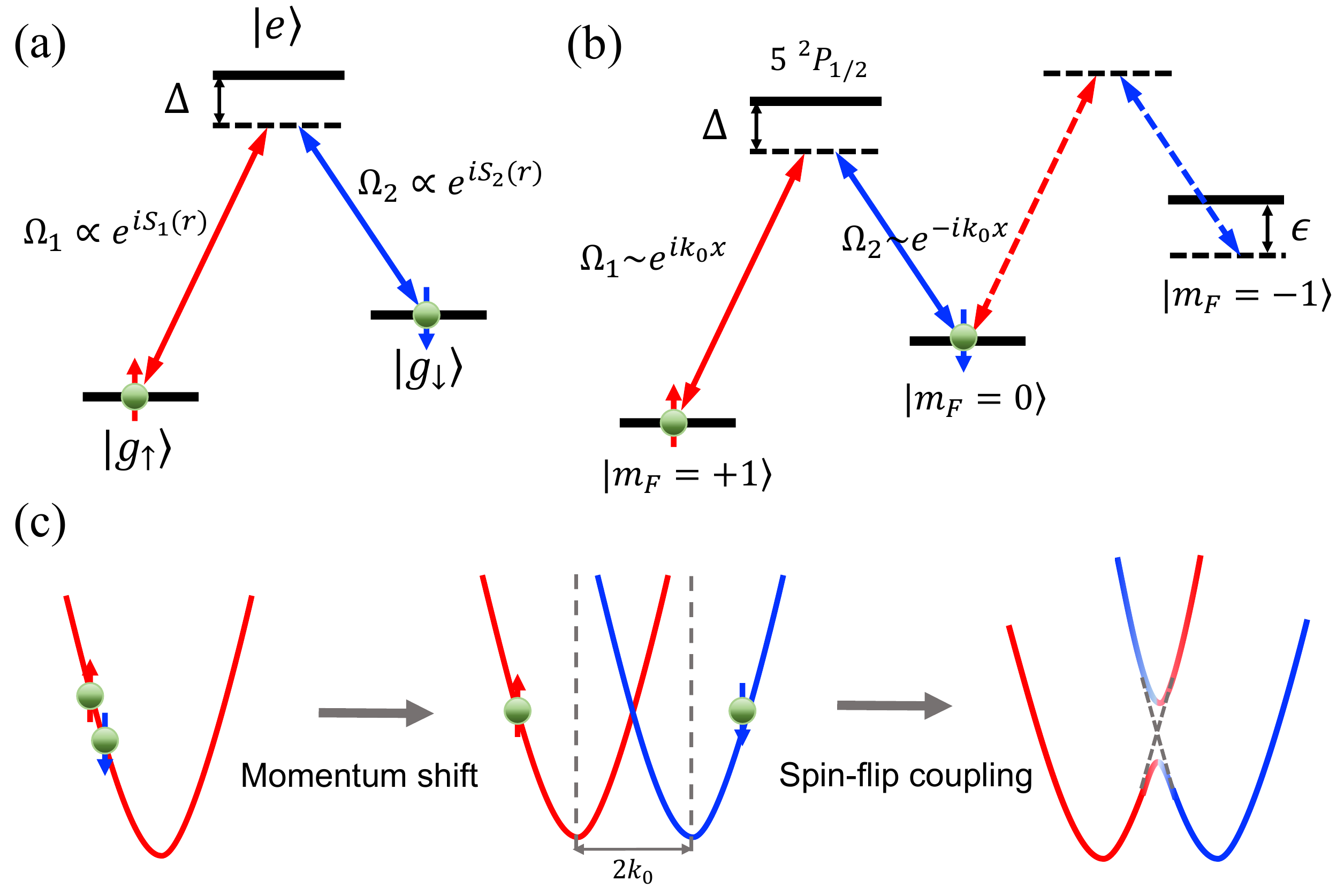}
\caption{ A minimal
scheme for 1D SO coupling. (a) The $\Lambda$-type configuration, with the atom levels (dubbed spin up and spin-down) coupled to two light beams. (b) A practical layout for $^{87}$Rb atoms, coupling to two beams propagating oppositely in $x$ direction.
The $F=1$ manifold is employed with one hyperfine state well separated by a quadratic Zeeman shift $\epsilon$.
(c) An intuitive picture for the 1D SO coupling. The two-photon Raman coupling imprints a momentum difference to the two spin states due to momentum transfer in the two-photon process, and induce a spin-flip transition which opens a Zeeman gap at zero momentum. The whole effects generate a 1D SO coupling together with a Zeeman term.
} \label{fig_1dsoc}
\end{figure}

{\it The Lambda-type configuration}.--A minimal scheme to realize SO coupling by generating spin-dependent gauge potential was proposed based on a $\Lambda$-type configuration~\cite{Liu2009}. As shown in Fig.~\ref{fig_1dsoc}(a),
the there-level atoms are coupled to
two laser beams, with the interaction Hamilonian
\begin{equation}
H_{\rm I}=\hbar\Delta-(\hbar\Omega_1|g_{\uparrow}\rangle\langle e|+\hbar\Omega_2|g_{\downarrow}\rangle\langle e|+{\rm h.c.}),
\end{equation}
where $\Omega_1=\Omega\sin\theta e^{iS_1}$ and $\Omega_2=\Omega\cos\theta e^{iS_2}$ with $\Omega=\sqrt{|\Omega_1|^2+|\Omega_2|^2}$, with $S_{1,2}$ being determined by the phases of laser beams.
When the detuning $\Delta$ is much larger than the coupling strength $\Omega$, the above Hamiltonian has two
nearly degenerate ground states
\begin{eqnarray}
|\chi_1(\bm r)\rangle&=&\cos\theta e^{iS_2({\bm r})}|g_{\uparrow}\rangle-\sin\theta e^{iS_1({\bm r})}|g_\downarrow\rangle,\\
|\chi_2(\bm r)\rangle&\simeq&\sin\theta e^{iS_2({\bm r})}|g_{\uparrow}\rangle+\cos\theta e^{iS_1({\bm r})}|g_\downarrow\rangle,
\end{eqnarray}
which make up a spin-$1/2$ system providing the realization of SO coupling with proper spatial-dependent phases $S_{1,2}(\bold r)$. The following is an alternative equivalent picture. Due to the condition that $|\Delta|\gg\Omega$, the single-photon transitions between
 the ground states $|g_{\uparrow,\downarrow}\rangle$ and the excited state $|e\rangle$ are greatly suppressed, while the two-photon transition between the two
 ground states becomes dominant. In this regime, by adiabatically removing the excited state the system is effectively regarded as a two-state ($|g_{\uparrow}\rangle$ and $|g_\downarrow\rangle$) configuration coupled by a two-photon Raman process. Consider a simple case where the laser fields are counter-propagating along $x$-direction so that the Rabi frequencies $\Omega_1=\Omega_0e^{ik_0x}$ and $\Omega_2=\Omega_0e^{-ik_0x}$ [see Fig.~\ref{fig_1dsoc}(b)], where $k_0$ denotes the wave number of the two lasers.
 As mentioned above, when $|\Delta|$ is large enough, one can adiabatically eliminate the excited state and obtain the effective spin-1/2 Hamiltonian in $x$-direction
 \begin{equation}
 H_{\rm eff}=\left(
 \begin{array}{lr}
 \frac{\hbar^2k_x^2}{2m}+\frac{\delta}{2} & \frac{\Omega_{\rm R}}{2}e^{i2k_0x}   \\
 &\\
\frac{\Omega_{\rm R}}{2}e^{-i2k_0x}  & \frac{\hbar^2k_x^2}{2m}-\frac{\delta}{2}
 \end{array} \right),
 \end{equation}
 where $\delta$ is a small two-photon detuning for the Raman coupling and the Raman Rabi frequency $\Omega_{\rm R}=\Omega_0^2/\Delta$.  After the transformation $U=\exp(-ik_0x\sigma_z)$,
 the Hamiltonian renders a 1D SO coupling (with equal amplitudes of Rashba and Dresselhaus terms)
 \begin{equation}\label{1DSOC}
 H_{\rm 1D}=UH_{\rm eff}U^\dagger=\frac{\hbar^2(k_x+k_0\sigma_z)^2}{2m}+\frac{\delta}{2}\sigma_z+\frac{\Omega_{\rm R}}{2}\sigma_x.
 \end{equation}
 From the above SO Hamiltonian one can see that the generated gauge potential $\bold A=\hbar k_0\sigma_z\hat e_x$ is spin-dependent, but still Abelian, rather than non-Abelian.

The SO coupling can naturally emerge from the view of momentum transfer, as illustrated in Fig.~\ref{fig_1dsoc}(c). Due to
the laser polarization, the transition from $|g_{\uparrow}\rangle$ to
$|g_{\downarrow}\rangle$ or the other way round amounts to absorbing a photon from one laser and
then emitting a photon to the other, accompanied with momentum transfer by $2k_0$. To be specific, the atomic momentum
increases by $k_0$, the so-called recoil momentum, after absorption of one photon; it gains another $k_0$ in
the subsequent emission since the two coupling beams are counter-propagating. As a result, by Raman process, the spectra of atoms in
the two states $|g_{\uparrow}\rangle$ and $|g_{\downarrow}\rangle$ exhibit a relative $2k_0$-momentum shift along $k_x$ direction [the former two pictures in Fig.~\ref{fig_1dsoc}(c)]. Furthermore, a finite Raman coupling strength leads to a spin-flip transition, which corresponds to the $\sigma_x$ term in Eq.~\eqref{1DSOC} and opens a gap at $k_x=0$, eventually leading to the spectra shown in the last picture of Fig.~\ref{fig_1dsoc}(c). It is the spin-flip transition associated with momentum transfer that accounts for the basic mechanism of generating SO couplings.

It is noteworthy that simply applying another pair of laser beams along say $\pm y$ directions to induce additional Raman coupling and momentum transfer along $y$ direction cannot yield 2D SO coupling. Instead, in this case, the both Raman couplings along $x$ and $y$ axes will combine and induce the 1D SO coupling along $\hat e_x+\hat e_y$ direction. As we shall see later, to realize high-dimensional SO couplings based on $\Lambda$-type configurations, one needs to consider optical lattices.

\begin{figure}[b]
\centering
\includegraphics[width=0.6\columnwidth]{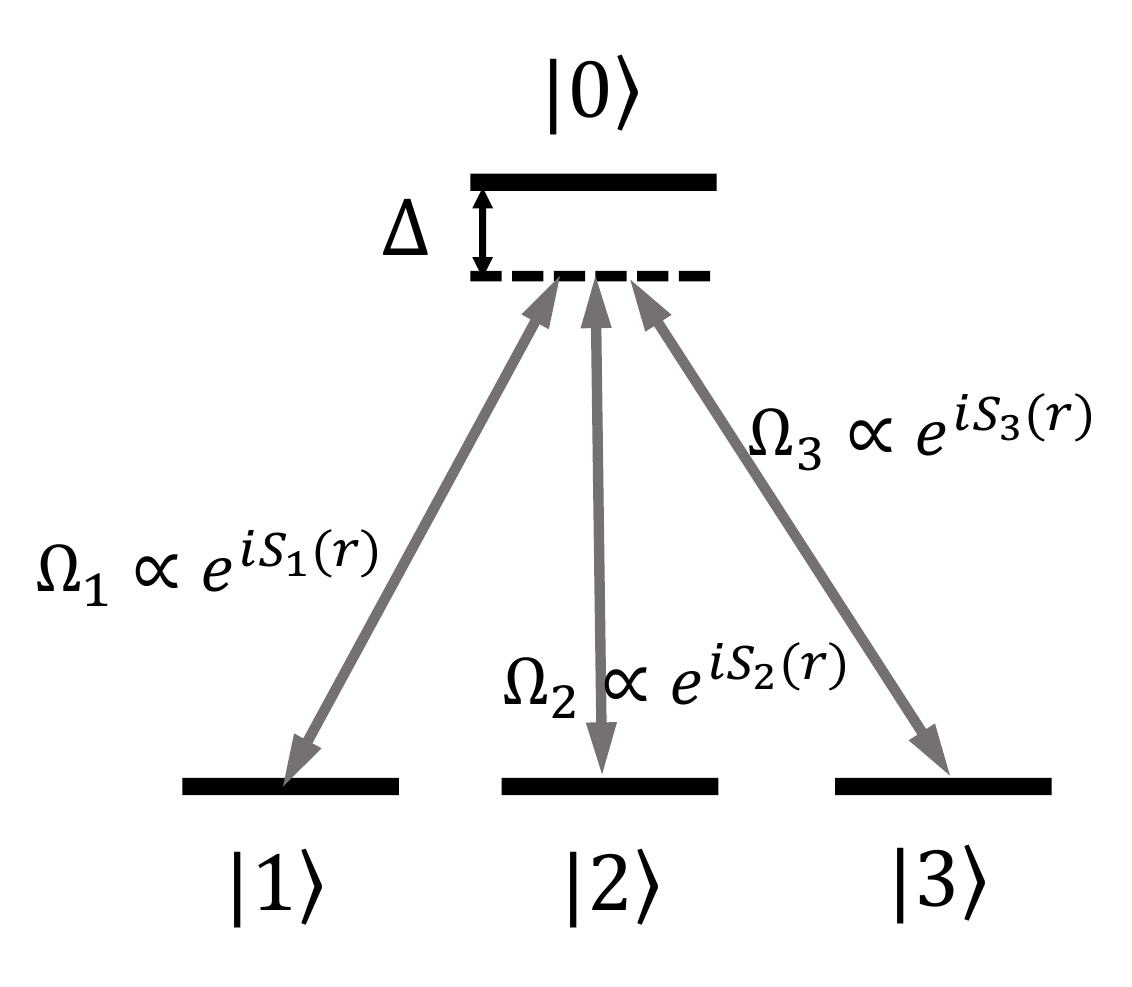}
\caption{The tripod configuration for non-Abelian gauge fields. The three ground states couple to three beams which drive transitions to the excited level.
} \label{fig_tripod}
\end{figure}

{\it The tripod-configuration}.--An early scheme to generate non-Abelian gauge potentials is based on a tripod-type configuration with four-level atoms coupled to spatially varying laser fields~\cite{Ruseckas2005}, as depicted in Fig.~\ref{fig_tripod}(a). The atom-light coupling Hamiltonian reads
\begin{equation}
H_{\rm I}=\hbar\Delta|0\rangle\langle0|-\hbar\sum_{j=1}^3(\Omega_1|0\rangle\langle j|+{\rm h.c.}),
\end{equation}
where the Rabi frequencies $\Omega_j$ characterize the corresponding atomic transition amplitudes and are defined by $\Omega_1=\Omega\sin\theta\cos\phi e^{iS_1}$, $\Omega_2=\Omega\sin\theta\sin\phi e^{iS_2}$,
and $\Omega_3=\Omega\cos\theta e^{iS_3}$, with $\Omega=\sqrt{|\Omega_1|^2+|\Omega_2|^2+|\Omega_3|^2}$. The above Hamiltonian has two degenerate dark states
\begin{eqnarray}
|\chi_1\rangle&=&\sin\phi e^{iS_{31}}|1\rangle-\cos\phi e^{iS_{32}}|2\rangle, \\
|\chi_2\rangle&=&\cos\theta\cos\phi e^{iS_{31}}|1\rangle+\nonumber\\
&&+\cos\theta\sin\phi e^{iS_{32}}|2\rangle-\sin\theta|3\rangle,
\end{eqnarray}
with $S_{ij}=S_i-S_j$. One can embed spatial dependence into dark states by using standing-wave fields (associated with the angles $\phi$ and $\theta$)
or non-parallel propagating running waves (associated with the relative phases $S_{ij}$) or vortex beams with orbital angular momentum (phases $S_{ij}$). The two
dark states form a pseudospin-$1/2$ system, for which a U(2) non-Abelian gauge potential $\bold A^{(2)}$ can arise in the adiabatic motion, with the elements $\bold A_{jk}^{(2)}=i\hbar\langle\chi_j|\nabla|\chi_k\rangle$ given by
\begin{eqnarray}\label{eqn:gauge13}
\bold A_{11}^{(2)}&=&\hbar(\cos^2\phi\nabla S_{23}+\sin^2\phi\nabla S_{13}),\nonumber\\
\bold A_{22}^{(2)}&=&\hbar\cos^2\theta(\cos^2\phi\nabla S_{13}+\sin^2\phi\nabla S_{23}),\\
\bold A_{12}^{(2)}&=&\hbar\cos\theta[{1}/{2}\sin(2\phi)\nabla S_{12}-i\nabla\phi].\nonumber
\end{eqnarray}
Properly choosing the parameters $\theta, \phi$ and/or $S_{ij}$ can readily yield Rashba or Dresselhaus type SO couplings. For example, let $\theta=\phi=\pi/4,S_{12}=k_0x, S_{23}=k_0x+k_0y$, and $S_{13}=-k_0x+k_0y$. The gauge potential becomes
\begin{eqnarray}\label{eqn:gauge14}
\bold A^{(2)}=(\sqrt{2}\hbar/4)k_0\sigma_x\hat e_x+(\hbar/4)k_0\sigma_z\hat e_y,
\end{eqnarray}
where a trivial constant has been neglected. The above gauge potential is of $U(2)$ form and leads to a 2D SO coupling. The tripod scheme provides a natural realization of non-Abelian gauge potentials for an effective pseudospin-$1/2$ system, while it also suffers challenges in experimental study. For example, the tripod configuration includes three ground states $|j\rangle$ ($j=1,2,3$), with one of the three states having to be a metastable state which may have relatively short lifetime, particularly in the presence of interactions~\cite{Gediminas2017}. On the other hand, the realization of tripod scheme necessitates the resonant Raman couplings between each two of the three ground states, which is vulnerable to the fluctuation of external magnetic field which is applied for experiments of generation SO couplings. Finally, unlike the $\Lambda$ scheme, in the tripod system the SO coupling is defined for pseudospin states (i.e. dark states), rather than real atom spins (hyperfine eigenstates), which could be hard to be directly measured or engineered in experiments. As a result, while the first proposal of tripod scheme was introduced over a decade ago~\cite{Ruseckas2005}, it was successfully demonstrated in experiment only recently by Shanxi University in collaboration with CUHK~\cite{Huang2016}.

\subsection{Gauge potential in optical lattices}

In optical lattices the motion of atoms can be described by hopping between lattice sites. Accordingly, the SO coupling in optical lattices corresponds to the spin-flip hopping between neighboring sites, similar to the spin-flip transition associated with momentum transfer in the continuum gas. For example, for a 2D square optical lattice, the tight-binding model for the case with a Rashba SO coupling can be written as
\begin{eqnarray}\label{eqn:latticegauge1}
H=\sum_{\langle\vec j,\vec l\rangle,s,s'}\bigr[-t_0c_{j,s}^\dag (e^{i\frac{\theta}{2}\bold A\cdot\vec d_{jl}})_{ss'} c_{l,s'}+h.c.],
\end{eqnarray}
where $\bold A=(\sigma_y,-\sigma_x)$, $c_{l,s}$ and $c_{l,s}^\dag$ are annihilation and creation operators, respectively, $\langle\vec j,\vec l\rangle$ represents the hopping between nearest-neighbor sites, and $\vec d_{jl}$ denotes the unit vector from $\vec l$-site to $\vec j$-site. The above tight-binding Hamiltonian describes a spin-conserved hopping with coefficient $t_s=t_0\cos\theta/2$ along $x$ and $y$ directions, and spin-conserved hoppings with coefficient $t_{\rm so}^x=t_0\sin\theta/2$ ($t_{\rm so}^y=-t_0\sin\theta/2$) coupling to $\sigma_y$ ($\sigma_x$) along $x$ ($y$) direction. Transforming the Hamiltonian into $\bold k$ space yields $H=\sum_{\bold k}c_{\bold k,s}^\dag{\cal H}_{\bold k}^{ss'}c_{\bold k,s}$, with the Bloch Hamiltonian
\begin{eqnarray}\label{eqn:latticegauge2}
{\cal H}(\bold k)&=&-2t_s(\cos k_xa+\cos k_ya)\hat I\nonumber\\
&&-2t_{\rm so}\bigr[\sin k_xa \sigma_y-\sin k_y\sigma_x\bigr],
\end{eqnarray}
where $\hat I$ is a $2\times2$ identity matrix and $t_{\rm so}=t_0\sin\theta/2$. While a purely Rashba SO coupling has not been realized in experiment in an optical lattice, Engels' group first demonstrated the 1D SO coupling for bosons in optical lattices~\cite{Hamner2015}.

The recent studies show that instead of realizing a purely Rashba SO coupling, it is more natural to realize the Dirac-type 2D SO coupling based on the so called optical Raman lattice scheme~\cite{Liu2013a,Liu2014a}, where besides a 2D Rashba-type SO term induced by Raman coupling lattice, the spin-conserved hopping couples to the $\sigma_z$-component rather than identity matrix $\hat I$. In the lowest $s$-band regime, the Bloch Hamiltonian becomes
\begin{eqnarray}\label{eqn:latticegauge3}
{\cal H}(\bold k)&=&\bigr[m_z-2t_s(\cos k_xa+\cos k_ya)\bigr]\sigma_z\nonumber\\
&&-2t_{\rm so}\bigr[\sin k_xa \sigma_y-\sin k_y\sigma_x\bigr],
\end{eqnarray}
which describes a minimal quantum anomalous Hall (QAH) model driven by SO coupling and has been first realized by the collaborating team at PKU and USTC~\cite{Wu2016}. The optical Raman lattice schemes exhibit high feasibility in experimental realization with multiple advantages, and are becoming a prevailing technique in investigating novel high-dimensional SO effects and topological physics in experiment. The detailed discussions will be presented in the next sections.

\section{Experimental issues of realizing the 1D SO coupling}

\subsection{Realization of $\Lambda$-type configuration}

The $\Lambda$-type coupling scheme can be readily realized in real cold atom systems. Fig.~\ref{fig_1dsoc}(b) illustrates a typical hyperfine-level configuration in $^{87}$Rb bosons, with ground manifold $F=1$. The excited state $|e\rangle$ indeed includes all relevant states corresponding to $D_1$ and $D_2$ transitions. To separate the $\Lambda$-type configuration from other state, one can apply an external bias magnetic field $\bold B=B_z\hat e_z$ to split up the ground states. The energy shift is nonlinear due to the quadratic Zeeman splitting
 \begin{eqnarray}
\Delta E_{|F,m_F\rangle}=\mu_Bg_Fm_FB_z+\frac{(g_J-g_I)^2\mu_B^2}{2\hbar\Delta E_{\rm hfs}}B^2,
 \end{eqnarray}
where $g_{I,J,F}$ are Land\'{e} factors and $\Delta E_{\rm hfs}$ denote hyperfine splitting. As a result,
when two of the three ground states are coupled in two-photon resonance, they are detuned from the third state, and an isolated $\Lambda$-system is resulted. For example, with a bias field of strength $14$G, the Zeeman splitting between two neighboring hyperfine states $|m_F=0\rangle$ and $m_F=-1$ is about $10.2$MHz, which has a discrepancy of $8E_r$ with respect to the Zeeman splitting between $|m_F=0\rangle$ and $m_F=1$ if applying $\lambda=786$nm Raman beams [Fig.~\ref{fig_1dsoc} (b)]. With the simple scheme the 1D SO coupling can now be routinely realized in both Bose-Einstein condensates (BECs)~\cite{Lin_SOC,USTC2012}  and Fermi gases~\cite{Shanxi2012,MIT2012}.

\subsection{Cancellation of Raman couplings through D1 and D2 transitions}

Note that the net Raman coupling is obtained by taking into account the contributions through both the $D_1$ and $D_2$ lines. The total Raman coupling between two ground states, e.g. $|m_F=+1\rangle$ and $|m_F=0\rangle$ for $^{87}$Rb, is given by
\begin{eqnarray}\label{sum1}
\Omega_R=\sum_{F}\frac{\Omega_{2F,D_1}^*\Omega_{1F,D_1}}{\Delta}+\sum_{F}\frac{\Omega_{2F,D_2}^*\Omega_{1F,D_2}}{\Delta+E_s},
\end{eqnarray}
where $\Delta$ is one-photon detuning for $D_1$ line, $E_s$ is the fine-structure splitting, $\Omega_{1F,D_j}$ and $\Omega_{2F,D_j}$ represent the one-photon Rabi-frequencies induced by the two laser beams corresponding to the transitions from $|m_F=+1\rangle$ and $|m_F=0\rangle$ to an excited state of quantum number $F$ in the $D_j$ ($j=1,2$) lines, respectively.

Let $|e_{F,D_j}\rangle$ denote an excited state corresponding to the $D_j$ $(j=1,2)$ line. We can verify the following identity
\begin{eqnarray}
&&\sum_{F}\Omega_{2F,D_1}^*\Omega_{1F,D_1}+\sum_{F}\Omega_{2F,D_2}^*\Omega_{1F,D_2}\nonumber\\
&&=\sum_{F,j=1,2}\langle g_\uparrow|\bold d_{2F}\cdot\bold E_2|e_{F,D_j}\rangle\langle e_{F,D_j}|\bold d_{1F}\cdot\bold E_1|g_\downarrow\rangle,\nonumber\\
&&=\langle g_\uparrow|(\bold d_{2F}\cdot\bold E_2)(\bold d_{1F}\cdot\bold E_1)|g_\downarrow\rangle,\nonumber\\
&&=0,
\end{eqnarray}
where $\bold d_{jF}$ is the corresponding dipole vector. In the above derivative we have applied the identity that $\sum_{F,j}|e_{F,D_j}\rangle\langle e_{F,D_j}|=I$. Thus the couplings through the $D_1$ and $D_2$ lines in alkali atoms
contribute oppositely to the Raman transition. We obtain that
\begin{eqnarray}\label{sum1}
\Omega_R=\sum_{F}\frac{E_s}{\Delta(\Delta+E_s)}\Omega_{2F,D_1}^*\Omega_{1F,D_1}.
\end{eqnarray}

Assuming that the deunings for both $D_1$ and $D_2$ lines are in the same sign, we consider the following two situations. First, consider the red-detuned regime (similar for blue detunings) and when $\Delta<E_s$, we have from the above result that
\begin{eqnarray}
\Omega_R\propto|\Omega_1\Omega_2/\Delta|\propto(1/\tau_{\rm life})(\Delta/\Gamma),
\end{eqnarray}
where the lifetime satisfies $1/\tau_{\rm life}\propto|\Omega_j|^2\Gamma/\Delta^2$ and $\Gamma$ denotes the natural linewidth of the relevant
transition~\cite{grimm2000optical}. Secondly, when $|\Delta|\gg E_s$, we have
\begin{eqnarray}
\Omega_R\propto|\Omega_1\Omega_2|/\Delta^2\propto(1/\tau_{\rm life})(E_s/\Gamma).
\end{eqnarray}
In this case, the Raman coupling strength $\Omega_R$ and life time $\tau_{\rm life}$ cannot be enhanced at the same time by increasing $\Delta$ and Rabi-frequencies $\Omega_j$. Thus we have the following conclusions: 1) To induce an appreciable Raman coupling strength $\Omega_R$, the detuning $\Delta$ cannot be much larger than fine structure splitting $E_s$ of the excited states.
2) A large enough life time $\tau_{\rm life}$, however, requires that $\Delta$ should be much larger than $\Omega_j$.
3) For alkali atoms, the proper parameter regime is that $|\Delta|\sim E_s$. 4) The atomic candidates with large fine structure splitting $E_s$ are preferred for the generation of spin-orbit coupling.

It is noteworthy that, if the optical transitions are applied between $D_1$ and $D_2$ lines, namely $0<-\Delta<E_s$, the both $D_{1,2}$ transitions contribute in the same sign to the Raman coupling. This implies that the Raman coupling can be largely enhanced in this regime. Nevertheless, the magnitude of $\Delta$ is restricted by the fine-structure splitting $E_s$. The typical magnitudes of $E_s$ for alkali atoms are that $E_s\sim7.1$THz for $^{87}$Rb~\cite{DataRb}, $E_s\sim1.8$THz for $^{40}$K~\cite{DataKLi}, and $E_s\sim10$GHz for $^{6}$Li~\cite{DataKLi}, implying that the realization is favorable for $^{87}$Rb, marginally feasible for $^{40}$K, while suffers strong heating for $^{6}$Li atoms.

\subsection{Realization of 1D SO coupling for Bosons}

\begin{figure}[btp]
\centering
\includegraphics[width=0.75\columnwidth]{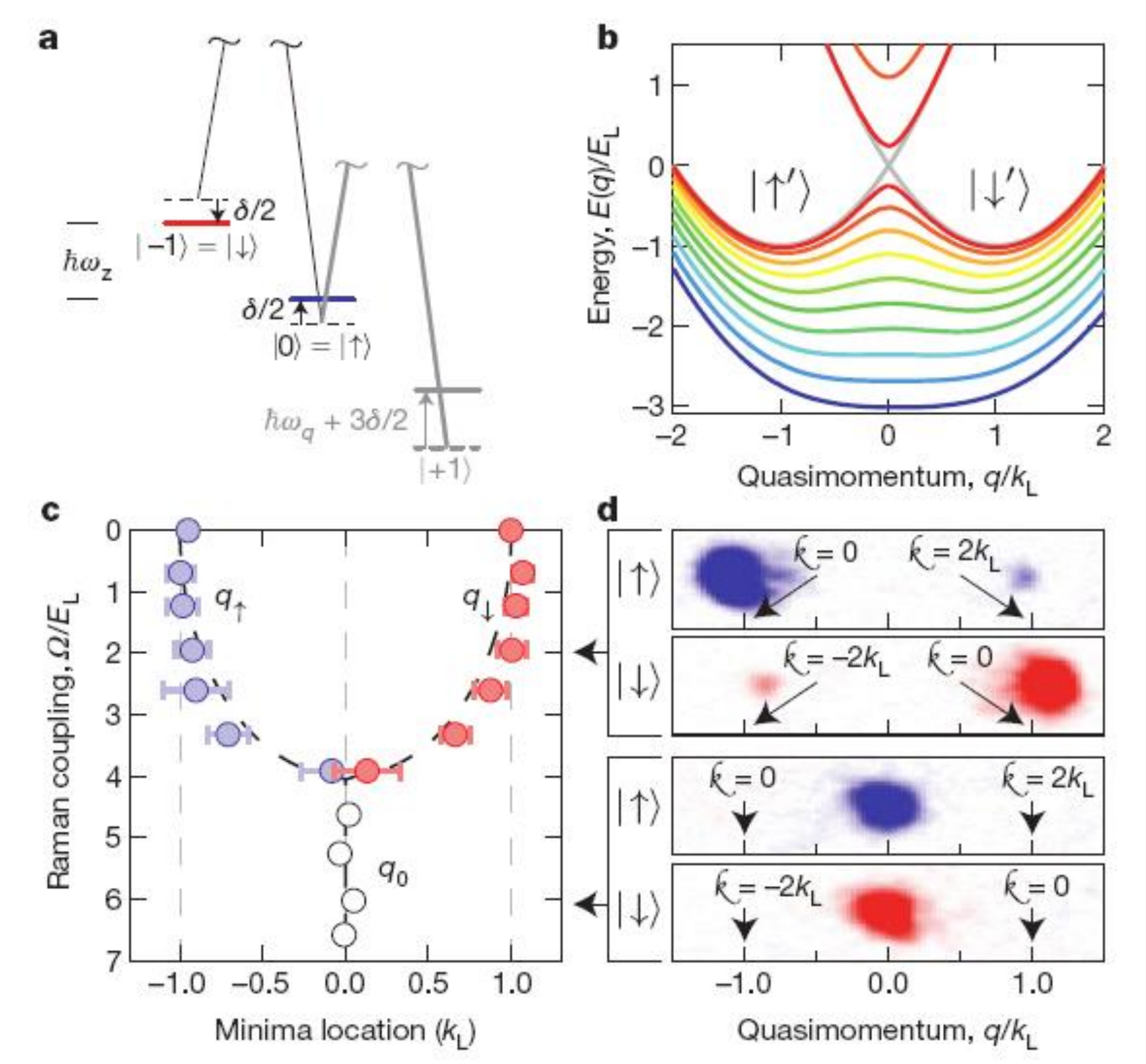}
\caption{The NIST experiment of the realization of 1D SO coupling with $^{87}$Rb bosons~\cite{Lin_SOC}. (a) The the Raman coupling configuration for the ground manifold. (b) The spectra of the SO coupled dressed states with different magnitudes of the Raman coupling strengths. (c) The band minima of the SO coupled spectra with different Raman coupling strengths. (d) The TOF imaging of atom clouds due to SO coupling.} \label{NIST}
\end{figure}
\begin{figure}[btp]
\centering
\includegraphics[width=0.99\columnwidth]{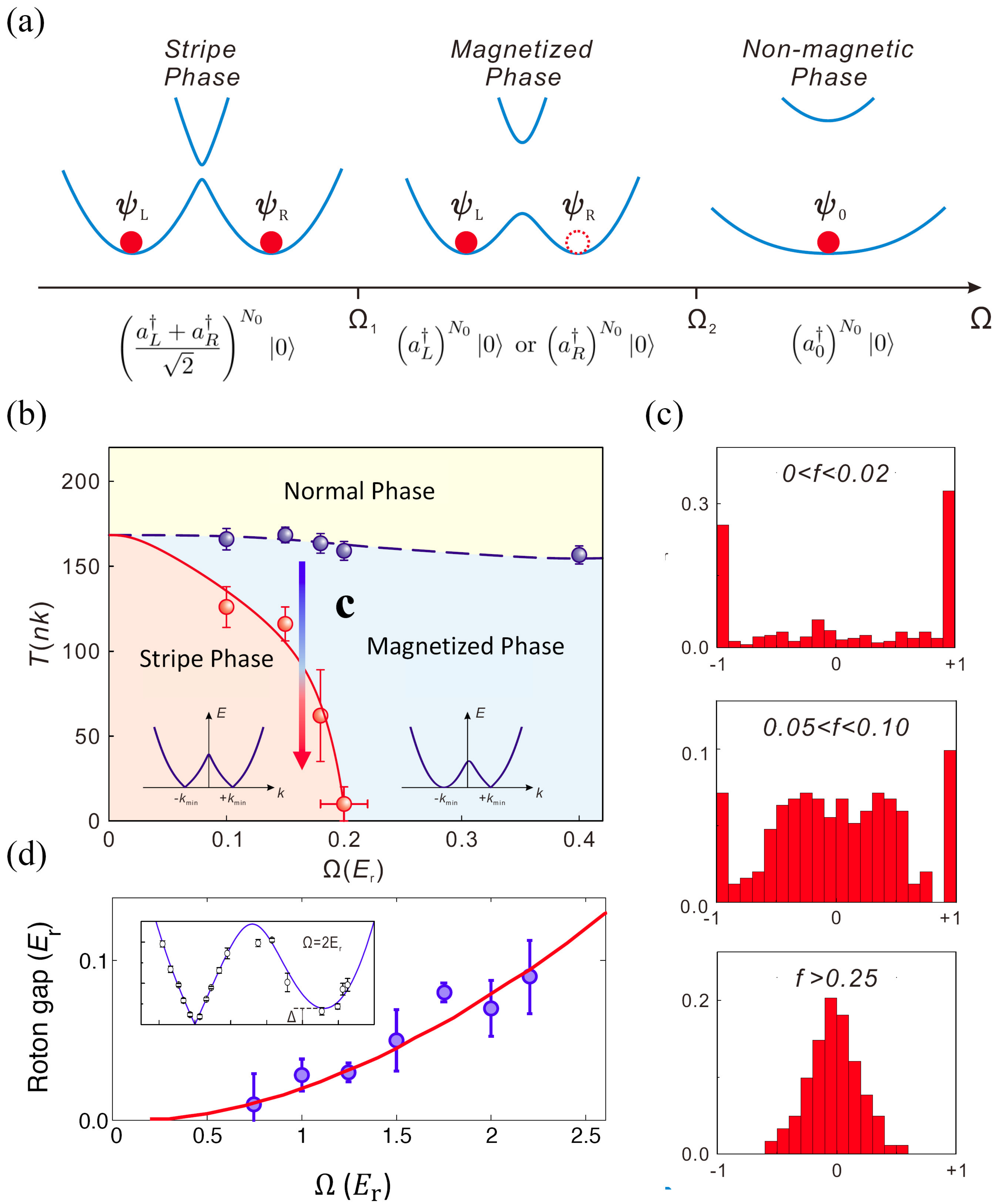}
\caption{Zero- and finite-temperature phase diagrams of $^{87}$Rb Bose gases with 1D SO coupling.
(a) Single-particle dispersion and zero-temperature phase diagram as a function of the Raman coupling $\Omega$, which shows a stripe phase, a magnetized phase and a non-magnetic phase as $\Omega$ increases.
(b) Finite-temperature phase diagram of spin-orbit coupled bosons. Insets: Schematic low-energy spectrum for the stripe and magnetized phases.
(c) Evolution of the magnetization histogram as the temperature is lowered (equivalent to increasing the condensate fraction $f$).
(d) Softening of roton mode. The measured roton gap becomes smaller as $\Omega$ decreases and the vanishing indicates the phase transition to the stripe phase.
Inset: The low-energy excitation spectrum, which clearly shows a roton-maxon structure.
} \label{fig_phases}
\end{figure}

Lin {\it et al.}~\cite{Lin_SOC} first reported the experimental realization of 1D SO coupling in a Bose-Einstein condensate of $^{87}$Rb atoms, and in particular,
demonstrated two quantum phase transitions driven by Raman coupling strength [Fig.~\ref{NIST} (a,d)]. One is two minima of the dressed energy dispersion merging into a single minimum [Fig.~\ref{NIST} (b,c)];
the other is a transition from a spatially spin-mixed state to a separated state. The latter one is induced by the modified
interactions between the two dressed states (the two minima of dispersion)~\cite{Zheng2013}.
The spin-mixed phase is also called ``stripe'' phase~\cite{Stringari2012a}, where atoms condense into a
superposition of the two minima [Fig.~\ref{fig_phases}(a)], thus exhibiting density fringes~\cite{Zhai2010,Wu2011,Ho2011}, while the separated phase is
known as ``plane-wave'' or ``magnetized'' phase since the condensation occurs at only one minimum [Fig.~\ref{fig_phases}(a)].

Ji {\it et al.}~\cite{USTC2014} extended the phase diagram of $^{87}$Rb BECs to the finite-temperature case
by observing the evolution of condensate magnetization with the temperature.
Unlike the ``magnetized" phase, the condensate in the stripe state should exhibit no magnetization at zero temperature; it is not always true in experiments due to
undesired circumstances (e.g. the forming of domain wall). However, the histogram of magnetization distribution should have a Gaussian-like peak around zero
magnetization after a large number of measurements, as shown in Fig.~\ref{fig_phases}(c).
By contrast, the histogram in the magnetized phase shows two sharp peaks around $\pm1$ [Fig.~\ref{fig_phases}(c)],
indicating the spontaneous $Z_2$-symmetry breaking in Bose condensation. Experimental measurements determines the finite-temperature phase diagram as in
Fig.~\ref{fig_phases}(b), where the stripe phase will first turn into the magnetized phase before becoming the normal state as the temperature increases.

Another related issue is about excitation spectrum. Because the stripe phase breaks both U(1) symmetry (superfluid phase) and translational symmetry, there will be two linear Goldstone modes in the spectrum~\cite{Stringari2013} [schematically shown in the insets of Fig.~\ref{fig_phases}(b)]. The magnetized phase, however, breaks only one continuous (phase) symmetry, and there is only one linear Goldstone mode. Furthermore,
the tendency towards periodic order (the stripe phase) near the transition indicates that the excitation spectrum in the magnetized phase has a roton-type mode~\cite{Zheng2013,Stringari2012b}, which was experimentally demonstrated by Ji {\it et al.}~\cite{USTC2015} [see inset of Fig.~\ref{fig_phases}(d)].
In Ref.~\cite{USTC2015}, roton-mode softening was observed by Bragg spectroscopy,
showing the roton gap is vanishingly small in the critical regime [Fig.~\ref{fig_phases}(d)]. Thus the magnetized phase has a greater low-energy density-of-states than the stripe phase, meaning the magnetized phase can gain more entropy from thermal fluctuations and become more favorable. This gives an explanation for why the phase boundary bends towards the stripe phase side in the finite-temperature phase diagram [Fig.~\ref{fig_phases}(b)].


\subsection{Realization of SO coupling for Fermions}

Wang {\it et al.}~\cite{Shanxi2012} at Shanxi group and Cheuk {\it et al.}~\cite{MIT2012} at MIT group respectively realized 1D SO coupling in atomic Fermi gases using $^{40}$K and $^6$Li atoms. The 1D SO coupling induced momentum transfer along $x$ direction leads to a spin-momentum lock along the this direction. This spin imbalance was observed by time-of-flight (TOF) expansion in the experiment with $^{40}$K atoms. On the other hand, note that $^6$Li fermions have a tiny fine structure splitting in excited levels. The strong heating brings about serious challenging in realizing the SO coupling with degenerate $^6$Li gas. Instead, the MIT group observe the SO coupling induced spin imbalance through spin injection radio-frequency (rf) spectroscopy. During the rf spectroscopy, the atoms are prepared in the reservoir states which are not involved in realizing the SO coupling. Then, a low-power rf pulse was applied to pump the atoms at a certain reservoir state to the target spin-$1/2$ subspace with the 1D SO coupling generated by Raman couplings. Being a function of the spin polarization of dressed states, from the resonant pumping rate of rf spectroscopy one can read out both the band structure and spin-polarization of SO coupled system.
Since the realization, SO coupling induced novel superfluids has been a hot issue. In particular, the interplay between SO coupling and strong interactions (BEC-BCS crossover) has been analyzed theoretically~\cite{Vyasanakere2011,Hu2011,YuZQ2011,He2012} and experimentally~\cite{Williams2013,Huang2015}.

The 1D SO coupling has also been realized in fermionic lanthanide and alkali earth atoms, including Dy fermions by
Lev's group~\cite{Burdick2016}, Yb atoms by Jo's group at HKUST~\cite{Song2016} and by Fallani's group~\cite{Livi2016}, and Sr atoms by Ye's group at JILA~\cite{Kolkowitz2017}. Compared with alkali atoms, the lanthanide and alkali earth atoms have an effective large fine structure splitting $E_s$ while a small natural linewidth of transition $\Gamma$. As a result, the SO coupling generated for such atom candidates in principle can have a long lifetime. In particular, for the $^{161}$Dy fermions, a lifetime up to $400$ms was observed in the experimental realization~\cite{Burdick2016}. Such life is limited by dipolar decay of the spin, rather than by light-induced heating.

%
%



\section{SO couplings beyond one dimension}

\begin{figure}[btp]
\centering
\includegraphics[width=0.75\columnwidth]{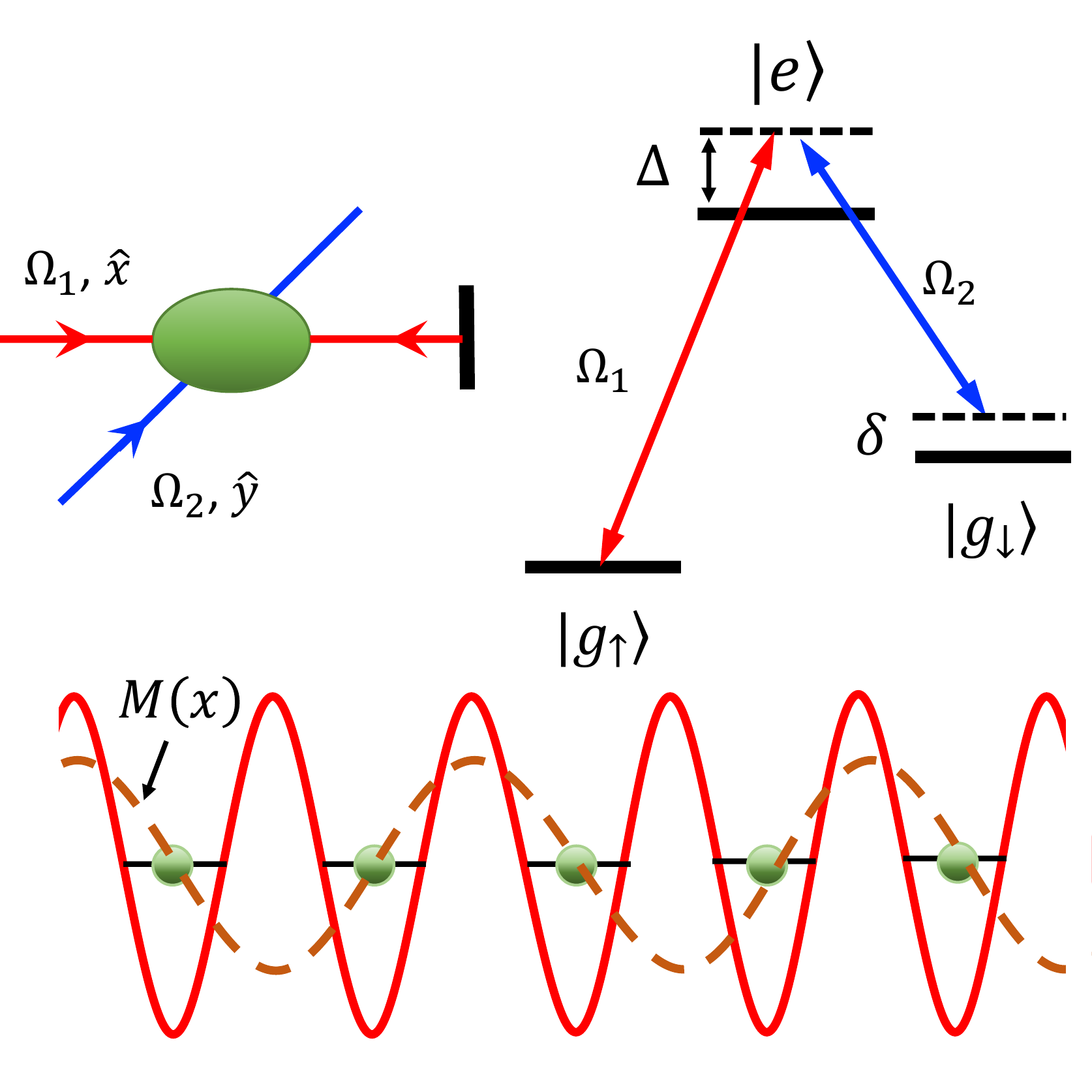}
\caption{The Raman lattice scheme in 1D case. Atoms are trapped in a 1D
optical lattice with an internal three-level $\Lambda$-type configuration
coupled to radiation by a standing wave (red line) and a running wave (blue line). The induced Raman potential $M(x)$ is
anti-symmetric with respect to each lattice site.} \label{fig_1dRaman}
\end{figure}

Until 2016, only the 1D SO couplings had been reported in experiments. Realization of SO couplings beyond 1D regime is much more important for quantum simulation~\cite{Zhang2008,Sato2009,Sau2011,Goldman2010,ZhuSL2011,He2012,Seo2012,Cui2014,Xu2015}. It is because a high-dimensional SO coupling (e.g. the 2D Rashba term) corresponds to a non-Abelian gauge potential, which is associated with nonzero Berry's curvature and nontrivial geometric or topological effects, while 1D SO coupling is an Abelian potential and does not give a nonzero Berry curvature. The study of broad classes
of novel topological states necessitates the realization of high-dimensional SO couplings, such as the 2D and 3D topological insulating states, Weyl semimetals, and topological superfluid phases. The previous schemes for realizing 2D and 3D SO couplings include the tripod scheme~\cite{Ruseckas2005} and its multipod extension~\cite{Juzeliunas2010}, ring-structure coupling scheme~\cite{Anderson2013a}, and realization with magnetic pulses~\cite{Anderson2013b,Xu2013}. Nevertheless, except for the tripod scheme successfully demonstrated in the recent experiment~\cite{Huang2016}, most of these schemes are hard to achieve in real cold atom experiments.
%
%

{\it Optical Raman lattice.}--Recently, the so called optical Raman lattice schemes are proposed to realize high-dimensional SO couplings and novel topological quantum phases. The essential idea of the optical Raman lattice is that one combines the realization of conventional optical lattice, which governs the normal hopping of atoms, and the periodic Raman lattice, which determines spin-flip hopping couplings, in the system. Moreover, such realization of optical and Raman lattices are achieved through transitions induced by {\it the same} optical fields. A consequence of this combination is that the generated conventional optical lattice and Raman lattice exhibit an automatically fixed relative spatial configuration, satisfying certain intrinsic relative space symmetries without any fine tunings. It turns out that, with such space symmetries, not only the realization of high-dimensional SO couplings can be greatly simplified, but also the resulted SO coupled systems naturally host various types of topological phases. The experimental realizations of optical Raman lattices have been successfully achieved in 1D and 2D regimes. In the following we shall review the details of optical Raman lattice schemes, and the recent important experimental progresses.

\subsection{Optical Raman lattice: 1D case}

\subsubsection{Model}

To illustrate how the optical Raman lattice scheme works, we first consider 1D regime, as proposed in Ref.~\cite{Liu2013a,Liu-note}. As shown in Fig.~\ref{fig_1dRaman}, the 1D blue-detuned optical Raman lattice is formed through a simple $\Lambda$-type configuration, with a standing-wave beam $\Omega_1$ (red line, along $x$ direction)
and a running wave beam $\Omega_2$ (blue line, along $y$ direction) being applied~\cite{Liu2013a}. The standing-wave beam applies to both the spin-up $|g_\uparrow\rangle$ and spin-down $|g_\downarrow\rangle$ states. For linearly polarized leaser beams, the standing wave field generates a 1D optical lattice which is spin-independent. Namely, the lattice potential reads $V_{\rm latt}(x)=V_0\cos^2k_0x$, with $V_0\propto|\Omega_1|^2/\Delta$. Further applying the running wave beam generates the Raman coupling potential through the two photon process, given by $M(x)=M_0\cos k_0x$, with the amplitude $M_0\propto|\Omega_1\Omega_2|/\Delta$. All the irrelevant phase factors, including the initial phases of beams and $e^{ik_0y}$ in $\Omega_2$, are ignored. The realized 1D Hamiltonian reads
\begin{equation}
H_{\rm 1D}=\frac{\hbar^2k_x^2}{2m}+V_0\cos^2k_0x+M_0\cos k_0x\sigma_x+\frac{\delta}{2}\sigma_z,
\end{equation}
where the tunable two-photon detuning $\delta$ plays the role of an effective Zeeman splitting. Before proceeding, we give a few remarks on the above realization. First, the whole setting applies only two laser beams which can be generated from a single laser source, and no fine tuning is required. Secondly, both the optical lattice and Raman lattice are generated through the same standing wave beam $\Omega_1$. This leads to a fixed relative configuration between $V_{\rm latt}$ and $M(x)$, namely, a) the periodicity of $M(x)$ is one-half of that of $V_{\rm latt}$; b) the Raman potential is antisymmetric with respect to each lattice site of $V_{\rm latt}$. This relative configuration is topologically stable, since any phase fluctuations, if existing in the laser beams, only lead to global shift of the optical Raman lattice, but cannot affect the relative configuration. We shall see that with this relative space symmetry the optical Raman lattice can naturally realizes nontrivial topological phases.

\begin{figure*}[t]
\includegraphics[width=1.8\columnwidth]{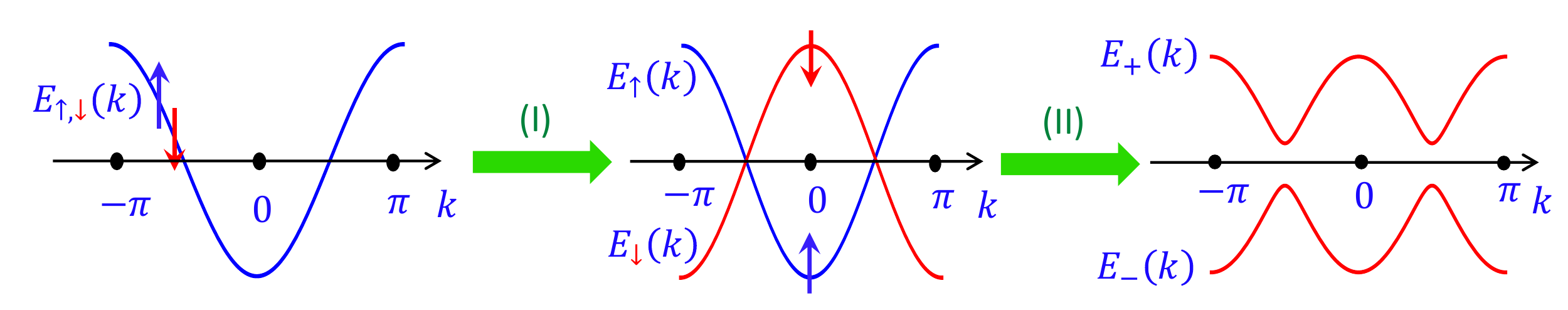}
\caption{(Color online) Illustration of band structure of 1D optical Raman lattice in the $s$-band regime. (I) The $\pi/a$-momentum transfer effectively reverses the sign of hopping coefficient of one spin state (spin-up); (II) The SO coupling opens a topological gap, leading to a 1D AIII class topological insulator.}\label{1Dpicture}
\end{figure*}
The topological physics can be best seen with the tight-binding model, while we emphasize that the following results are not restricted in the tight-binding regime. Consider the lowest $s$-band model, the tight-binding Hamiltonian generally takes the form
\begin{eqnarray}\label{TB1d0}
H_{\rm TB}&=&\sum_{\langle i,j\rangle,\sigma}t_0^{ij}c^{\dagger}_{i\sigma}c_{j\sigma}+\sum_i\frac{\delta}{2}(n_{i\uparrow}-n_{i\downarrow}) \nonumber\\
&+&\sum_{\langle i,j\rangle}(t_{\rm so}^{ij}c^\dagger_{j\uparrow}c_{j+1\downarrow}+{\rm h.c.}),
\end{eqnarray}
where $c_{j\sigma}$ ($c_{j\sigma}^\dagger$) are the annihilation (creation) operators of $s$-orbit for
spin $\sigma=\uparrow,\downarrow$ at lattice site $j$ and the atom number operator
$n_{j\sigma}\equiv c_{j\sigma}^\dagger c_{j\sigma}$. Here the spin-conserved hopping couplings $t_0^{ij}$
are induced by the lattice potential, given by
\begin{equation}
t_0^{ij}=\int dx\phi_{s\sigma}^{(i)}(x)\left[\frac{\hbar^2k_x^2}{2m}+V_{\rm latt}(x)\right]\phi_{s\sigma}^{(j)}(x),
\end{equation}
with $\phi_{s\sigma}^{(j)}(x)$ being the Wannier functions for s-bands,
and the Raman-coupling driven spin-flip hopping coefficients $t_{\rm so}^{ij}$ are
\begin{equation}
t_{\rm so}^{ij}=\int dx\phi_{s\uparrow}^{(i)}(x)M(x)\phi_{s\downarrow}^{(j)}(x).
\end{equation}
Since the wavefunctions $\phi_{s\sigma}^{(i)}(x)$ are spin-independent (due to the spin-independent lattice) and
satisfy $\phi_{s\sigma}^{(j)}(x)=\phi_{s\sigma}^{(0)}(x-x_j)$, it follows that $t_0^{ij}=-t_0$ with
\begin{equation}
t_0=-\int dx\phi_{s}^{(0)}(x)\left[\frac{\hbar^2k_x^2}{2m}+V_{\rm latt}(x)\right]\phi_{s}^{(0)}(x-a).
\end{equation}
On the other hand, from the relative space antisymmetry of Raman coupling potential, one can find that
\begin{eqnarray}
t_{\rm so}^{j,j\pm1}=\pm(-1)^jt_{\rm so},
\end{eqnarray}
with
\begin{equation}\label{tso}
t_{\rm so}=M_0\int dx\phi_{s}^{(0)}(x)\cos(k_0x)\phi_{s}^{(0)}(x-a).
\end{equation}
The staggered property of the spin-flip hopping can be absorbed by a gauge transformation that $c_{j\downarrow}\to-e^{i\pi x_j/a}c_{j\downarrow}$, where $a$ denotes the lattice constant.
The tight-binding model is finally written as
\begin{eqnarray}\label{TB1d}
H_{\rm TB}&=&-t_0\sum_{\langle i,j\rangle}(c^{\dagger}_{i\uparrow}c_{j\uparrow}-c^{\dagger}_{i\downarrow}c_{j\downarrow})+\sum_i\frac{\delta}{2}(n_{i\uparrow}-n_{i\downarrow}) \nonumber\\
&+&\left[\sum_jt_{\rm so}(c^\dagger_{j\uparrow}c_{j+1\downarrow}-c^\dagger_{j\uparrow}c_{j-1\downarrow})+{\rm h.c.}\right].
\end{eqnarray}
It can be seen that the gauge transformation effectively reverses the sign of the hopping coefficient of spin-down states. Transforming the Hamiltonian to momentum space yields (with $m_z=\frac{\delta}{2}$)
\begin{eqnarray}
H_{\rm TB}&=&\sum_{k}c_{k,s}^\dag{\cal H}_{ss'}( k_x)c_{\bold k,s'},\\
{\cal H}(k_x)&=&(m_z-2t_0\cos k_xa)\sigma_z+2t_{\rm so}\sin k_xa\sigma_y.\nonumber
\end{eqnarray}
From the above results one can see that the Raman coupling has two novel effects: a) it induces $\pi/a$-momentum transfer between spin-up and spin-down states; b) it further induces the SO coupling. This is a direct consequence of the relative configuration between Raman and optical lattices. The two effects can also be pictorially described in Fig.~\ref{1Dpicture}. In particular, the $\pi/a$-momenta transfer effectively reverses the sign of hopping coefficients of one spin states (say spin-up), and the remaining SO coupling further opens a gap, giving an insulating phase with nontrivial topology.

\subsubsection{1D AIII class topological insulator}

\begin{figure}[b]
\includegraphics[width=1.0\columnwidth]{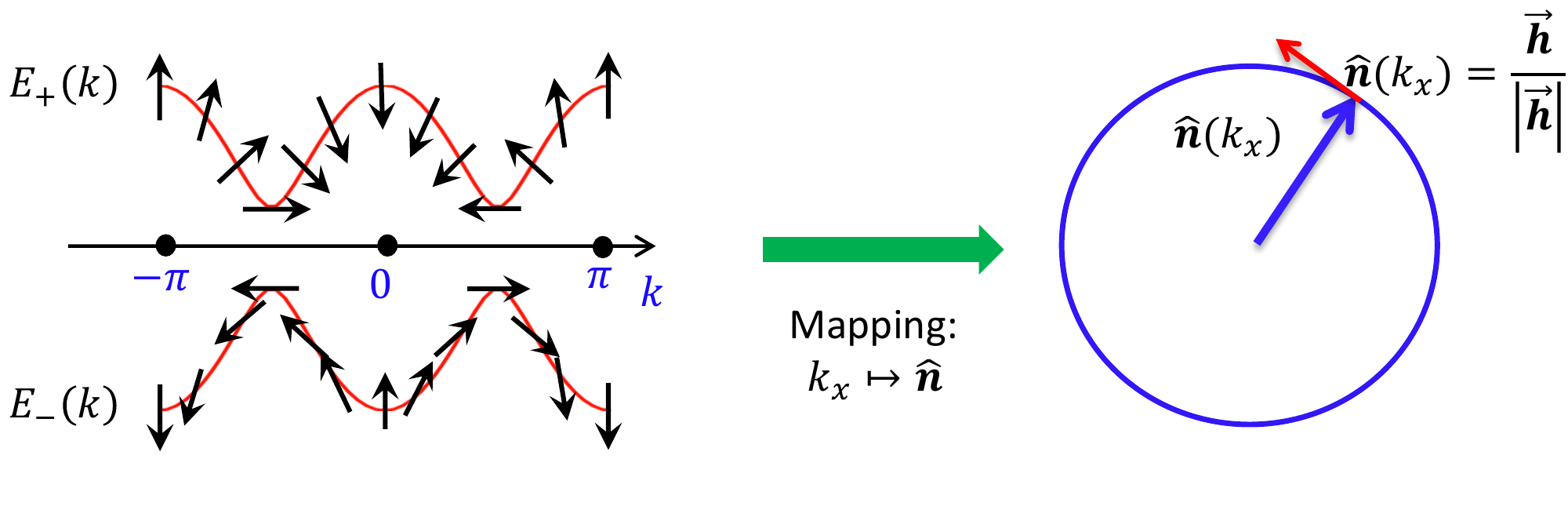}
\caption{\label{fig:winding} The nonzero winding number implies that the spin polarization axis $\vec h$ of the Bloch states winds over one circle when $k$ runs over the FBZ. This lead to a mapping between $k$ space and the $S^1$ circle and covers the $S^1$ completely.}
\end{figure}

By a direct check one can see that the time-reversal symmetry ${\cal T}=i\sigma_yK$ is broken for the Hamiltonian, where $K$ is the complex conjugate. On the other hand, the charge-conjugation (particle-hole) symmetry, defined by ${\cal C}: (c_s,c_s^\dag)\rightarrow(\sigma_z)_{ss'}(c^\dag_{s'},c_{s'})$, is also broken. The chiral symmetry, defined as the product of the charge-conjugation and time-reversal symmetry is conserved, with $({\cal C}{\cal T})H({\cal C}{\cal T})^{-1}=H$, and $({\cal C}{\cal T})^2=1$. Note that the charge-conjugation is a unitary symmetry, and the chiral symmetry is anti-unitary. The complete symmetry group of the Hamiltonian is $U(1)\times Z_2^T$, with $Z_2^T$ being an anti-unitary group formed by $\{I,{\cal C}{\cal T}\}$. In the first quantization picture the chiral symmetry is imply $S=\sigma_x$, and is satisfied even when a $m_y\sigma_y$ term exists in the Hamiltonian. From the symmetry analysis we know that the Hamiltonian $H$ belongs to chiral unitary (AIII) class according to the Altland-Zirnbauer ten-fold classification~\cite{AZ1997}, and its topology is characterized by an integer $Z$, namely the 1D winding number, given by
\begin{eqnarray}
N_{1D}&=&\frac{1}{4\pi i}\int dk_x{\rm Tr}\bigr[\sigma_x{\cal H}^{-1}(k_x)\partial_{k_x} {\cal H}(k_x)\bigr]\nonumber\\
&=&\frac{1}{2\pi}\int dk_x\hat h_z\partial_{k_x}\hat h_y,
\end{eqnarray}
where $(\hat h_y,\hat h_z)=(h_y,h_z)/h$, with $h_y=2t_{\rm so}\sin k_xa, h_z=m_z-2t_0\cos k_xa$, and $h=(h_y^2+h_z^2)^{1/2}$. By a straightforward calculation one can find that
\begin{eqnarray}
N_{1D}= \left\{ \begin{array}{ll}
        1, \ {\rm for}\ |m_z|<2t_0, \\
             0, \ {\rm for}\ |m_z|>2t_0. \\
        \end{array} \right.
\end{eqnarray}
The nonzero winding number can simply interpreted that the spin axis of the lower subband winds one complete circle when the momentum $k$ runs over the FBZ, as sketched in Fig.~\ref{fig:winding}.

The nontrivial topology with $|m_z|<2t_0$ can support degenerate boundary modes. Considering open boundaries located at $x=0,L$, respectively and diagonalizing $H$ in position space $H=\sum_{x_i}\mathcal{H}(x_i)$ with $\mathcal{H}(x_i)=-(t_s\sigma_z+it^{(0)}_{\rm so}\sigma_y)\hat c_{x_i}^\dag\hat
c_{x_i+a}+m_z\sigma_z\hat c_{x_i}^\dag\hat
c_{x_i}+h.c.$, we obtain the edge state for the
boundary $x=0$ as
\begin{eqnarray}\label{eqn:gapstate1}
\psi_L(x_i)=\frac{1}{\sqrt{{\cal N}}}[(\lambda_+)^{x_i/a}-(\lambda_-)^{x_i/a}]|\chi_+\rangle,
\end{eqnarray}
and accordingly the one on $x=L$ by $\psi_R(x_i)=\frac{1}{\sqrt{{\cal N}}}[(\lambda_+)^{(L-x_i)/a}-(\lambda_-)^{(L-x_i)/a}]|\chi_-\rangle$, with ${\cal N}$ being the normalization factor, the spin eigensates $\sigma_x|\chi_\pm\rangle=\pm|\chi_\pm\rangle$, and $\lambda_\pm=(m_z\pm\sqrt{m_z^2-4t_s^2+4|t^{(0)}_{\rm so}|^2})/(2t_s+2|t^{(0)}_{\rm so}|)$. Thus the two edge modes are polarized to the opposite $\pm x$ directions. Note $\psi_{L}$ and $\psi_{R}$ span the complete Hilbert space of {\it one single} $1/2$-spin or spin-qubit. Each edge state equals {\it one-half} of a single spin, namely, a $1/4$-spin. Furthermore, the edge state to $1/2$-fractionalization. A convenient way is to consider the semi-infinite geometry which has the open boundary at $x=0$. We then calculate the particle number of the zero mode localized on this boundary. Note the total number of quantum states in the system is given by
\begin{eqnarray}\label{eqn:apparticle1}
N=\sum_{E<0}\langle\psi_E|\hat n_E|\psi_E\rangle+\sum_{E>0}\langle\psi_E|\hat n_E|\psi_E\rangle+\langle\psi_0|\hat n_E|\psi_0\rangle,\nonumber\\
\end{eqnarray}
where we denote by $\hat n_E=\mathbb{I}$ the state number operator and $|\psi_E\rangle$ is the eigenstate with energy $E$. Since the Hamiltonian satisfies $\{{\cal H},\sigma_x\}=0$, the energy spectrum is symmetric. We have then $\sum_{E<0}\langle\psi_E|\hat n_E|\psi_E\rangle=\sum_{E>0}\langle\psi_E|\hat n_E|\psi_E\rangle$. It follows that
\begin{eqnarray}\label{eqn:apparticle2}
\sum_{E<0}\langle\psi_E|\hat n_E|\psi_E\rangle=\frac{1}{2}[N-\langle\psi_0|\hat n_E|\psi_0\rangle].
\end{eqnarray}
The particle number of the zero mode depends on its occupation. If the zero mode is unoccupied, the particle number of it is given by
\begin{eqnarray}\label{eqn:apparticle3}
n_0=\sum_{E<0}\bigr[\langle\psi_E|\hat n_E|\psi_E\rangle_1-\langle\psi_E|\hat n_E|\psi_E\rangle_0\bigr].
\end{eqnarray}
Here $\langle\rangle_1$ and $\langle\rangle_0$ represents the cases with one (topological phase) and zero (trivial phase) bound modes, respectively. Using the Eq.~\eqref{eqn:apparticle2} one finds directly
\begin{eqnarray}\label{eqn:apparticle3}
n_0=-\frac{1}{2}\langle\psi_0|\hat n_E|\psi_0\rangle=-\frac{1}{2}.
\end{eqnarray}
Similarly, if the zero mode is occupied, the particle number is $n_0=1/2$. It is trivial to know that this result can be applied to the case with two boundaries located far away from each other, say respectively at $x=0$ and $x=L$. Since the two zero modes are obtained independently, each of them carries $+1/2$ ($-1/2$) particle if it is occupied (unoccupied).

\subsubsection{Interacting effects on topology}

The $Z$ classification implies that single-particle couplings respecting $U(1)$ and ${\cal CT}$ cannot gap out the edge modes in arbitrary $N$-chain system of 1D lattices. The topological classification can be reduced with interactions. In other words, in the presence of interaction, the edge states may split out even the system respects the $U(1)$ and ${\cal CT}$ symmetry. It was shown the Ref.~\cite{Liu2013a} that with the following generic interactions between edge modes at one side (left or right) of $N$ chains
\begin{eqnarray}
H_{\rm int}&=&\sum_{j,j'}^N\sum_{\{12...j;1'2'...j'\}}V_{12...j}^{1'2'...j'}c_1^\dag c_2^\dag...c_j^\dag c_{j'}...c_{2'}c_{1'}\nonumber\\
&&+{\rm H.c.},
\end{eqnarray}
where $V_{12...j}^{1'2'...j'}=(V_{12...j}^{1'2'...j'})^*$ to ensure the chiral symmetry, the edge-state degeneracy of this side is still protected until the number of chains reaches $N=4$. This implies that under (arbitrarily small) interactions the topological classification of 1D AIII insulator reduces from $Z$ to $Z_4$. This result can also be understood from the projective representations of the $U(1)\times Z_2^T$ symmetry group. It can be shown that the $U(1)\times Z_2^T$ group has $4$ inequivalent projective representations, with three being topological and one trivial~\cite{LiuYiZhang2017}. This also confirms that the topological classification reduces to $Z_4$ by interactions.

On the other hand, for a single chain of 1D AIII class insulator, increasing interaction can lead to topological phase transition. The relevant interaction includes the onsite Hubbard interaction
\begin{eqnarray}
H_U=U\sum_{j}n_{j\uparrow}n_{j\downarrow},
\end{eqnarray}
which is shown to renormalize the mass $u=2t_{0}$ and magnetization $w=m_z$, and thus can induce phase transition when the interaction exceeds critical value~\cite{Liu2013a}. From the previous discussion we know that in the single-particle regime the critical point of phase transition is described by the scaling relation $u=w$. Under repulsive interaction, one can expect that $u$ increases compared with the linear scaling due to the detrimental effects of interaction. The new phase transition scaling under interaction can be studied with standard bosonization method, together with renormalization group (RG) flow~\cite{Giamarchi_book}. For convenience we rotate that ${\sigma_y,\sigma_z}\rightarrow(\sigma_z,-\sigma_y)$. The Hamiltonian can then be written as
$H\rightarrow H'=-t_{so}^{(0)}\sum_{<i,j>,\sigma}\hat c_{i\sigma}^{\dag}\hat
c_{j\sigma}+H_\text{SO}+H_\text{Z}$, with $H_\text{SO}=t_{s}\sum_j (-1)^j(c_{j\uparrow}^\dag c_{j+1,\downarrow}-c_{j\uparrow}^\dag c_{j-1,\downarrow}+\text{h.c.})$ and $H_\text{Z}=\Gamma_y\sum_j (-1)^j(ic_{j\downarrow}^\dag c_{j\uparrow}-\text{h.c.})$. Note that the SO term and spin-conserved hopping term are exchanged. The low-energy physics can be well captured by the continuum approximation ($x=ja$):
\begin{equation}
  c_{j\sigma}\approx \sqrt{a}[\psi_{R\sigma}(x)e^{ik_Fx}+\psi_{L\sigma}(x)e^{-ik_Fx}],
  \label{}
\end{equation}
with $k_F=\pi/2$. Neglecting the fast oscillating terms we obtain the continuum representation of the two mass terms by:
\begin{eqnarray} \label{eqn:mass}
	  H_\text{SO}&\approx& iu\int dx\,(\psi_{L}^\dag\sigma^x\psi_R-\psi_{R}^\dag\sigma^x\psi_{L}), \\
	  H_\text{Z}&=&w\int d x\,(\psi_R^\dag \sigma^y\psi_L+\psi_L^\dag\sigma^y\psi_R).
\end{eqnarray}
Here $u=2t_{\text{so}}, w=\Gamma_y$. Using the standard bosonization formula $\psi_{rs}=\frac{1}{\sqrt{2\pi a}}e^{-\frac{i}{\sqrt{2}}[r\phi_\rho-\theta_\rho+s(r\phi_\sigma-\theta_\sigma)]}$ with $r=R,L$ and $s=\uparrow,\downarrow$,
we reach the bosonized Hamiltonian densities
\begin{eqnarray}
	\mathcal{H}_\text{SO}&=&\frac{u}{\pi a}\sin\sqrt{2}\phi_\rho\cos\sqrt{2}\theta_\sigma, \\
	\mathcal{H}_\text{Z}&=&\frac{w}{\pi a}\cos\sqrt{2}\phi_\rho\sin\sqrt{2}\theta_\sigma.
	\label{eqn:mass_bosonization}
\end{eqnarray}

The topology of the system depends on which of $u$ and $w$ flows to the strong-coupling regime first under RG. A direct power counting shows the same RG flow for the masses $u$ and $w$ in the first-order perturbation. Therefore the next-order perturbation expansion is necessary to capture correctly the fate of the topological phase transition. By deriving the RG flow equations up to one-loop order~\cite{Liu2013a}, the renormalization to $u, w$, the umklapp scattering $g_\rho$ and spin backscattering $g_\sigma$ by:
\begin{equation}
  \begin{split}
	\frac{\mathrm{d}u}{\mathrm{d} l}&=\frac{3-K_\rho}{2}u-\frac{g_\rho u}{4\pi v_F}+\frac{g_\sigma u}{4\pi v_F},\\
	\frac{\mathrm{d}w}{\mathrm{d} l}&=\frac{3-K_\rho}{2}w +\frac{g_\rho w}{4\pi v_F}+\frac{g_\sigma w}{4\pi v_F},\\
	\frac{\mathrm{d}g_\rho}{\mathrm{d} l}&=\frac{g_\rho^2}{\pi v_F}, \ \ \
	\frac{\mathrm{d}g_\sigma}{\mathrm{d} l}=\frac{g_\sigma^2}{\pi v_F},
\end{split}
  \label{}
\end{equation}
where the bare values of the coupling constants $g_\rho=-g_\sigma=U, u=2t^{(0)}_\text{so}, w=\Gamma_y$,
and $l$ is the logarithm of the length scale. Being a higher order correction, the renormalization of Luttinger parameter $K_\rho$ has been neglected.
For $U>0$, $g_\sigma$ marginally flows to zero and can be dropped off. This is consistent with the result that repulsive interaction cannot gap out the spin sector in the 1D Hubbard model. $g_\rho$ is marginally relevant and can be solved by $g_\rho(l)=\frac{\pi v_F g_\rho(0)}{\pi v_F-g_\rho(0)l}$.
Substituting this result into RG equations of $u$ and $w$ yields after integration $u(l)=u(0)[1-\frac{g_\rho(0)l}{\pi v_F}]^{\frac{1}{4}}e^{(3-K_\rho)l/2},
w(l)=w(0)[1+\frac{g_\rho(0)l}{\pi v_F}]^{\frac{1}{4}}e^{(3-K_\rho)l/2}$. As having been analyzed previously, the repulsive interaction ($g_\rho>0$) suppresses SO induced mass term $u$ while enhances the trivial mass term $w$. With these results one can find that
that the scaling of topological phase transition is given by~\cite{Liu2013a}
\begin{equation}
  u(0)=\bigr[w(0)\bigr]^\gamma, \: \gamma\approx 1-\frac{g_\rho(0)}{\pi v_F(3-K_\rho)},
  \label{}
\end{equation}
where $u(0)$ and $w(0)$ denote their bare values (before having interactions), $g_\rho(0)$ is the umklapp scattering coefficient, and $K_\rho$ is the Luttinger coefficient of the charge sector. Note that in the original paper of Ref.~\cite{Liu2013a} a ``$1/4$"-factor was wrongly included in the second term of $\gamma$ by carelessness. For repulsive interaction $U>0$, one has $\gamma<1$ . The above scaling relation implies that a repulsive interaction suppresses the topological phase (note that the bare value $u(0),w(0)<1$ for the weak coupling regime). Accordingly, if initially the noninteracting system is topologically nontrivial with $u(0) > w(0) > 0$, increasing $U$ to the regime $u(0) < [w(0)]^\gamma$ can drive the system into a trivial phase.

\subsubsection{Further studies}

With optical Raman lattice many interesting physics can be explored besides the studies introduced above. In particular, it was shown that if adding an $s$-wave pairing potential, the AIII class topological phase enters into a 1D BDI class topological SC/superfluid. A resonant cross Andreev reflection was predicted in such topological SC~\cite{He2014}, as a consequence of oppositely spin-polarized edge modes in the left and right hand ends of the 1D system. On the other hand, by putting the current system in a cavity a novel topological phase, called topological superradiant phase~\cite{Pan2015}, was proposed, and was further generalized to the superfluid regimes~\cite{Yu2018}. Finally, with the relative configuration between Raman and optical lattices a Hidden nonsymmorphic symmetry was pointed out for the optical Raman lattice system, and was shown to be responsible for the degeneracy at the first Brillouin zone~\cite{ChenHua2016,Lang2017}.

\subsubsection{Experimental realization}

The optical Raman lattice scheme for 1D topological state was realized very recently by HKUST group collaborating with PKU group with the alkali earth $^{173}$Yb fermions~\cite{SongBo2017}. The realization with alkali earth atoms is of explicit advantage in having a long lifetime due to the absence of relatively small fine-structure splitting between $D_1$ and $D_2$ lines which limits the realization of SO coupling in alkali atoms to apply near-resonant optical transitions. Another peculiar property of the realization in $^{173}$Yb atoms is that the optical lattice is spin-dependent, rather than spin-independent, namely $V_{\rm latt}^\uparrow(x)\neq V_{\rm latt}^\downarrow(x)$. In this case, the chiral symmetry discussed in the above subsection is generically not satisfied. Interestingly, it was shown in the work that the 1D topological phase and degenerate end states can still be obtained, and are protected by two hidden symmetries called magnetic group and non-local chiral symmetries~\cite{SongBo2017}. The quenching dynamics are also investigated in the experiment, with a nontrivial topology-dependent spin relaxation dynamics being observed.

\subsection{Optical Raman lattice: 2D case}

\subsubsection{Model}

The optical Raman lattice scheme can be generalized to the 2D case. As was proposed in~\cite{Liu2014a}, a basic model of 2D optical Raman lattice for spin-$1/2$ system reads
\begin{eqnarray}\label{2Dmodel}
H_{\rm 2D}&=&\frac{\hbar^2{\bf k}^2}{2m}+V_{\rm latt}(x,y)+m_z\sigma_z+M_x(x,y)\sigma_x\nonumber\\
&&+M_y(x,y)\sigma_y,
\end{eqnarray}
where the spin-independent lattice potential $V_{\rm latt}(x,y)=V_0(\cos^2k_0x+\cos^2k_0y)$, the Raman lattice potentials $M_x=M_0\cos k_0x\sin k_0y$ and $M_y=M_0\cos k_0y\sin k_0x$, with $(V_0,M_0)$ being the amplitudes, and the constant Zeeman term $m_z$ relates to the two-photon detuning by $m_z=\delta/2$. Similar to the 1D model, in the present 2D optical Raman lattice each Raman potential ($M_{x}$ or $M_y$) is antisymmetric with respect to lattice potential $V_{\rm latt}(x,y)$ along one direction (e.g. the $x$ direction for $M_{x}$), while it is symmetric along another direction (the $y$ direction for $M_{x}$). We shall see that this fundamental model Hamiltonian naturally realizes 2D SO coupling and can bring about rich topological phases~\cite{Liu2014a,Wu2016}.


\begin{figure}[h]
\includegraphics[width=0.9\columnwidth]{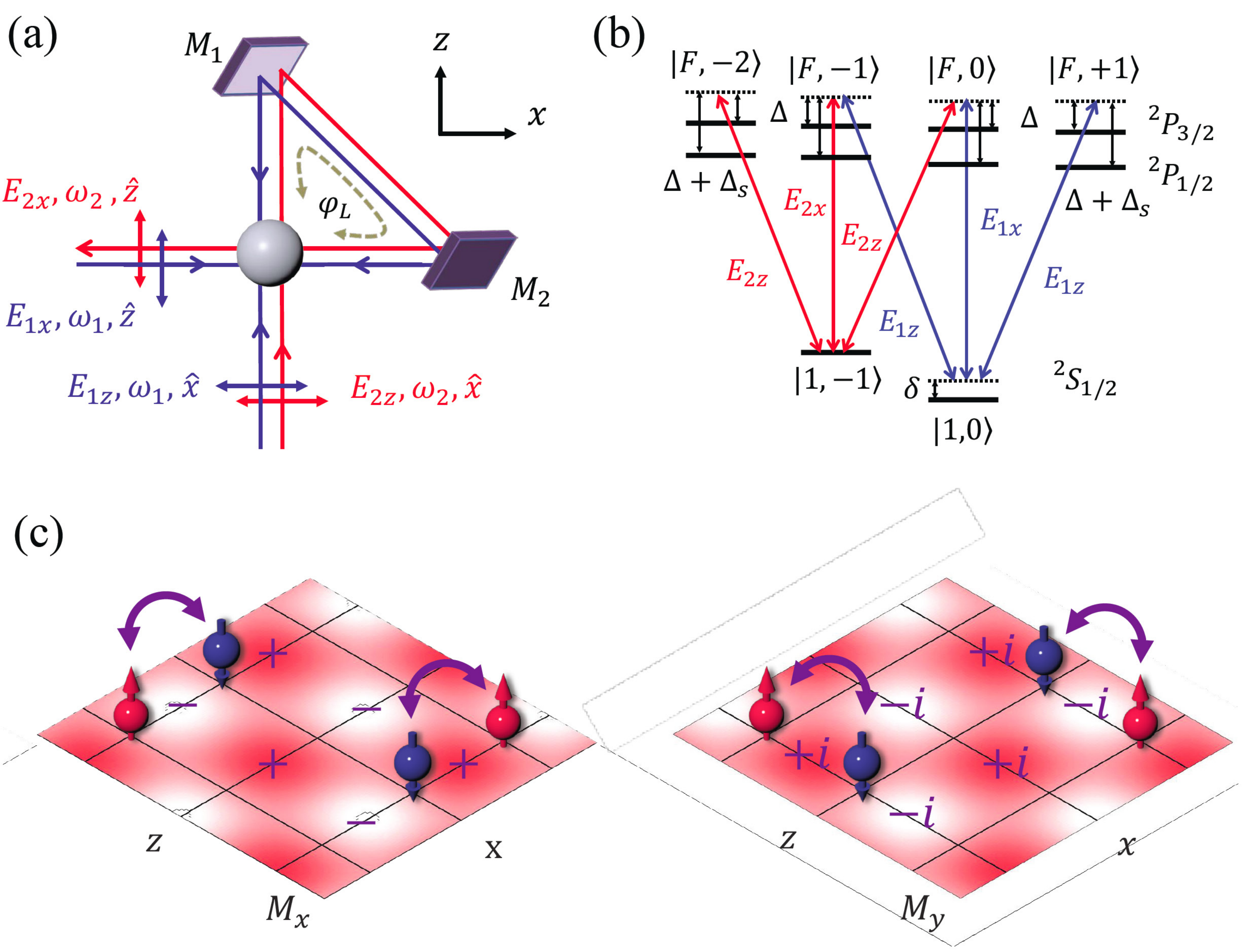}
\caption{The Raman lattice scheme in 2D case~\cite{Wu2016}. (a) Sketch of the setup for realization. The light components $E_{1x,1z}$ (blue lines) form
a spin-independent square optical lattice in the intersecting area and generate two periodic Raman potentials, together with the light components $E_{2x,2z}$ (red lines). (b) All the relevant optical transitions for $^{87}$Rb atoms, including the $D_2$ ($5^2S_{1/2}\to5^2P_{3/2}$) and $D_1$ ($5^2S_{1/2}\to5^2P_{1/2}$) lines of transitions. The spin is defined by $|\uparrow\rangle=|1,-1\rangle$ and $|\downarrow\rangle=|1,0\rangle$, $\Delta_s$ denotes the fine-structure splitting, and $\delta$ is a two-photon detuning. (c) Spin-flip hopping driven by Raman potential $M_x$ (left), and $M_y$ (right).  The Raman potential $M_x$ ($M_y$) drives spin-flip hopping along $x$ ($z$) direction, with the forwarding and backward hopping from a site having opposite signs.} \label{fig_2dRaman}
\end{figure}

We introduce now how to realize the Hamiltonian~\eqref{2Dmodel}. The following realization is based on $^{87}$Rb bosons, as proposed in Ref.~\cite{Wu2016}. However, the results are generically valid for both bosons and fermions. The 2D plane is set in the $x-z$ plane, as considered in Ref.~\cite{Wu2016}. The 2D Raman lattice scheme, sketched in Fig.~\ref{fig_2dRaman}(a), consists of a blue-detuned square lattice created by two standing-wave light components (blue lines),
and two periodic Raman potentials generated with additional plane-wave lights (red lines). The standing waves in the intersecting area are
${\bm E}_{1x}=\hat{z}\bar E_{1x}e^{i\varphi_{L}/2}\cos(k_0x-\varphi_{L}/2)$ and
${\bm E}_{1z}=\hat{x}\bar E_{1z}e^{i\varphi_{L}/2}\cos(k_0z-\varphi_{L}/2)$, where $\bar E_{1x/1z}$ are amplitudes and the phase
$\varphi_{L}=k_0L$ is acquired through the optical path $L$ from intersecting point to mirror $M_1$, then to $M_2$, and back to the intersecting point [Fig.~\ref{fig_2dRaman}(a)].
Here the initial phases have been ignored.
For alkali atoms,
the optical potential generated by linearly polarized lights is spin-independent when the detuning $\Delta$
is much larger than the hyperfine structure splittings.
The square lattice potential then takes the form
\begin{equation}\label{lattice}
V_{\rm latt}(x,z)=V_{0x}\cos^2(k_0x-\varphi_{L}/2)+V_{0z}\cos^2(k_0z-\varphi_{L}/2),
\end{equation}
where the amplitudes are given by
\begin{equation}\label{V_amp}
V_{0x/0z}=\frac{3\Delta+2\Delta_s}{3\Delta(\Delta+\Delta_s)}\alpha_{D_1}^2\bar{E}_{1x/1z}^2.
\end{equation}
Here $\Delta_s$ denotes the energy splitting between the $D_1$ and $D_2$ lines, as discussed in section III. The dipole matrix elements $\alpha_{D_2}=|\langle J=1/2||e{\bf r}||J'=3/2\rangle|=\sqrt{2}\alpha_{D_1}=\sqrt{2}|\langle J=1/2||e{\bf r}||J'=1/2\rangle|=4.227ea_0$, with $a_0$ the Bohr radius. The lattice potentials have taken into account all relevant transitions from $D_1$ and $D_2$ lines for the alkali atoms. It can be seen that by simply tuning the strengths $|\bar{E}_{1x}|=|\bar{E}_{1z}|$, we reach a square lattice with the same depths along $x$ and $z$ directions.

The Raman couplings are induced by applying another beam of frequency $\omega_2$, incident from the $z$ direction. This light along with lattice beams can be generated from a single laser source via an acoustic-optic modulator which controls their frequency difference $\delta\omega=\omega_2-\omega_1$ and amplitude ratio. Two plane-wave fields are formed at
the intersecting point ${\bm E}_{2z}=\hat{x}\bar{E}_{2z}e^{ik_0z}$ and
${\bm E}_{2x}=\hat{z}\bar{E}_{2x}e^{i(-k_0x+\varphi_L+\delta\varphi_L)}$.
The relative phase $\delta\varphi_L=L\delta\omega/c$, acquired by ${\bm E}_{2x}$, can be precisely manipulated by changing the optical path $L$ or $\delta\omega$, and it controls the dimensionality of the realized SO coupling.
The standing-wave and plane-wave beams form a {\it double-$\Lambda$} type configuration as shown in Fig.~\ref{fig_2dRaman}(b), with
${\bm E}_{1x}$ and ${\bm E}_{2z}$ generating one Raman potential via $|e_1\rangle$ in the form
$M_{0x}\cos(k_0x-\varphi_L/2)e^{i(k_0z-\varphi_L/2)}$,
and $E_{1z}$ and $E_{2x}$ producing another one via $|e_2\rangle$ as
$M_{0y}\cos(k_0x-\varphi_L/2)e^{-i(k_0z-\varphi_L/2)+i\delta\varphi_L}$, with
 $M_{0x(0y)}\propto\bar{E}_{1x(1z)}\bar{E}_{2z(2x)}$.
Note that for the present blue-detuned lattice, atoms are located in the region of minimum intensity of lattice fields. It follows that terms like $\cos(k_0x-\varphi_L/2)\cos(k_0z-\varphi_L/2)$, which are antisymmetric with respect to each lattice site in both $x$ and $z$ directions, have negligible contribution to the low-band physics. Neglecting such terms yields the Raman coupling potentials
\begin{eqnarray}\label{Raman}
{\cal M}_x(x,z)\sigma_x&=&(M_{x}-M_{y}\cos\delta\varphi_L)\sigma_x, \nonumber \\
{\cal M}_{y}(x,z)\sigma_y&=&M_y\sin\delta\varphi_L\sigma_y.
\end{eqnarray}
Here $M_x=M_{0x}\cos(k_0x-\varphi_L/2)\sin(k_0z-\varphi_L/2)$ and $M_{y}=M_{0y}\cos(k_0z-\varphi_L/2)\sin(k_0x-\varphi_L/2)$. Similar to the calculation of the lattice potentials given above, the amplitudes of the Raman potentials are obtained straightforwardly that
\begin{eqnarray}\label{M0xy}
M_{0x/y}=\frac{\bar{E}_{1x/z}\bar{E}_{2z/x}\alpha_{D_1}^2}{6\sqrt{2}}(\frac{1}{\Delta}-\frac{1}{\Delta+\Delta_s}).
\end{eqnarray}

With the above results for square lattice and Raman potentials, the total Hamiltonian is followed by
\begin{eqnarray}\label{Ham}
H_{\rm 2D}&=&\frac{\hbar^2{\bf k}^2}{2m}+V_{\rm latt}(x,z)+{\cal M}_x(x,z)\sigma_x\nonumber\\
&&+{\cal M}_y(x,z)\sigma_y+m_z\sigma_z,
\end{eqnarray}
where the effective Zeeman term $m_z=\delta/2$ is considered [Fig.~\ref{fig_2dRaman}(b)]. The above Hamiltonian returns to that in Eq.~\eqref{2Dmodel} when the phase difference takes the optimal value $\delta\varphi_L=\pi/2$, and replace $z\rightarrow y$. The present scheme is of topological stability, namely, the realization is intrinsically immune to any phase fluctuations in the setting. Moreover, the relative phase $\delta\varphi_L$ determines the relative strength of $\sigma_x$ and $\sigma_y$ terms in the Raman potential, which generates SO coupling along $x$ and $y$ directions, respectively. Tuning $\delta\varphi$ between $\pi/2$ and $\pi$ can reach a crossover between the 2D and 1D SO couplings. If tuning $\delta\omega=50$MHz, we have $\delta\varphi_L=\pi/2$ for $L=1.5$m, which gives a 2D SO coupling, while $\delta\varphi_L=\pi$ if increasing to $L=3.0$m, and the SO coupling becomes 1D form. Note that the fluctuations, e.g. due to the mirror oscillations, have very tiny effect on $L$ and thus the relative phase. These advantages ensure that this proposal can be realized in experiment.

\subsubsection{Chern insulator for $s$-bands}

Similar to the 1D regime, the present 2D optical Raman lattice can bring about nontrivial topological quantum states. We consider the lowest $s$-band model for the current consideration, while we emphasize that for the lattice system, the generic high-band model can be naturally obtained and may give rise to novel new topological physics.
For convenience, we focus our study on the isotropic case with $V_{0x}=V_{0z}=V_{0}$, $M_{0x}=M_{0y}=M_{0}$ and $\delta\phi_L=\pi/2$.
The topological physics can be best understood with tight-binding model, while all the results derived below in this section is not restricted by tight-binding approximation, namely, they are valid for non-tight-binding regime. Taking into account the spin-conserved and spin-flip hopping terms between the nearest-neighbor sites, we obtain the tight-binding Hamiltonian for the $s$-bands by
\begin{eqnarray}\label{eqn:tightbinding1}
H_{\rm TI}&=&-\sum_{<\vec{i},\vec{j}>,\sigma}t_0^{\vec{i}\vec{j}}\hat c_{\vec{i}\sigma}^{\dag}\hat
c_{\vec{j}\sigma}+\sum_{<\vec{i},\vec{j}>}\bigr(t_{\rm so}^{\vec{i}\vec{j}}\hat c_{\vec{i}\uparrow}^\dag\hat c_{\vec{j}\downarrow}+{\rm H.c.}\bigr)\nonumber\\
&&+\sum_{\vec{i}}m_z\bigr(\hat n_{\vec{i}\uparrow}-\hat n_{\vec{i}\downarrow}\bigr),
\end{eqnarray}
where $\vec i=(i_x,i_z)$ is the 2D lattice-site index, the particle number operators $\hat n_{\vec{i}\sigma}=\hat c_{\vec{i}\sigma}^\dag\hat c_{\vec{i}\sigma}$. The spin-conserved hopping couplings $t_0^{\vec{i}\vec{j}}$ are induced by the lattice potential, calculated by
\begin{eqnarray}\label{eqn:hoping1}
t_0^{\vec{i}\vec{j}}=\int d^2\bold r\phi^{(\vec i)}_{s\sigma}(\bold r)\bigr[\frac{p_x^2+p_z^2}{2m}+V_{\rm latt}(\bold r)\bigr]\phi^{(\vec j)}_{s\sigma}(\bold r),
\end{eqnarray}
where $\phi^{(\vec j)}_{s\sigma}(\bold r)$ denotes the $s$-orbital Wannier function at the $\vec j$-th site at the spin state $\sigma$.
On the other hand, the spin-flip hopping couplings $t_{\rm so}^{\vec{i}\vec{j}}$ are driven by the Raman potentials ${M}_x(x,z)$ and $M_y(x,z)$, which are antisymmetric with respect to each square lattice site along $x$ and $z$ directions, respectively, and are obtained by
\begin{eqnarray}\label{eqn:hoping2}
t_{\rm so}^{\vec i\vec j}=\int d^2\bold r\phi^{(\vec i)}_{s\uparrow}(\bold r)[M_x(x,z)\sigma_x+M_y(x,z)\sigma_y]\phi^{(\vec j)}_{s\downarrow}(\bold r).
\end{eqnarray}
It is trivial to know that $t_0^{\vec{i}\vec{j}}$ for the nearest-neighbor hopping is a constant in all directions, namely $t_0^{\vec{i}\vec{i}\pm\vec1}=t_0$. On the other hand, from the Raman potential configurations we have that
\begin{widetext}
\begin{eqnarray}\label{eqn:hoping3}
\int d^2\bold r\phi^{(\vec i)}_{s\uparrow}(\bold r)M_x(x,z)\phi^{(\vec j)}_{s\downarrow}(\bold r)&=&\int d^2\bold r\phi^{(\vec i)}_{s\uparrow}(\bold r)M_0\cos(k_0x)\sin(k_0z)\phi^{(\vec j)}_{s\downarrow}(\bold r)\nonumber\\
&=&(-1)^{i_x+i_z}M_0\int d^2\bold r\phi^{0,0}_{s\uparrow}(\bold r)\cos(k_0x)\sin(k_0z)\phi^{j_x-i_x,j_z-i_z}_{s\downarrow}(\bold r)\delta_{i_zj_z}(1-\delta_{i_x,j_x}),
\end{eqnarray}
\begin{eqnarray}\label{eqn:hoping4}
\int d^2\bold r\phi^{(\vec i)}_{s\uparrow}(\bold r)M_y(x,z)\phi^{(\vec j)}_{s\downarrow}(\bold r)&=&\int d^2\bold r\phi^{(\vec i)}_{s\uparrow}(\bold r)M_0\cos(k_0z)\sin(k_0x)\phi^{(\vec j)}_{s\downarrow}(\bold r)\nonumber\\
&=&(-1)^{i_x+i_z}M_0\int d^2\bold r\phi^{0,0}_{s\uparrow}(\bold r)\cos(k_0z)\sin(k_0x)\phi^{j_x-i_x,j_z-i_z}_{s\downarrow}(\bold r)\delta_{i_xj_x}(1-\delta_{i_z,j_z}).
\end{eqnarray}
\end{widetext}
The above formulas show that the Raman potential $M_x(x,z)$ only induces the spin-flip hopping along $x$ direction, while $M_y(x,z)$ only induces the spin-flip hopping along $z$ direction, both of them cannot drive onsite spin-flip transitions. These properties are resulted from the antisymmetry of $M_x(x,z)$ ($M_y(x,z)$) along $x$ ($z$) direction [Fig.~\ref{fig_2dRaman}(c)]. Note that we have neglected the term $M_0\cos(k_0x)\cos(k_0z)$ in the originally realized Hamiltonian, which is antisymmetric with respect to lattice site along both $x$ and $z$ directions, cannot induce hopping along either $x$ or $z$ direction, nor induce the onsite spin-flip coupling within the $s$-band. The leading-order nonzero contribution of this term is the onsite transition between $s$-band and $p_xp_z$-band (not $p_x$ or $p_z$ band), which is negligible when the Raman lattice is weak compared with the square lattice potential $V_{\rm latt}$.

From the Eqs.~\eqref{eqn:hoping3} and~\eqref{eqn:hoping4}, and together with the antisymmetry of Raman potentials, it can be read that the spin-flip hopping terms between two neighboring sites satisfy
\begin{eqnarray}\label{eqn:hoping5}
t_{\rm so}^{j_x,j_x\pm1}&=&\pm(-1)^{j_x+j_z}t_{\rm so},\nonumber\\
t_{\rm so}^{j_z,j_z\pm1}&=&\pm i(-1)^{j_x+j_z}t_{\rm so},
\end{eqnarray}
where $t_{\rm so}=M_0\int d^2\vec r\phi^{0,0}_{s}(x,z)\cos(k_0x)\cos(k_0z)\phi^{0,0}_{s}(x-a,z)$ is proportional to Raman coupling strength $M_0$. Note that the staggered property of the SO terms are a consequence of the relative spatial configuration of the lattice and the Raman potentials, namely, the periodicity (unit cell) of the Raman lattice is one-half (double) of the square lattice periodicity (unit cell), and the Raman potential $M_x(x,z)$ ($M_y(x,z)$) is antisymmetric along $x$ ($z$) direction, while symmetric along $z$ ($x$) direction [Fig.~\ref{fig_2dRaman}(c)]. Since the relative spatial profiles of the Raman potentials and the square lattice are determined by the same standing wave lights and are automatically fixed, these properties are topologically stable against any kind of fluctuations in the system.

The staggered spin-flip hopping terms described in Eq.~\eqref{eqn:hoping5} bring about two important effects. First, the staggered property implies that the coupling between spin-up and spin-down states transfers $\pi/a$ momentum between them along both $x$ and $z$ direction, which effectively shifts the Brillouin zone by half for the spin-down relative to spin-up states. This is exactly similar to the situation in the 1D optical Raman lattice, as discussed previously. Moreover, in additional to the relative half Brillouin zone shift, the remaining effect of the spin-flip hopping leads to the normal Rashba type SO coupling in the $x-z$ plane. Note that the former effect can be absorbed by redefining the spin-down operator $\hat c_{\vec{j}\downarrow}\rightarrow e^{i\pi\vec r_{\vec j}/a}\hat c_{\vec{j}\downarrow}$. We recast the tight-binding Hamiltonian into
\begin{eqnarray}
H_{\rm TB}
&=&-t_0\sum_{\langle\vec{i}\,\vec{j}\rangle}\left(c^\dagger_{\vec{i}\uparrow}c_{\vec{j}\uparrow}-c^\dagger_{\vec{i}\downarrow}c_{\vec{j}\downarrow}\right)+\sum_{\vec{i}}m_z(n
_{\vec{i}\uparrow}-n_{\vec{i}\downarrow}) \nonumber\\
&+&\bigr[\sum_{j_x}t_{\rm so}(c^\dagger_{j_x\uparrow}c_{j_x+1\downarrow}-c^\dagger_{j_x\uparrow}c_{j_x-1\downarrow})+{\rm h. c.}\bigr] \nonumber\\
&+&\bigr[\sum_{j_z}it_{\rm so}(c^\dagger_{j_z\uparrow}c_{j_z+1\downarrow}-c^\dagger_{j_z\uparrow}c_{j_z-1\downarrow})+{\rm h. c.}\bigr].
\end{eqnarray}
Note that the Raman potential $M_x(x,z)$ only induces the spin-flip hopping along $x$ direction, while $M_y(x,z)$ only induces the spin-flip hopping along $z$ direction. These properties are resulted from the antisymmetry of $M_x(x,z)$ [$M_y(x,z)$] along $x$ ($z$) direction.
Just like the 1D Hamiltonian [Eqs.~(\ref{TB1d})], the opposite signs in $t_0$ terms are due to the
relative $(\pi,\pi)$-momentum transfer between spin-up and spin-down Bloch states.
The staggered property of the SO terms are a consequence of the relative spatial configuration of the lattice and the Raman potentials, namely, the Raman potential $M_x(x,z)$ [$M_y(x,z)$] is antisymmetric along $x$ ($z$) direction, while symmetric along $z$ ($x$) direction [Fig.~\ref{fig_2dRaman}(c)].

Transforming the tight-binding Hamiltonian into momentum space
yields $H_{\rm TB}=\sum_{\bold q,s\sigma,\sigma'}c^\dag_{\bold q\sigma}{\cal H}_{\sigma\sigma'}({\bm q})c_{\bold q\sigma'}$, with the Bloch Hamiltonian
\begin{eqnarray}\label{Hq}
{\cal H}({\bm q})&=&[m_z-2t_0(\cos q_xa+\cos q_z a)]\sigma_z \nonumber\\
&+&2t_{\rm so}\sin q_xa\sigma_y+2t_{\rm so}\sin q_za\sigma_x,
\end{eqnarray}
which is a typically two-band model of the form
${\cal H}({\bm q})=\sum_{\alpha=x,y,z}h_\alpha({\bm q})\sigma_\alpha$,
with $h_x({\bm q})=2t_{\rm so}\sin q_za$, $h_y({\bm q})=2t_{\rm so}\sin q_xa$ and $h_z({\bm q})=m_z-2t_0(\cos q_xa+\cos q_z a)$,
giving the spectra $E_\pm=\pm\sqrt{\sum_\alpha h^2_\alpha({\bm q})}$. When $|m_z|\neq 4t_0$ or $0$, the system is fully gapped in the bulk; otherwise,
the system has one or two Dirac points.
This Bloch Hamiltonian
around $\Gamma$ point takes the form
\begin{eqnarray}\label{Hq}
{\cal H}({\bm q})&\approx&[m_z-4t_0+t_0(q^2_xa+q^2_z a)]\hat{\sigma}_z\nonumber\\
&&+\lambda_{\rm so}q_x\hat{\sigma}_y+\lambda_{\rm so}q_z\hat{\sigma}_x,
\end{eqnarray}
with $\lambda_{\rm so}=2at_{\rm so}$.
Thus the lowest $s$-band model is described by a normal Rashba type SO coupling in a square lattice, plus the kinetic energy
term coupling to spin component $\sigma_z$ term. This is in sharp contrast to a purely Rashba SO coupled system which is topologically trivial in the single-particle regime. The present realization gives a minimal two-band QAH model driven by 2D SO coupling, which was shown to exhibit novel topological features in both the bulk and the edge~\cite{Liu2014b}. We note that this model cannot be precisely realized with solid state materials, where the minimal case is a four-band model~\cite{QAHE2013}, but has been firstly achieved in the ultracold atoms.

The topology of the present QAH model is characterized by Chern number, which can be calculated by
\begin{eqnarray}
{\rm Ch}_{1}=-\frac{1}{24\pi^2}\int d\omega dk_x dk_y{\rm Tr}\bigr[{G}d {G}^{-1}\bigr]^3,
\end{eqnarray}
where the Green's function ${G}^{-1}(\omega,\bold q)=\omega+i\delta^+-{\cal H}(\bold q)$, and the trace is operated on the spin space. Integrating over the frequency space yields that
\begin{eqnarray}
{\rm Ch}_{1}=\frac{1}{4\pi}\int dq_xdq_z \bold {\hat h}\cdot\partial_{q_x}\bold {\hat h}\times\partial_{q_z}\bold {\hat h},
\end{eqnarray}
 where $\bold {\hat h}=(h_x,h_y,h_z)/|\bold h(\bold q)|$. By a straightforward calculation one can find that
\begin{eqnarray}
{\rm Ch}_{1}= \left\{ \begin{array}{ll}
        {\rm sgn}(m_z), \ {\rm for}\ 0<|m_z|<4t_0, \\
             0, \ \ \ \ \ {\rm for}\ |m_z|>4t_0, {\rm or} \ m_z=0. \\
        \end{array} \right.
\end{eqnarray}
While the Chern number can in principle be determined by measuring the Berry's curvature at the whole $\bold q$ space, the precise measurement could be quite challenging in the real experiment. Interestingly, the Chern number of the current system can be measured by a much simpler way as introduced below, due the symmetry of the present optical Raman lattice.
\begin{figure}[btp]
\centering
\includegraphics[width=0.98\columnwidth]{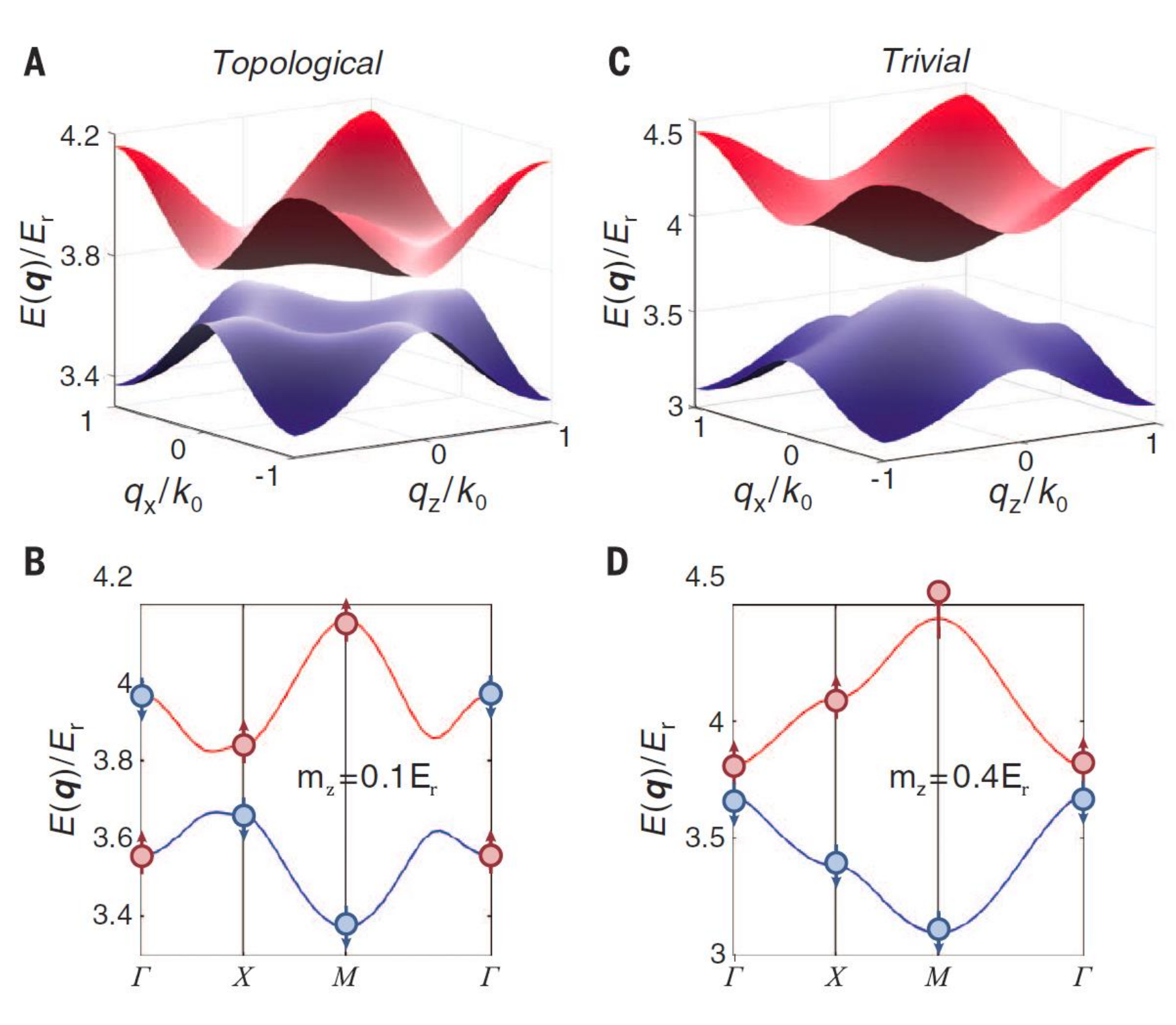}
\caption{Band structure and spin texture of the $s$-bands~\cite{Wu2016}. (a-c) Two examples of gapped band structure with nontrivial (a) or trivial (c) topology.
(b,d) The band structure and spin polarization distributions $\langle\sigma_z\rangle$ for the topological (b)
and trivial (d) cases.
Here $V_{0}=5E_{\rm r}$, $M_{0}=1.2E_{\rm r}$ and $m_z=0.1E_{\rm r}$ (a,b),
or $0.4E_{\rm r}$ (c,d).} \label{fig_topoband}
\end{figure}

\subsubsection{Detecting band topology by minimal measurements}

As the topology of a Chern band is classified by Chern numbers, and also characterized by chiral edge states in the boundary, the detection of Chern bands can in principle be performed by measuring the bulk Chern number or chiral edge states. For the edge states, the measurement strategy includes the light-Bragg scattering which can detect the spectra of the edge modes~\cite{Liu2010,Douglas2011,Lewenstein2011,Goldman2012}. On the other hand, for a Chern band in a synthetic dimension, the edge states can be easily imagined directly~\cite{Mancini2015,Stuhl2015}. The bulk Chern number can be in principle be measured by imaging the Berry curvature over the Brillouin zone~\cite{Alba2011,Price2012}, e.g. through the Hall transport measurement. The Chern number is given by the integral of the Berry curvature over Brillouin zone. The measurements of Bulk Chern number were performed in the recent experiments~\cite{Jotzu2014,Duca2015,Aidelsburger2015,Flaschner2016}.

In the present Chern insulator realized with 2D optical Raman lattice, a minimal measurement of the Chern number can be performed due to the novel inversion symmetry existing in the model, as proposed in Ref.~\cite{Liu2013b}.
The inversion symmetry of the Hamiltonian $H_{\rm 2D}$ is defined by ${\cal P}=\sigma_z\otimes R_{\rm 2D}$, giving ${\cal P}H_{\rm 2D}{\cal P}^{-1}=H_{\rm 2D}$,
where the 2D spatial operator $R_{\rm 2D}$ transforms the Bravais lattice vector $\bf R\to-\bf R$. Thus, the Bloch Hamiltonian satisfies ${\cal P}{\cal H}({\bm q}){\cal P}^{-1}={\cal H}(-{\bm q})$, which gives $[\sigma_z , {\cal H}({\bm \Lambda}_j)] = 0$ at four highly symmetric momenta $\{{\bm \Lambda}_j\}=\{\Gamma(0,0), X_1(0,\pi), X_2(\pi,0), M(\pi,\pi)\}$.
It is then indicated that in the tight-binding regime the Bloch states $|u_{\pm}({\bm \Lambda}_j)\rangle$ are also eigenstates of the parity operator $\sigma_z$ with eigenvalues $\xi^{(\pm)}=+1$ or $-1$. Therefore, similar to topological insulators~\cite{Fu2007},  one can define the following invariant
\begin{eqnarray}\label{eqn:parity}
\Theta=\prod_j{\rm sgn}\left[\xi^{(-)}({\bm \Lambda}_j)\right].
\end{eqnarray}
It has been proven rigorously that $\Theta=+1$ when the band is in trivial, and $\Theta=-1$ when the band is topological~\cite{Liu2013b}. Such generic proof needs to construct a nontrivial connection between the inversion symmetric QAH system and the time-reversal invariant topological insulating system.
Moreover, the Chern number
of the lower band
is given by~\cite{Liu2013b}
\begin{eqnarray}\label{Ch2d}
{\rm Ch}_1=-\frac{1-\Theta}{4}\sum_i{\rm sgn}\left[\xi^{(-)}({\bm \Lambda}_j)\right].
\end{eqnarray}
For the present specific system, it is straightforward to check that when the Zeeman term varies from $m_z\gtrsim0$ to $m_z\lesssim0$, two parity eigenvalues $\xi^{(-)}(X_{1,2})$ change sign and then ${\rm Ch}_1$ changes from $1$ to $-1$. Also, the Chern number vanishes for $|m_z|>4t_0$.

The fact that the topology of the inversion symmetric bands can be determined by only the Bloch states at four symmetric momenta can greatly simplify the experimental detection of the topological bands. In the experiment at low but finite temperature $T$,
the parity eigenvalues can be measured through the spin polarizations at the corresponding Bloch momenta.
The measured spin polarization at momentum ${\bm q}$ is given by
\begin{eqnarray}\label{polarization1}
P({\bm q})=\frac{n_\uparrow({\bf q},T)-n_\downarrow({\bf q},T)}{n_\uparrow({\bf q},T)+n_\downarrow({\bf q},T)},
\end{eqnarray}
with $n_{\uparrow,\downarrow}({\bf q},T)$ being the density of atoms of the corresponding spin state at temperature $T$ in the first Brillouin zone. At the four symmetric momenta, the spin-polarization measured in the experiment is
\begin{eqnarray}\label{polarization1}
P({\bm \Lambda_j})&\approx& \xi^{(-)}({\bf \Lambda}_j)f(E^{(-)}({\bf \Lambda}_j),T)+\xi^{(+)}({\bf \Lambda}_j) \nonumber\\
&\times&f(E^{(+)}({\bf \Lambda}_j),T),\nonumber\\
&\approx& \xi^{(-)}({\bf \Lambda}_j)\bigr[f(E^{(-)}({\bf \Lambda}_j),T)-f(E^{(+)}({\bf \Lambda}_j),T)\bigr],\nonumber\\
\end{eqnarray}
where the distribution function $f(E)=1/[e^{(E-\mu)/k_{\rm B}T}-1]$ if the topological band is simulated with bosons, and $f(E)=1/[e^{(E-\mu)/k_{\rm B}T}+1]$ if it is simulated with fermions, with $\mu$ being the chemical potential, and $E^{(-)}(\bold q)$ and $E^{(+)}(\bold q)$ are the energy of the lower and upper $s$-bands, respectively. Note that $f(E^{(-)}({\bf \Lambda}_j),T)\geqslant f(E^{(+)}({\bf \Lambda}_j),T)$, which ensures
\begin{eqnarray}
{\rm sgn}[\xi^{(-)}({\bm \Lambda}_j)]={\rm sgn}[P({\bm \Lambda}_j)].
\end{eqnarray}
The above result implies that the measurement of the spin-polarization at low but finite temperature can be applied to determine the Chern number of the Bloch bands. From the expectation values of $\sigma_z$ one can find that the Bloch bands are topological in Fig.~\ref{fig_topoband}(a,b) and trivial in Fig.~\ref{fig_topoband}(c,d). Further, together with the Eq.~(\ref{Ch2d}), in the former topological regime we have the Chern number ${\rm Ch}_1=+1$ for the lower band and ${\rm Ch}_1=-1$ for the upper band [Fig.~\ref{fig_topoband}(b)].

\begin{figure*}[btp]
\centering
\includegraphics[width=0.98\textwidth]{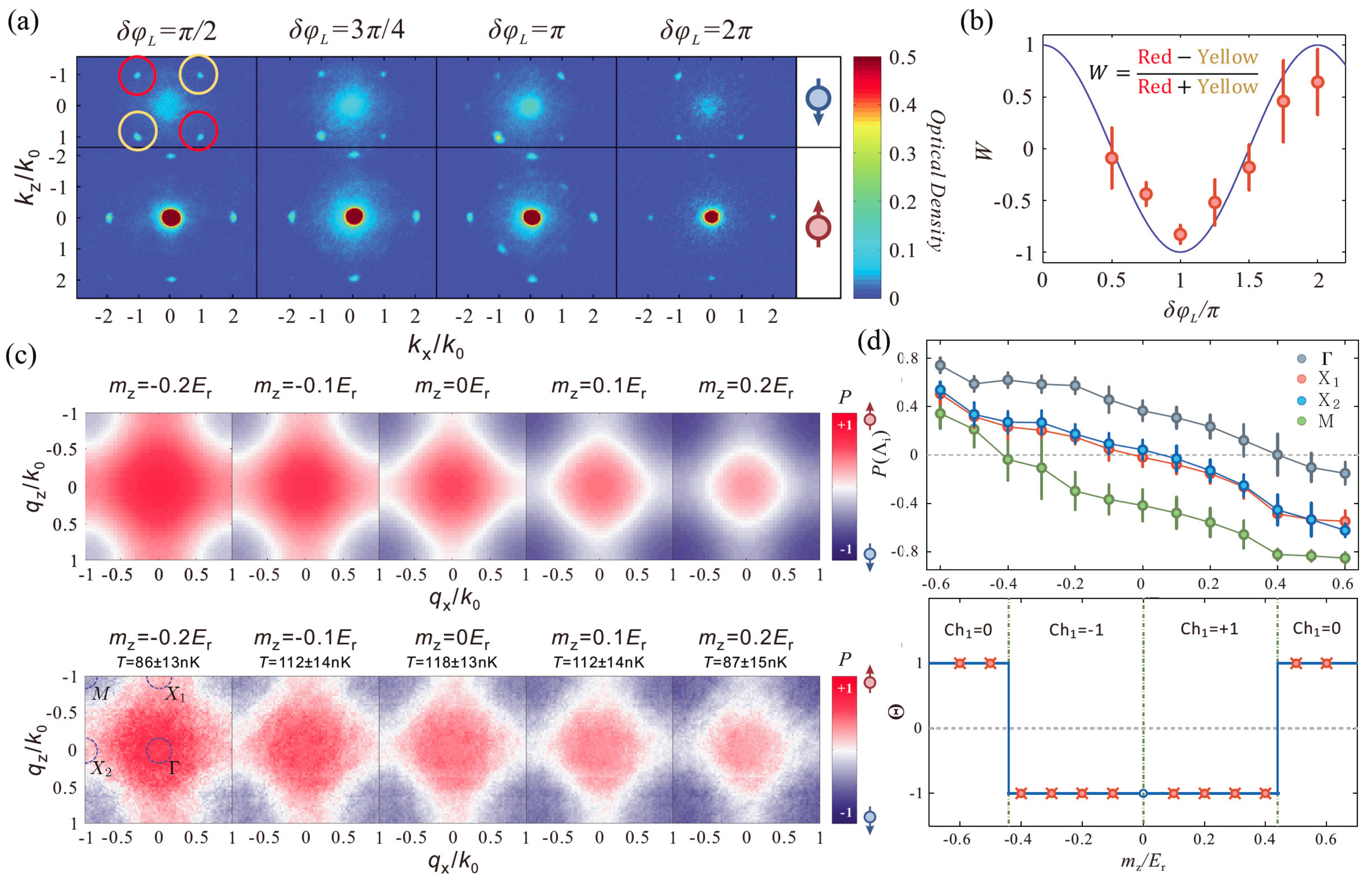}
\caption{Experimental demonstration~\cite{Wu2016}. (a-b) The crossover between 1D and 2D SO couplings.
Spin-resolved TOF images of BEC atoms are shown for different relative phase
$\delta\varphi_L$ (a) and the imbalance $W$
between the Raman coupling induced atoms in the two diagonal directions is measured as a function of $\delta\varphi_L$ (b).
(c-d) Spin texture and band topology. Spin texture at different $m_z$ is shown by tuning the two-photon detuning (c).
Experimental measurements (lower row) are compared to numerical calculations at $T=100$nK (upper row).
Spin polarization $P({\bf\Lambda}_j)$ at the four symmetric momenta
$\{{\bf\Lambda}_j\}=\{{\it \Gamma},X_1,X_2,M\}$ can be readily read as a function of $m_z$ (c),
which determines the product $\Theta=\Pi_{j=1}^4{\rm sgn}[P({\bf \Lambda}_j)]$ and
the Chern number ${\rm Ch}_1$.
In all the cases, $V_{0}=4.16E_{\rm r}$ and $M_{0}=1.32E_{\rm r}$.} \label{fig_exp2d}
\end{figure*}

\subsection{Experimental realization of 2D SO coupling and topological bands}

Wu {\em et al.}~\cite{Wu2016} performed the first experimental realization of the 2D Raman lattice scheme in a Bose gas
of $^{87}$Rb atoms. In the setup, the frequency difference is set at $\delta\omega=35$MHz, and the relative phase $\delta\varphi_L$
is controlled by the propagating length between the two mirrors [Fig.~\ref{fig_2dRaman}(a)]. Two observations are mainly preformed: (i) The
crossover between 1D and 2D SO couplings, which is reflected by the atom distributions in spin-resolved TOF images.
(ii) The measurements of spin texture and band topology.
The Chern number ${\rm Ch}_1$ [Eq.~(\ref{Ch2d})] can be readily read from the measured spin texture at a well-chosen temperature.

\subsubsection{1D-2D crossover}

The atoms are first prepared in the spin-up state and then
adiabatically loaded into the ${\it \Gamma}$ point. The
spin-resolved TOF expansion is performed to projects Bloch states onto free momentum states with fixed spin polarizations.
Figure~\ref{fig_exp2d}(a) shows the TOF images for various values of $\delta\varphi_L$.
For spin-up ($|\uparrow\rangle$) state, five atom clouds are observed:
besides the major BEC cloud retained at momentum $(k_x,k_z)=(0,0)$, four small fractions of BEC
clouds are transferred to momenta $(\pm2k_0,0)$ and $(0,\pm2k_0)$ by the first-order transition
due to the lattice potential $V_{\rm latt}$.
The SO coupling is reflected by the two or four small BEC clouds in the state $|\downarrow\rangle$,
depending on $\delta\varphi_L$, at the four diagonal corners with momenta $(\pm k_0,\pm k_0)$.
These atom clouds are generated by the Raman transitions, which flip spin and transfer momenta
of magnitude $\sqrt{2} k_0$ along the diagonal directions.
As given in Eq.~(\ref{Raman}), the Raman terms ${\cal M}_x$ and ${\cal M}_y$ depend on $\delta\varphi_L$.
For $\delta\varphi_L=\pi/2$, four small clouds in state $|\downarrow\rangle$ with TOF
momentum $\vec k=(\pm k_0,\pm k_0)$ are observed, reflecting the
2D SO coupling. On the other hand, by tuning the relative phase to $\delta\varphi_L=3\pi/4$,
the population of atom clouds in the two diagonal directions becomes imbalanced. Furthermore,
the system reduces to 1D SO couplings when $\delta\varphi_L= \pi$ and $2 \pi$, with
${\cal M}_x=M_x \pm M_y$ and ${\cal M}_y=0$.
In this case, the Raman pumping only generates a single diagonal pair of BEC clouds, as shown
in Fig.~\ref{fig_exp2d}a for $\delta\varphi_L = \pi, 2\pi$. This is similar to the 1D SO
coupling in the free space, where
the Raman coupling flips the atom spin and generates a pair of atom clouds with opposite momenta.
Fig.~\ref{fig_exp2d}(a) also shows a difference of distribution between lower left and upper right
BEC clouds at $|\downarrow\rangle$, which is due to non-tight-binding correction. A simple analysis
reveals that while the fully antisymmetric Raman terms $\cos(k_0x+\alpha)\cos(k_0z+\beta)$ have
negligible effects in the tight-binding limit of the lattice, they give finite contributions in the
moderate lattice regime and are responsible for such difference of distribution.
To quantify crossover effect, we define $W=({\cal N}_{\hat{x}-\hat{z}}-{\cal N}_{\hat{x}+
\hat{z}})/ ({\cal N}_{\hat{x}-\hat{z}}+{\cal N}_{\hat{x}+\hat{z}})$
to characterize the imbalance of the Raman coupling induced atom clouds,
with ${\cal N}_{\hat{x} \pm \hat{z}}$ denoting the atom number of the two BEC clouds
along the diagonal $\hat{x} \pm \hat{z}$ direction.
The result of $W$ is shown in Fig.~\ref{fig_exp2d}(b) and is characterized by a simple
cosine curve $\cos\delta\varphi_L$, signifying the crossover between 2D and 1D SO couplings
realized in the present BEC regime.

\subsubsection{Band topology}

To detect the topology of the bands, the spin distribution in the first Brillouin zone is measured
for the 2D isotropic SO coupling with $\delta\varphi_L=\pi/2$. For this purpose, a cloud of atoms needs to
prepared at a temperature such that the lowest band is occupied by a sufficient number of atoms,
whereas the population of atoms in the higher bands is small. Measurements of spin texture
at different temperatures~\cite{Wu2016} suggest that a temperature around $T=100$nK is
preferred to extract the spin texture information of the lowest band. In comparison,
if the temperature is too high, atoms are distributed over several bands and the visibility of
the spin polarization will be greatly reduced, while a too low temperature can also reduce
the experimental resolution since the atoms will be mostly condensed at the band bottom.

The spin polarization is then measured as a function of detuning $m_z$ to reveal the topology
of the lowest energy band, with $V_{0}=4.16E_{\rm r}$ and $M_{0}=1.32E_{\rm r}$.
The numerical calculations and TOF measured images of $P({\bf q})$ are given in Fig.~\ref{fig_exp2d}(c),
which also show agreement between the theoretical and experimental results.
In Fig.~\ref{fig_exp2d}(d),  the values of polarization $P({\bf \Lambda}_j)$ are plotted for the four
highly symmetric momenta ${\it \Gamma}$, $X_1$, $M$ and $X_2$.
It can be seen that $P(X_1)$ and $P(X_2)$ always have the same sign, while the signs of
$P({\it \Gamma})$ and $P(M)$ are opposite for small $|m_z|$ and the same for large $|m_z|$,
with a transition occurring at the critical value of $|m_z^c|$ which is a bit larger than $0.4E_r$.
At transition points the spin-polarization $P(X_1)$ or $P(X_2)$ vanishes due to the gap closing and thermal equilibrium.
From the measured spin polarizations, the product $\Theta$ and the corresponding Chern number
are readily read off and also plotted [Fig.~\ref{fig_exp2d}(d)].
The results agree well with numerical calculations which predict two transitions between the topologically trivial and nontrivial bands
near $m^c_z \approx \pm 0.44E_{\rm r}$. Note that around $m_z=0$, the spin-polarizations at $X_{1,2}$
change sign through zero, implying the gap closing at $X_{1,2}$ and a change of Chern number by $2$.
This confirms that for the 2D SO coupled system realized in the present experiment, the energy band
is topologically nontrivial when $0<|m_z|<|m_z^c|$, while it is trivial for $|m_z|>|m_z^c|$.

\subsection{Recent improvement of the realization and generalization to 3D SO coupling}

\begin{figure}[h]
\centerline{\includegraphics[width=0.9\columnwidth]{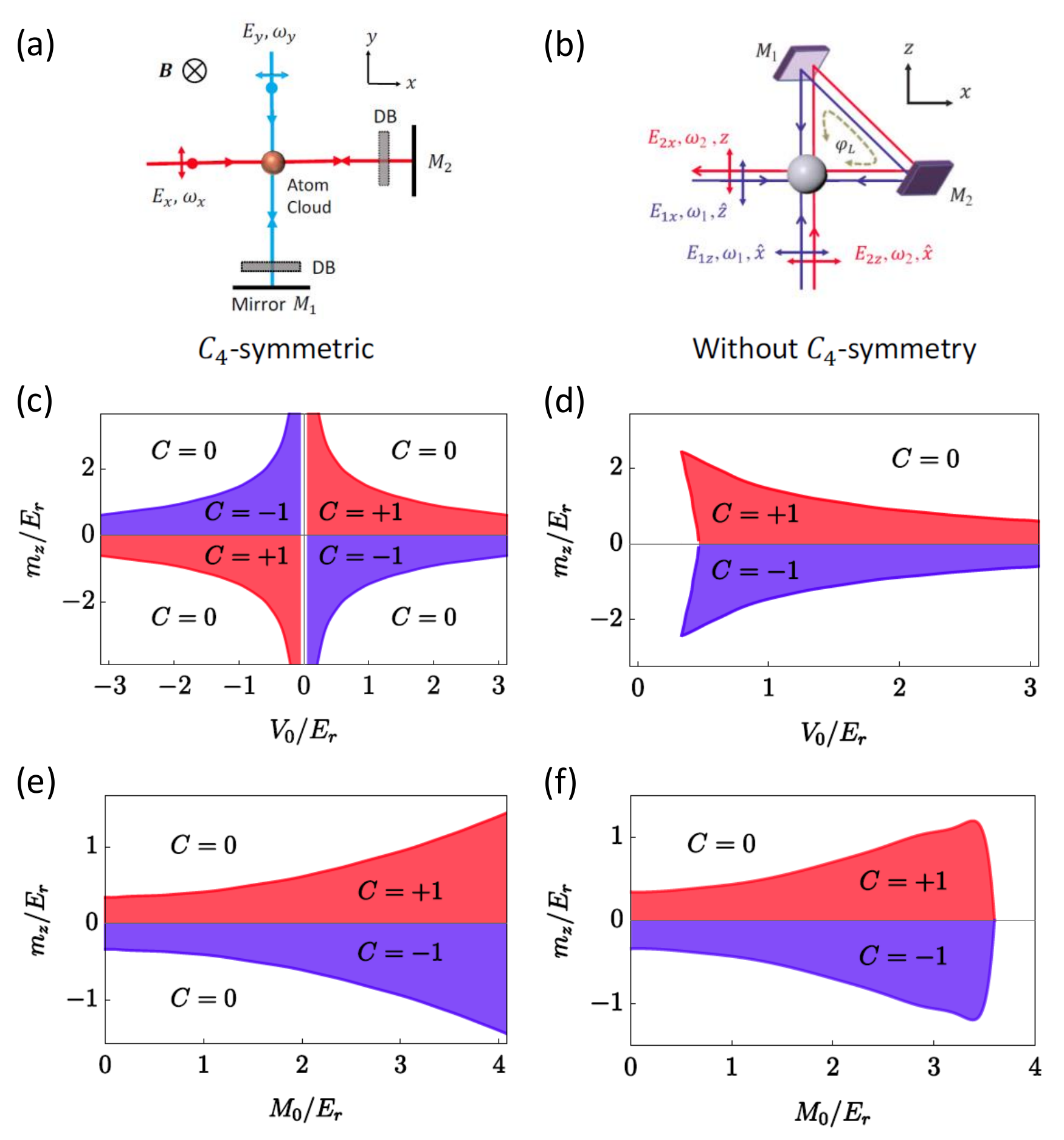}}
\caption{The new optical Raman lattice scheme for 2D Dirac type SO coupling~\cite{Wang2018}, and comparison to the previous realization~\cite{Wu2016}. (a) The new scheme applies two independent laser beams incident along $x$ and $y$ directions, through which the lattice and Raman potentials are generated simultaneously. The complete setup is as simple as the generation of the conventional square lattice for ultracold atoms, and the system is of inversion and $C_4$ symmetries. (b) The previous realization adopts the laser beams long $x$ and $z$ directions which are correlated by a long triangle loop formed though lattice area, mirrors $M_1$ and $M_2$. This system does not have precise inversion symmetry or $C_4$ symmetry, which are approximately valid only in the weak Raman coupling regime~\cite{Wu2016}. (c,e) With the new scheme the broad topological regime is obtained in the parameter space. (d,f) In the previous realization only a finite relatively narrow area of the parameter space supports topological phase~\cite{PanJS2016}.}
\label{newscheme}
\end{figure}

While the blue-detuned optical Raman lattice exhibits high feasibility in realizing the 2D SO coupling and topological physics, it still suffers a couple of limitations for the study. First, the realization is restricted in the blue-deunted regime, which has a limited tunability in the relative strength of the lattice and Raman potentials, since the optical transitions from $D_1$ and $D_2$ lines cancel out the Raman potentials. Moreover, the present scheme lacks the precise $C_4$ and inversion symmetry ${\cal P}$ due to the presence of the term like $\cos(k_0x-\varphi_L/2)\cos(k_0z-\varphi_L/2)$ in the Raman potentials [ [Fig.~\ref{newscheme}(b)]]. Such symmetry-breaking term, negligible only when the Raman potential is weak compared with the square lattice depth, can generically induce the coupling between the $s$-band and higher band states. This effect is shown to be detrimental and can reduce the topological region of the $s$-band~\cite{PanJS2016} [Fig.~\ref{newscheme}(d,f)]. Finally, the realization necessitates a relatively large bias magnetic field to split the hyperfine levels, so that the phase difference $\delta\varphi_L$ with required magnitudes can be achieved. Such a strong bias magnetic field can bring up heating due to the uncontrollable fluctuations.

To solve the challenges pointed above, a new optical Raman lattice scheme was proposed very recently for the realization of high-dimensional SO couplings with high controllability~\cite{Wang2018}. In the new realization, the system has a rigorous relative reflection symmetry/antisymmetry between lattice and Raman potentials, rendering a precise inversion symmetry of the QAH model [Fig.~\ref{newscheme}(a)]. This symmetry is irrespective of the type of detuning of optical transitions, namely, the realization is valid for both blue and red optical transitions, and also for transitions between $D_1$ and $D_2$ lines which render the optimal regime to realize SO couplings. In particular, in the case with $\delta\varphi=\pi/2$ and $M_{01}=M_{02}=M_0$, the system exhibits a precise $C_4$ symmetry defined in the 2D lattice plane: $(x,y;\sigma_x,\sigma_y) \to (y,-x;\sigma_y,-\sigma_x)$, giving $C_4HC_4^{-1}=H$. The precisely controllable symmetry, the phase diagram with topological region much broader than that in the previous case without exact inversion or $C_4$ symmetry has been predicted [Fig.~\ref{newscheme}(c-f)]. The new scheme was realized in a latest experiment with $^{87}$Rb atoms, where all the key advantages of the new scheme have been confirmed~\cite{Sun2017}. Especially, a long lifetime up to several seconds is observed in the realized 2D SO coupled gas.

Moreover, the new optical Raman lattice scheme can be extended to the realization of 2D Rashba type and 3D Weyl type SO couplings. Compared with the scheme for the 2D Dirac type SO coupling which has been focused on in the above discussions, the The generation of 2D Rashba and 3D Weyl types has an exquisite request that two sources of laser beams have distinct frequencies of factor-two difference. Interestingly, it was found that the $^{133}$Cs atoms provide an ideal candidate for such realization~\cite{Wang2018}. The new nontrivial topological physics were also predicted. The new optical Raman lattice schemes solve the essential challenges in exploring high-dimensional SO coupled quantum gases and shall advance the research in this direction, particularly in the quantum many-body physics and quantum far-from-equilibrium dynamics with novel topology for ultracold atoms.

\section{Topological superfluid and Majorana zero modes}

The realization of SO couplings beyond 1D regime for ultracold atoms advances an important step to explore in experiment the topological superfluid, which is a highly-sought-after phase hosting the exotic Majorana modes, as introduced in this section.

The discussion in this section is organized as follow. First we introduce the basics of Majorana modes, clarifying what kind of systems can host such modes. Then we introduce the spinless 1D $p$-wave and 2D chiral $p+ip$ SCs, and discuss the emergence of the Majorana modes in such intrinsic topological SCs. The realization of topological SC/superfluid from a hybrid system formed by SO coupled system with $s$-wave pairing order is briefly discussed. After the introduction to the background, we turn to showing a generic theory for the 2D chiral SCs/superfluids, with which the topology of a 2D SC/superfluid, characterized by Chern numbers, can be simply determined by Fermi surface (FS) properties and pairing symmetry. This theorem provides a simple but generic criteria to identify the topology of a 2D SC/superfluid phase, even the real system might be complicated. The application of this generic theory to various SO coupled systems is considered. Moreover, we introduce a generic theory for the existence of non-Abelian Majorana zero modes (MZMs), from which we show that the MZMs can exist in 2D trivial SCs. The results can be further generalized to 3D Weyl semimetal with superconductivity/superfluidity.

\subsection{Basics of Majorana zero modes}

Eighty years ago, Ettore Majorana proposed a new fermion which is a real solution to the Dirac equation, and
identical to its antiparticle and now called Majorana fermion (MF)
\cite{Majorana1981}, and speculated that it might interpret neutrinos. The self-hermitian property of the Majorana particle indicates that
the operator of a MF in the real space satisfies
\begin{eqnarray}
\gamma(\vec x)=\gamma^\dag(\vec x).
\end{eqnarray}
A direct consequence is that the Majorana field breaks $U(1)$ gauge symmetry and conserves no electric charge. Alternatively, the expectation value of a MF is zero, namely, it is charge neutral.

While the evidence of MFs as elementary particles in high energy physics
is yet elusive, the search for MFs, or MZMs, has energetically revived in condensed
matter physics and become an exciting pursuit in recent years~\cite{Wilczek2009,Alicea2012RPP,Franz2013a}. The quest for Majorana modes in solid state physics is mostly driven by both the exploration of the fundamental physics and the promising applications of such modes, obeying non-Abelian statistics, to a building block for fault-tolerant topological quantum computer \cite{non-Abelian1,non-Abelian2,non-Abelian3,non-Abelian4,non-Abelian5,TQC1,TQC2,TQC3}. Note that in condensed matter materials the only elementary particle is the electron which has an effective `antiparticle' called hole in solid state physics. A MF can then emerge as a quasiparticle in a solid state material, e.g. such quasiparticle can be formed as a superposition of electron and hole
\begin{eqnarray}
\gamma=uc+vc^\dag, \ \ u=v^*.
\end{eqnarray}
A natural hosting material for the Majorana-like quasiparticles could be SCs (or superfluids), where the superconducting pairing couples the electron and hole, leading to the qusiparticles in SCs being superpositions of electrons and holes. Nevertheless, for an $s$-wave SC, the pairing occurs between spin-up and spin-down electrons
\begin{eqnarray}
H^s_{\rm pair}=\sum_{\bold k}\Delta_sc_{\bold k,\uparrow}c_{-\bold k,\downarrow}+h.c.,
\end{eqnarray}
or equivalently, in the Nambu space it couples the spin-up electron and spin-down hole in the form $H_{\rm pair}=\sum_{\bold k}\Delta_sd^\dag_{\bold k,\uparrow}c_{\bold k,\downarrow}+h.c.$, where the hole operator is defined as $d_{\bold k,\uparrow}=c_{-\bold k,\uparrow}^\dag$. Thus the quasiparticle in the $s$-wave SC in general renders the superposition form of the electron and hole $b_{\bold k}=uc_{\bold k,\uparrow}+vd_{\bold k,\downarrow}=uc_{\bold k,\uparrow}+vc_{-\bold k,\downarrow}^\dag$, which is not a MF due to the distinction of the spin states of the electron and hole. One can soon find that the Majorana quasiparticle can be realized once the spin degree of freedom can be effectively removed in the SC. For a (an effective) spinless (or spin-polarized) fermion system, the basic superconducting pairing order is $p$-wave and
\begin{eqnarray}
H^p_{\rm pair}=\sum_{\bold k}\Delta_p(\bold k)c_{\bold k}c_{-\bold k}+h.c.,
\end{eqnarray}
with the parity-odd pairing $\Delta_p(\bold k)=-\Delta_p(-\bold k)$. Similar to the analysis for the $s$-wave SC, the quasiparticle in this case takes the form $\gamma_{\bold k}=uc_{\bold k}+vc_{-\bold k}^\dag$, which in real space gives $\gamma(\vec x)=u(\vec x)c(\vec x)+v(\vec x)c^\dag(\vec x)$. The MF is then resulted for $u=v^*$.

Due to the self-hermitian property, a single MZM has no well defined Hilbert space spanned by usual complex fermion quantum states. Instead, a complex fermion mode, whose Hilbert space defines a single qubit and is spanned by two fermionic quantum states $|0\rangle$ and $|1\rangle$, can be formed by two independent Majorana quasiparticles. This follows that the Hilbert space of two MFs equals to that of a single complex fermion mode, namely, the quantum dimension of a MF $d_\gamma^2=2$, which leads to a highly unusual consequence that a single Majorana mode has an irrational quantum dimension~\cite{TQC2}
\begin{eqnarray}
d_\gamma=\sqrt{2}.
\end{eqnarray}
The non-integer quantum dimension leads to an exotic property, namely, the non-Abelian statistics for the MZMs, which is the essential motivation in the recent years of extensive studies of topological SCs and in condensed matter physics and related topics in ultracold atoms.

\subsection{Intrinsic $p$-wave SCs}

\subsubsection{1D spinless $p$-wave SC}

The simplest toy model hosting Majorana modes is the 1D spinless $p$-wave SC, as proposed by Kitaev~\cite{Kitaev2001}. In the topologically nontrivial phase, at each end of the 1D system is located a Majorana zero bound mode. The Hamiltonian of the model is given below
\begin{eqnarray}\label{Kitaev1}
H=-\mu\sum_{j}c^\dag_jc_j-\frac{1}{2}\sum_{j}(tc_j^\dag c_{j+1}+\Delta e^{\phi}c_{j}c_{j+1}+h.c.),\nonumber\\
\end{eqnarray}
where $\mu$ is chemical potential, $t$ is hopping coefficient, and $\Delta$ is the $p$-wave pairing with a phase $\phi$. Transforming the above Hamiltonian into momentum space yields the Bogoliubov每
de Gennes (BdG) form
\begin{eqnarray}\label{Kitaev2}
H&=&\frac{1}{2}\sum_{k}{\cal C}_k^\dag{\cal H}_k{\cal C}_k,\\
{\cal H}_k&=&(-t\cos k-\mu)\tau_z-\sin k(\cos\phi\tau_y-\sin\phi\tau_x),\nonumber
\end{eqnarray}
where the operator ${\cal C}_k=[c_k, c_{-k}^\dag]^T$ in the Nambu space, and the Pauli matrices $\tau_{x,y,z}$ act on the Nambu space. It is convenient to rotate $\cos\phi\tau_y-\sin\phi\tau_x\rightarrow\tau_y$, so that ${\cal H}_k=-(t\cos k+\mu)\tau_z-\sin k\tau_y$. The present Hamiltonian is very similar to the case for the 1D AIII class topological insulator, as obtained in the 1D optical Raman lattice. The topology of the present system can then be studied in the similar way. First, the bulk of the present 1D superconductor is gapped when $\mu\neq\pm t$, and is gapless at $\mu=\pm t$, which corresponds to transition between topological and trivial phases. Note the Hamiltonian ${\cal H}_k$ has time-reversal symmetry ($T$) and charge-conjugation (particle-hole ${\cal C}$) symmetry, defined by
\begin{eqnarray}\label{symmetry1}
T=K, \ \ {\cal C}=\tau_xK,
\end{eqnarray}
with $K$ the complex conjugate. It follows that symmetries transform $T{\cal H}(k)T^{-1}={\cal H}(-k)$ and ${\cal C}{\cal H}(k){\cal C}^{-1}=-{\cal H}^*(-k)$. Thus the present system belongs to the so-called BDI class, with the topology being classified by 1D winding number. Similar to the 1D AIII class topological insulator, the 1D winding number is obtained straightforwardly by
\begin{eqnarray}
N_{1D}&=&\frac{1}{4\pi i}\int dk{\rm Tr}\bigr[\tau_x{\cal H}^{-1}(k)\partial_{k} {\cal H}(k)\bigr]\nonumber\\
&=&\frac{1}{2\pi}\int dk\hat h_z\partial_{k}\hat h_y,
\end{eqnarray}
where $(\hat h_y,\hat h_z)=(h_y,h_z)/h$, with $h_y=-\sin k, h_z=-t\cos k-\mu$, and $h=(h_y^2+h_z^2)^{1/2}$. It further gives that
\begin{eqnarray}
N_{1D}= \left\{ \begin{array}{ll}
        1, \ {\rm for}\ |\mu|<t, \\
             0, \ {\rm for}\ |\mu|>t. \\
        \end{array} \right.
\end{eqnarray}
The topological number has a simple intuitive picture that the vector $\vec h=(\hat h_y,\hat h_z)$ winds $2\pi$ over a circle when $k$ runs over the FBZ.

Similar to the insulating phase, the nontrivial topology with $|\mu|<t$ can a boundary mode at each end of the 1D system if considering open boundary condition. The only difference is that here the boundary mode is a MF, rather than a Dirac fermion mode. Let the open boundaries locate at $x=0,N$, respectively and diagonalizing $H$ in position space, we obtain the Majorana edge modes for the
boundaries $x=0,N$ as
\begin{eqnarray}\label{eqn:MZM1}
\gamma_L(x_j)&=&\frac{1}{\sqrt{{\cal N}}}[(\lambda_+)^{x_j/a}-(\lambda_-)^{x_j/a}]\bigr[c(x_j)e^{i\phi/2}\nonumber\\
&&+c^\dag(x_j)e^{-i\phi/2}\bigr],\\
\gamma_R(x_j)&=&\frac{i}{\sqrt{{\cal N}}}[(\lambda_+)^{(L-x_j)/a}-(\lambda_-)^{(L-x_j)/a}]\bigr[c(x_j)e^{i\phi/2}\nonumber\\
&&-c^\dag(x_j)e^{-i\phi/2}\bigr],
\end{eqnarray}
with ${\cal N}$ being the normalization factor and $\lambda_\pm=(\mu \pm\sqrt{\mu^2-t^2+|\Delta|^2})/(t+|\Delta|)$.  It is easy to verify that the both MZMs satisfy the self-hermitian property: $\gamma_{L,R}=\gamma_{L,R}^\dag$. Another important property of the MZMs is that when the phase factor $\phi$ varies by $2\pi$, one gets $\gamma_{L,R}\rightarrow-\gamma_{L,R}$. say each MZM acquires only $\pi$ phase. This property closely related to the fractional Josephson effect for $p$-wave SCs and the non-Abelian statistics of MZMs~\cite{Alicea2012RPP}.

\subsubsection{2D chiral $p+ip$ SC}

Similar to the connection between the 1D $p$-wave SC and the 1D AIII class insulator, the 2D $p+ip$ SC is a {\it superconducting version} of the quantum Hall effect~\cite{Read2000}. The Hamiltonian of the 2D $p+ip$ SC can be described by
\begin{eqnarray}\label{p+ip1}
H&=&\int d^2\bold r\biggr\{\psi^\dag(\bold r)\bigr(-\frac{\hbar^2}{2m}\nabla^2-\mu\bigr)\psi(\bold r)\nonumber\\
&&+\frac{\Delta}{2}\bigr[e^{i\phi}\psi(\partial_x+i\partial_y)\psi+h.c.\bigr],
\end{eqnarray}
where $\psi(\bold r)$ denotes the spinless fermion field operator at position $\bold r$ in the 2D space, $m$ is the mass of the fermion, and $\phi$ is the phase of the SC order $\Delta$. Again, transforming the Hamiltonian to $\bold k$ space we obtain
\begin{eqnarray}\label{p+ip2}
H&=&\frac{1}{2}\int d^2{\bold k}\Phi_{\bold k}^\dag{\cal H}_{\bold k}\Phi_{\bold k},\\
{\cal H}_{\bold k}&=&\xi(\bold k)\tau_z+k_x\Delta (\cos\phi\tau_y-\sin\phi\tau_x)\nonumber\\
&&-k_y\Delta(\cos\phi\tau_x+\sin\phi\tau_y),\nonumber
\end{eqnarray}
where $\Phi_{\bold k}=[\psi_{\bold k},\psi^\dag_{-\bold k}]^T$ and $\xi(\bold k)=\frac{\hbar^2k^2}{2m}-\mu$.
It is also convenient to rotate $\cos\phi\tau_y-\sin\phi\tau_x\rightarrow \tau_y$ and $\cos\phi\tau_x+\sin\phi\tau_y\rightarrow\tau_x$, so that ${\cal H}_{\bold k}=\xi(\bold k)\tau_z+k_x\Delta \tau_y-k_y\Delta\tau_x$. It is straightforward to know that for nonzero $\Delta$, the bulk of the SC is gapped when $\mu\neq0$. Accordingly, the gap is closed at $\bold k=0$ when $\mu=0$, which implies that the phase transition occurs, with the topology of the two regions with $\mu>0$ and $\mu<0$ being different. The regime of $\mu>0$ is called the `BCS' type weakly paired phase, while $\mu<0$ corresponds to the `BEC' type strongly paired phase~\cite{Read2000}.

It can be verified that the time-reversal symmetry $T=K$ is broken for the Hamiltonian, while the charge conjugation symmetry ${\cal C}=\tau_xK$ keeps. Thus the $p+ip$ SC belongs to the D class in the AZ ten-fold classification.
The topology of the present $p+ip$ SC is then characterized by Chern number, which can be calculated in the same way as done for QAH effect
\begin{eqnarray}
{\rm Ch}_{1}=-\frac{1}{24\pi^2}\int d\omega dk_x dk_y{\rm Tr}\bigr[{G}d {G}^{-1}\bigr]^3,
\end{eqnarray}
where the Green's function ${G}^{-1}(\omega,\bold q)=\omega+i\delta^+-{\cal H}(\bold k)$, and the trace is operated on the Nambu space. Integrating over the frequency space yields that
\begin{eqnarray}
{\rm Ch}_{1}=\frac{1}{4\pi}\int dk_xdk_y \bold {\hat h}\cdot\partial_{k_x}\bold {\hat h}\times\partial_{k_y}\bold {\hat h},
\end{eqnarray}
 where $\bold {\hat h}=(h_x,h_y,h_z)/|\bold h(\bold k)|$, with $h_x=-k_y\Delta, h_y=k_x\Delta$ and $h_z=\xi(\bold k)$. By a straightforward calculation one can find that
\begin{eqnarray}
{\rm Ch}_{1}= \left\{ \begin{array}{ll}
        +1, \ {\rm for}\ \mu>0, \\
             0, \ \ \  {\rm for}\ \mu<0. \\
        \end{array} \right.
\end{eqnarray}
In the topologically nontrivial phase $\mu>0$, the 2D SC supports chiral edge states in the boundary, which are analogy to the chiral edge states obtained in the boundary of the 2D quantum Hall effect. Nevertheless, in the present $p+ip$ SC, the edge states are Majorana modes, while in quantum Hall effect they are chiral Dirac fermions. Furthermore, when the SC order is attached with a vortex with $\Delta\rightarrow\Delta e^{i\theta(\bold r)}$, where $\theta(\bold r)$ is the azimuthal angle, a Majorana zero mode can be obtained in the vortex core. It is also noteworthy that the Pfaffian state of the $\nu=5/2$ fractional quantum Hall state can be mapped to a $p+ip$ SC ground state, hence hosting the MZMs~\cite{Moore-Read}. For more discussions about the MZMs localized in SC vortices the readers can refer to the nice review article by Alicea~\cite{Alicea2012RPP}.

\subsection{Topological superconductor/superfluid from a conventional $s$-wave pairing order}

While the intrinsic $p$-wave SCs naturally host the Majorana modes in the boundary and vortex cores, the materials with such intrinsic superconducting pairings are delicate and hard to synthesize.
More recently, it has been proposed that hybrid systems of $s$-wave SC and SO coupled matters with odd number of FSs can favor effective
$p$-wave pairing states, bringing the realization of
MZMs in realistic solid state experiments~\cite{Fu2008Majorana,Zhang2008,Sato2009,Sau2010Majorana,Lutchyn2010Majorana,Oreg2010Majorana,Choy2011,Nadj-Perge2013,Franz2015RevModPhys,Wilczek2009,Alicea2012RPP,Franz2013a}. In such hybrid systems, the the superconductivity is induced on the SO coupled material by proximity effect. Due to the presence of SO coupling, the parity-even ($s$-wave) and parity-odd ($p$-wave) pairing orders generically mix in the helical (eigenstate) bases at the interface. Under proper condition, e.g. by applying an external Zeeman field which kills the $s$-wave pairing while keeps the $p$-wave, the purely effective $p$-wave SC and Majorana modes can be obtained (More details for the topological superconductivity from proximity effect will be introduced in the later section after we present a generic theory for chiral topological superconductors).
Motivated by these proposals, experimental studies have been performed to observe Majorana induced zero bias conductance anomalies with different heterostructures formed by $s$-wave SCs and semiconductor nanowires~\cite{Mourik1003,Deng2012,Das2012,Finck2012}, magnetic
chains~\cite{Perge2014,Ruby2015,Meyer2016}, or topological insulators~\cite{JiaJin-Feng2015,JiaJin-Feng2016,HeQL2017}. Nevertheless, the current experimental observations are not fully unambiguous, and the rigorous proof of Majorana modes in experiment is yet to be available.

Along with the exciting progresses made in the solid state physics, the exploration of MFs with ultracold atom systems has been also proposed and extensively studied, see e.g.~\cite{Zhang2008,Sato2009,Zhu2011,Zhang2013,Qu2013,Hu2014}. The motivation is quite straightforward. The $s$-wave superfluid phase can be achieved in ultracold fermions by tuning the $s$-wave Feshbach resonance, which is a mature technology in ultracold atoms~\cite{Chin2010}. Together with the SO coupling synthesised for the cold atom systems, the effective $p$-wave superfluid from an $s$-wave Feshbach resonance can be obtained. Nevertheless, for ultracold atoms the superfluid should be realized intrinsically, rather than by proximity effect. As such an intrinsic $s$-wave superfluid cannot be achieved in 1D system, but at least for 2D or 3D systems. As a result, to observe MZMs in a topological superfluid from $s$-wave Feshbach resonance, to realize a 2D SO coupling for Fermi gas is necessary~\cite{Liu2014a}.

\subsection{Chiral topological superfluids/superconductors: a generic theory}

\begin{figure}[h]
\centerline{\includegraphics[width=0.9\columnwidth]{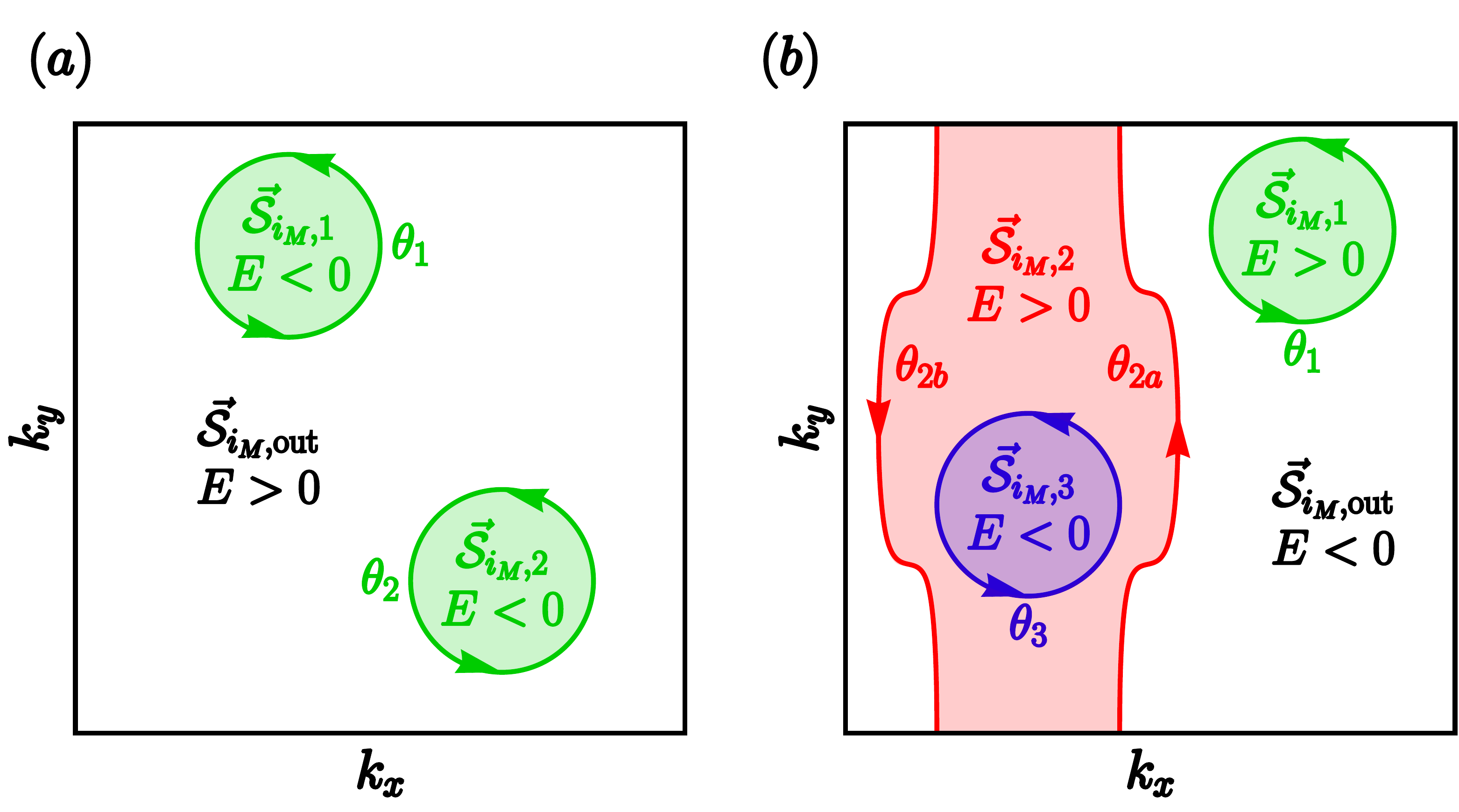}}
\caption{(a)-(b) Configurations of FSs of normal bands~\cite{Jeffrey2018PRB}. The solid lines with arrows denote FSs, which are boundaries of the vector areas $\vec S_{i_M,j}$ in $\bold k$ space. The blue and green colors denote the closed FSs, and the red color denotes open FSs. From the definition for the formula~\eqref{eqn:generalGenT}, we have for (a) that $q_{i_M,1}=q_{i_M,2}=q_{i_M}=1$, and for (b) that $q_{i_M}=q_{i_M,1}=q_{i_M,2}=0$, while $q_{i_M,3}=1$.}
\label{Fig:GenT}
\end{figure}

Instead of studying the topological superconductivity realized with various different platforms through proximity effect, we introduce in this subsection a generic theorem to determine the topology of a generic 2D system, as characterized by Chern numbers, after opening a gap through having superfluid/superconductivity~\cite{Jeffrey2018PRB}. Then we shall investigate the application of this theorem to various experimental systems. To simplify the description, we first classify the normal bands of the system without pairing into three groups: 1) the upper bands which are above the Fermi energy; 2) the lower bands which are below the Fermi energy; 3) the middle bands which are crossed by Fermi energy. In the most generic case, each middle band may have multiple FSs (Fermi loops), and we denote by $(i_M,j)$ the $j$-th FS loop of the $i_M$-th middle band. Let the total Chern number of the upper (lower) normal bands be $n_U$ ($n_L$). It can be shown that the Chern number the superfluid pairing phase induced in the system is given by~\cite{Jeffrey2018PRB}
\begin{eqnarray}\label{Chern1}
{\rm Ch}_1 &=& n_L - n_U + \sum_{i_M}\biggr[(-1)^{q_{i_M}}n_F^{(i_M)}- \nonumber \\
&&\frac{1}{2\pi}\sum_j (-1)^{q_{i_M,j}}\oint_{\partial \mathcal{\vec S}_{i_M,j}} \nabla_{\bold k} \theta_{\bold k}^{(i_M,j)} \cdot d{\bold k} \biggr]. \label{eqn:generalGenT}
\end{eqnarray}
Here $n_F^{(i_M)}$ is the Chern number of the $i_M$-th middle band and $\theta_{\bold k}^{(i_M,j)}={\rm arg}[\Delta^{(i_M,j)}_{\bold k}]$ is the phase of the pairing order projected onto the $(i_M,j)$-th FS loop. Note that the pairing can occur between two different Fermi surfaces, say between $(i_M,j)$ and $(i_M',j')$. In this case the Eq.~\eqref{Chern1} is still valid, but the integral will be performed on both Fermi surfaces at the same time. More details can be found in Ref.~\cite{Jeffrey2018PRB}. The integral direction is specified by arrows along FS lines in Fig.~\ref{Fig:GenT} (a,b), which defines the boundary $\partial \mathcal{\vec S}_{i_M,j}$ of the vector area $\vec S_{i_M,j}$ in $\bold k$ space. The quantities $\{q_{i_m,j},q_{i_M}\}=\{0,1\}$ are then determined by ``right-hand rule" specified below. The quantity $q_{i_M,j}=0$ (or $1$) if the energy of normal states within the area $\mathcal{\vec S}_{i_M,j}$ is positive (or negative), while $q_{i_M}=1$ (or $0$) if the region $\vec S_{i_M,\rm out}$, which is complementary to the sum of $\mathcal{\vec S}_{i_M,j}$, have positive (or negative) energy. Some typical examples are shown in Fig.~\ref{Fig:GenT}.
With this theorem the Chern number of the superfluid phase can be simply determined once we know the properties of lower and upper bands, and the normal states at the Fermi energy, which govern the phases of $\Delta^{(i_M,j)}_{\bold k}$.

We introduce the proof of the above theorem for a multiband system. However, to facilitate the discussion, here we focus on the case with a single FS. The generalization to the multiple FSs will be assessed, with the detailed generic proof can be found in Ref.~~\cite{Jeffrey2018PRB} (while some typos are corrected here). We write down the BdG Hamiltonian for a generic multi-band system by
\begin{eqnarray}
{\cal H}_{\text{\text{BdG}}}(\bold k)= \left(\begin{matrix}
{\cal H}(\bold k)-\mu & \Delta(\bold k) \\
\Delta^\dag(\bold k) & -{\cal H}^T(-\bold k)+\mu
\end{matrix}\right),
\end{eqnarray}
where ${\cal H}(\bold k)$ is the normal Hamiltonian, $\Delta(\bold k)$ is the pairing order matrix, with $$\Delta_{\bold k}^{(j_1,j_2)} = \langle u_{\bold k}^{(j_1)}| \Delta({\bold k}) | u_{-{\bold k}}^{(j_2)*}\rangle$$ for the two normal bands $j_1$ and $j_2$, and $\bold k$ is the local momentum measured from FS center. Note that if FS is not symmetric with respect to its center, one can continuously deform it to be symmetric without closing the gap. In this process the topology of the system is not changed. Finally we can always write down ${\cal H}_{\rm BdG}$ in the above form to study the topology. Denote by $u_{\bold k}^{(i_M)}$ the eigenvector of the normal band crossing Fermi energy and
\begin{eqnarray}
\mathcal{H}(\bold k) u_{\bold k}^{(i_M)} = \epsilon_{\bold k}^{(i_M)} u_{\bold k}^{(i_M)},
\end{eqnarray}
where $\epsilon_k^{(i_M)} $ denotes the normal dispersion relation.
Focusing on the pairing on FS, the eiegenstates of ${\cal H}_{\rm BdG}$ takes the generic form
\begin{eqnarray}
|u(\bold k)\rangle=[..., \alpha_k^{(i_M)}u_k^{(i_M)}; \beta_k^{(i_M)}u_{-k}^{(i_M)*}, ...]^T,
\end{eqnarray}
where `$...$' denotes the components from the subbands other than $i_M$-th one. The eigenstate is associated with a Berry's connection calculated by
\begin{eqnarray}
\mathcal{A}_{\bold k}=i\hbar\langle u(\bold k)|\nabla_{\bold k}|u(\bold k)\rangle.
\end{eqnarray}

The key process is that we consider the weak pairing potential limit with $\Delta(k) \rightarrow \gamma \Delta(k)$ with $\gamma \rightarrow 0^+$ (without closing bulk gap), in which case we can expect that the contribution from superconducting pairing to the Berry connection is fully dominated by the states right at FS. We shall then extract the $\mathcal{A}_{\bold k}^{(i_M)}$ component our of $\mathcal{A}_{\bold k}$, namely, the Berry connection corresponding to the middle $i_M$-th band, given by
\begin{eqnarray}\label{Berryconnection}
\mathcal{A}_{\bold k}^{(i_M)} &=& i [\alpha^{(i_M)}({\bold k}) u_{\bold k}^{(i_M)}]^\dag\nabla_{\bold k}\bigr[\alpha^{(i_M)}({\bold k}) u_{\bold k}^{(i_M)}\bigr]+\nonumber\\
&&i[\beta^{(i_M)}({\bold k}) u_{-{\bold k}}^{(i_M)*}]^\dag\nabla_{\bold k}\bigr[\beta^{(i_M)}({\bold k}) u_{-{\bold k}}^{(i_M)*}\bigr].
\end{eqnarray}
From the generic BdG equation we can obtain the coefficients in the weak pairing order regime by
\begin{eqnarray}
\alpha^{(i_M)}&=&\frac{\bar\epsilon_{\bold k}^{(i_M)}-\mu - \bigr[\bigr|\gamma\Delta_k^{(i_M)}\bigr|^2+\bigr(\frac{\epsilon_k^{(i_M)} + \epsilon_{-k}^{(i_M)}}{2}-\mu \bigr)^2\bigr]^{1/2}}{N^{(i_M)}(k)}, \nonumber\\
\beta^{(i_M)}&=&\frac{\gamma\Delta_k^{(i_M)*}}{N^{(i_M)}(k)}, \nonumber\\
N^{(i_M)}&=& \sqrt{\left| \alpha^{(i_M)}(k) \right|^2 + \left| \beta_{\pm}^{(i_M)}(k) \right|^2}, \nonumber
\end{eqnarray}
where $\bar\epsilon_{\bold k}^{(i_M)}=(\epsilon_{\bold k}^{(i_M)} + \epsilon_{-\bold k}^{(i_M)})/2$ and $\Delta_k^{(i_M)}=\Delta_k^{(i_M,i_M)}$. The Chern number of the superfluid is then obtained by
\begin{eqnarray}
{\rm Ch}_1=n_L-n_U+(2\pi)^{-1}\oint\nabla_{\bold k} \times \mathcal{A}_k^{(i_M)}d^2k.
\end{eqnarray}
By a straightforward calculation we obtain the Berry curvature associated with $\mathcal{A}_k^{(i_M)}$ in the weak pairing order limit by
\begin{eqnarray}
\mathcal{B}^{(i_M)}_{k} &=& \left[(2\Theta_{\vec{\mathcal{S}}}-1)\widetilde{\mathcal{B}}_k^{(i_M)} + 2\nabla_k \Theta_{\vec{\mathcal{S}}} \times \widetilde{\mathcal{A}}_k^{(i_M)} \right] \nonumber\\
 &-&\nabla_{\bold k} \times  \frac{\Theta_{\vec{\mathcal{S}}}\Im\bigr[\Delta_k^{(i_M)}* \nabla_k \Delta_k^{(i_M)}\bigr]}{\bigr| \Delta_k^{(i_M)} \bigr|^2}.
\end{eqnarray}
Here $\Theta_{\vec{\mathcal{S}}}$ denotes the step function that is $1$ for the area where the normal states have positive energies, and $0$ for the area where the normal states have negative energies, representing a step change when crossing the FS.
Let $\theta_k^{(i_M)}= {\rm arg}\bigr[\langle u_k^{(i_M)}|\Delta(k)| u_{-k}^{(i_M)*}\rangle\bigr]$ be the phase of order parameter on FS. With the above results we get further the Chern number of the superfluid phase by
\begin{eqnarray}
{\rm Ch}_1 &=&n_L-n_U +(-1)^{q_{i_M}} \biggr[n_F^{(i_M)}+\frac{1}{\pi} \biggr(\oint_{\partial{\vec S}_{i_M}} \widetilde{\mathcal{A}}_k^{(i_M)} \cdot dk \nonumber\\
& -&\int_{\mathcal{\vec S}_{i_M}} \nabla_{\bold k}\times\widetilde{\mathcal{A}}_k^{(i_M)}d^2k - \frac{1}{2}\oint_{\partial \mathcal{\vec S}_{i_M}} \nabla_k \theta_k^{(i_M)}\cdot dk\biggr) \biggr]\nonumber\\ \label{eqn:gaugeGenT}
\end{eqnarray}
with $\widetilde{\mathcal{A}}^{(i_M)}_k=i [u_{\bold k}^{(i_M)}]^\dag\nabla_{\bold k}u_{\bold k}^{(i_M)}$ being the Berry connection for the normal band states. If we choose a gauge so that $\widetilde{\mathcal{A}}^{(i_M)}_k$ is smooth on $\mathcal{\vec S}_{i_M}$, the two terms regarding $\widetilde{\mathcal{A}}^{(i_M)}_k$ in the right hand side of Eq.~\eqref{eqn:gaugeGenT} cancels. We then reach the formula~\eqref{Chern1} for the case with a single FS.

The proof can be generalized to the case with generic multiple FSs which may be closed or open~[Fig.~\ref{Fig:GenT}(b)], with multiple bands crossed by Fermi energy, and with the pairing within each FS or between two different FSs, given that the pairing fully gaps out the bulk~\cite{Jeffrey2018PRB}. Actually, for the case with multiple FSs, if the pairing occurs within each FS, which renders the Fulde-Ferrell-Larkin-Ovchinnikov (FFLO) or pair density wave (PDW) type orders~\cite{FF1964,LO1964}, the totoal Chern number is simply a summation of the contribution from all FSs. On the other hand, if the pairing occurs between two different FSs, the integration in the formula~\eqref{Chern1} is performed on both FSs simultaneously. The generic theorem introduced here is not restricted by pairing types, is powerful to quantitatively determine the topology of the superfluid phases, and can be particularly useful for condensed matter materials when the system is complicated. We also note that this theorem is best applied to judge the topology of phases with relatively weak pairing orders. This is because a strong pairing order may fully deform FSs of the original system and then change Chern number governed by Eq.~\eqref{Chern1}, driving a topological phase transition~\cite{Jeffrey2018PRB}. However, monitoring such phase transitions with increasing pairing orders can determine the whole topological phase diagram.

\subsection{Applications to topological superfluids (superconductors) for SO coupled systems}

In this subsection we introduce the applications of the generic theorem given in formula~\eqref{Chern1} to various types of SO coupled systems with $s$-wave superfluid/supercontuctor pairing.

\subsubsection{TI \& $s$-wave SC}

\begin{figure}
\includegraphics[width=0.6\columnwidth]{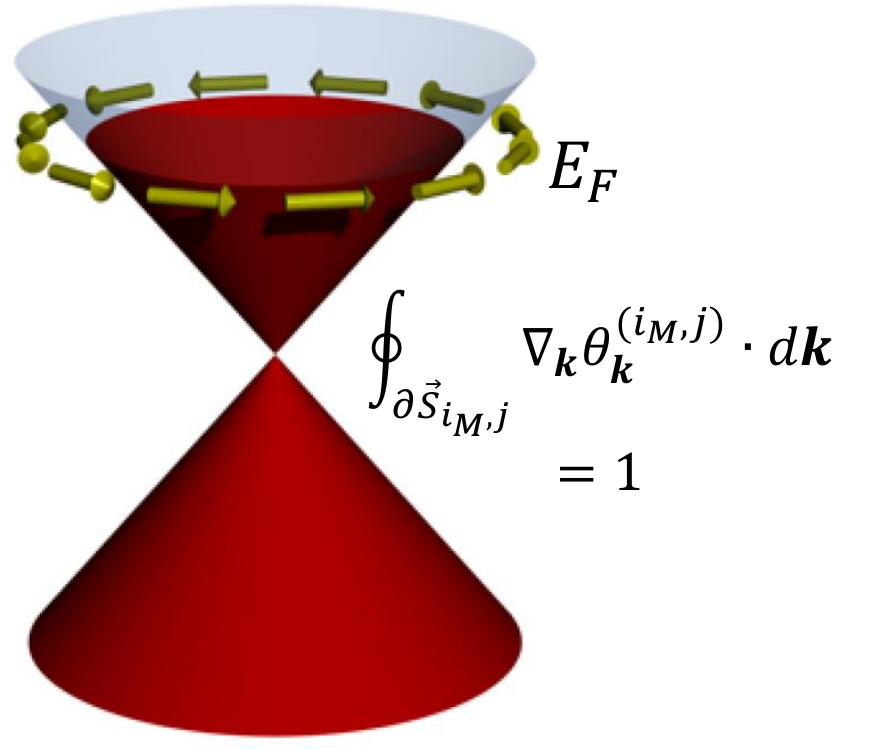}
\caption{\label{TIsurface} Surface states of the 3D topological insulator, with spin-momentum locking being shown at the Fermi surface. The phase winding of the pairing order projected onto the Fermi surface gives $+1$.}
\end{figure}
We first consider the 2D hybrid system formed by TI surface states and an $s$-wave SC (or superfluid)~\cite{Fu2008Majorana}. The TI surface states are described by (2+1)d Dirac Hamiltonian. Together with the $s$-wave pairing order induced by the substrate SC through proximity effect, the effective Hamiltonian can be written down as
\begin{eqnarray}\label{TI-SC1}
H&=&H_{\rm TI-surf}+H^s_{\rm pair},\\
H_{\rm TI-surface}&=&\int d^2{\bold k}c_{\bold k,s}^\dag\bigr[v_F(k_x\sigma_y-k_y\sigma_x)-\mu\bigr]_{ss'}c_{\bold k,s'},\nonumber\\
H^s_{\rm pair}&=&\int d^2{\bold k}\Delta_sc_{\bold k,\uparrow}c_{-\bold k,\downarrow}+h.c.,\nonumber
\end{eqnarray}
where $v_F$ is the Fermi velocity of the surface states. The normal states of the surface Hamiltonian ${\cal H}_{0}(\bold k)=v_F(k_x\sigma_y-k_y\sigma_x)$ read $|u_{+}(\bold k)\rangle=[1,e^{i\varphi(\bold k)}]/\sqrt{2}$ and $|u_{-}(\bold k)\rangle=[-e^{-i\varphi(\bold k)},1]/\sqrt{2}$, with $\varphi(\bold k)=\tan^{-1}(-k_x/k_y)$, and the energy $\epsilon_\pm=\pm v_F(k_x^2+k_y^2)^{1/2}$. If the Fermi energy is located at the upper band (Fig.~\ref{TIsurface}), the projected pairing on the FS is given by
\begin{eqnarray}\label{TI-SC2}
\Delta_+(\bold k)&=&\Delta_s\langle u_+(\bold k)|T|u_+(-\bold k)\rangle\nonumber\\
&=&\frac{\Delta_s}{2}\bigr[-e^{-i\varphi(-\bold k)}+e^{-i\varphi(\bold k)}\bigr]\nonumber\\
&=&\Delta_s e^{-i\varphi(\bold k)}.
\end{eqnarray}

We can apply the formula~\eqref{Chern1} to obtain the Chern number of the gapped superconducting phase. Note that here we only have a single band crossing the Fermi energy (i.e. one middle band). We have that $n_U=n_L=0$, and $n_F=0$. For the Fermi energy located above the Dirac point, we have all the factors $q_{i_M}=q_{i_M,j}=1$. Thus the Chern number
\begin{eqnarray}\label{Chern1-TI-SC}
{\rm Ch}_1 &=&-\frac{1}{2\pi}\oint_{\partial \mathcal{\vec S}_{i_M,j}} \nabla_{\bold k} \varphi_{\bold k} \cdot d{\bold k} \nonumber\\
&=&-1.
\label{eqn:generalGenT}
\end{eqnarray}
It is easy to confirm that when the Fermi energy is located below the Dirac point, the Chern number is again ${\rm Ch}_1=-1$, so the phase is the same.

\subsubsection{Rashba SO system \& $s$-wave SC \& Zeeman splitting}

Now we consider the 2D hybrid system formed by 2D Rashba SO coupled system with an external Zeeman field and the $s$-wave SC~\cite{Sato2009,Sau2011}. The effective Hamiltonian can be written down as
\begin{eqnarray}\label{Rashba-SC1}
H&=&H_{0}+H^s_{\rm pair},\\
H_{0}&=&\int d^2{\bold k}c_{\bold k,s}^\dag\bigr[\frac{\hbar^2\bold k^2}{2m}-\mu+\lambda_R(k_x\sigma_y-k_y\sigma_x)\nonumber\\
&&+V_z\sigma_z\bigr]_{ss'}c_{\bold k,s'},\nonumber\\
H^s_{\rm pair}&=&\int d^2{\bold k}\Delta_sc_{\bold k,\uparrow}c_{-\bold k,\downarrow}+h.c.,\nonumber
\end{eqnarray}
where $V_z$ is the external Zeeman field along the $z$ direction, and $\lambda_R$ is the Rashba SO coefficient. The normal states of the surface Hamiltonian ${\cal H}_{0}=\frac{\hbar^2\bold k^2}{2m}-\mu+\lambda_R(k_x\sigma_y-k_y\sigma_x)+V_z\sigma_z$ read $|u_{+}(\bold k)\rangle=[\cos(\theta/2),\sin(\theta/2) e^{i\varphi(\bold k)}]$ and $|u_{-}(\bold k)\rangle=[-\sin(\theta/2) e^{-i\varphi(\bold k)},\cos(\theta/2)]$, with $\varphi(\bold k)=\tan^{-1}(-k_x/k_y)$ and $\theta=\tan^{-1}(\lambda_R k/V_z)$, and the energy $\epsilon_\pm=\frac{\hbar^2\bold k^2}{2m}-\mu\pm \lambda_R(k_x^2+k_y^2)^{1/2}$. Assume that the Fermi energy is located at the lower band. The projected pairing on the FS is given by
\begin{eqnarray}\label{Rashba-SC2}
\Delta_-(\bold k)&=&\Delta_s\langle u_-(\bold k)|T|u_-(-\bold k)\rangle\nonumber\\
&=&{\Delta_s}\sin(\theta/2)\cos(\theta/2)\bigr[-e^{i\varphi(\bold k)}+e^{i\varphi(-\bold k)}\bigr]\nonumber\\
&=&-\Delta_s \sin(\theta/2)\cos(\theta/2) e^{i\varphi(\bold k)}.
\end{eqnarray}

\begin{figure}[b]
\includegraphics[width=1.0\columnwidth]{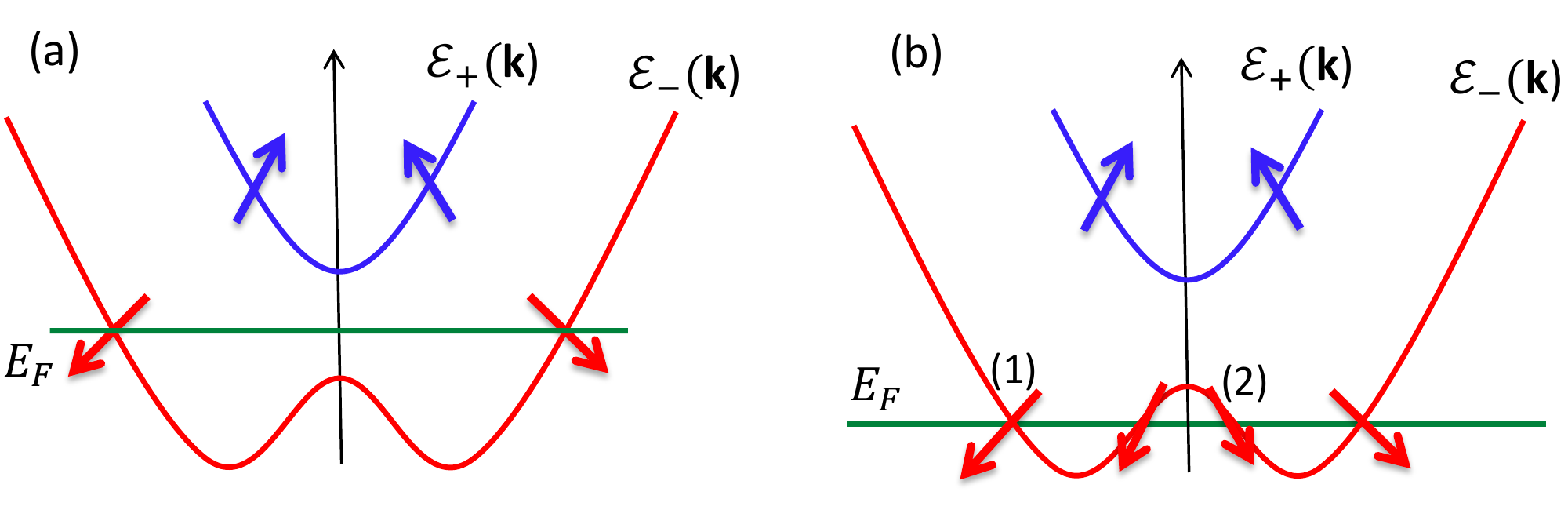}
\caption{\label{Rashbafigure} Sketch of the normal bands for a Rashba SO coupled system. (a) The Fermi energy locates within the Zeeman gap. (b) The Fermi energy locates below the Zeeman gap and crosses the lower subband at four Fermi points, leading to two Fermi surfaces: (1) the outer one and (2) the inner one.}
\end{figure}
We can also apply the formula~\eqref{Chern1} to obtain the Chern number of the gapped superconducting phase with different parameter conditions by considering the weak pairing order regime. First we take that the Fermi energy is located inside the Zeeman gap opened at $\bold k=0$, so that there is only one middle band [Fig.~\ref{Rashbafigure} (a)]. For the Rashba system we have that $n_U=n_L=0$, and $n_F=0$. For the present Fermi energy configuration, we have all the factors $q_{i_M}=q_{i_M,j}=1$. Thus the Chern number
\begin{eqnarray}\label{Chern1-Rashba-SC}
{\rm Ch}_1 &=&-\frac{1}{2\pi}\oint_{\partial \mathcal{\vec S}_{i_M,j}} \nabla_{\bold k} \varphi_{\bold k} \cdot d{\bold k} \nonumber\\
&=&+1.
\label{eqn:generalGenT}
\end{eqnarray}
On the other hand, if the Fermi energy crosses two FSs, as in the configuration shown in Fig.~\ref{Rashbafigure} (b), we have for the two FSs that have that $n_U=n_L=0$, and $n_F=0$, while $q_{i_M}=q_{i_M,1}=1$ and $q_{i_M,2}=0$. Then the Chern number reads
\begin{eqnarray}\label{Chern1-Rashba-SC2}
{\rm Ch}_1 &=&\frac{1}{2\pi}\oint_{\partial \mathcal{\vec S}_{i_M,2}} \nabla_{\bold k} \varphi_{\bold k} \cdot d{\bold k} -\frac{1}{2\pi}\oint_{\partial \mathcal{\vec S}_{i_M,1}} \nabla_{\bold k} \varphi_{\bold k} \cdot d{\bold k}\nonumber\\
&=&1-1=0.
\label{eqn:generalGenT}
\end{eqnarray}
Thus the cases in Fig.~\ref{Rashbafigure} (a) and Fig.~\ref{Rashbafigure} (b) correspond to two different basic phases, with one being topologically nontrivial and one trivial (similarly, if the Fermi energy is located above the Zeeman gap, the phase is also trivial). The quantitative transition between the two different phase occurs when the gap is closed. This can be obtained by examining the bulk gap, giving the critical point equation by $V_z^2=\mu^2+\Delta_s^2$. The topologically nontrivial regime corresponds to $V_z^2>\mu^2+\Delta_s^2$.

\begin{figure}[h]
\centerline{\includegraphics[width=1\columnwidth]{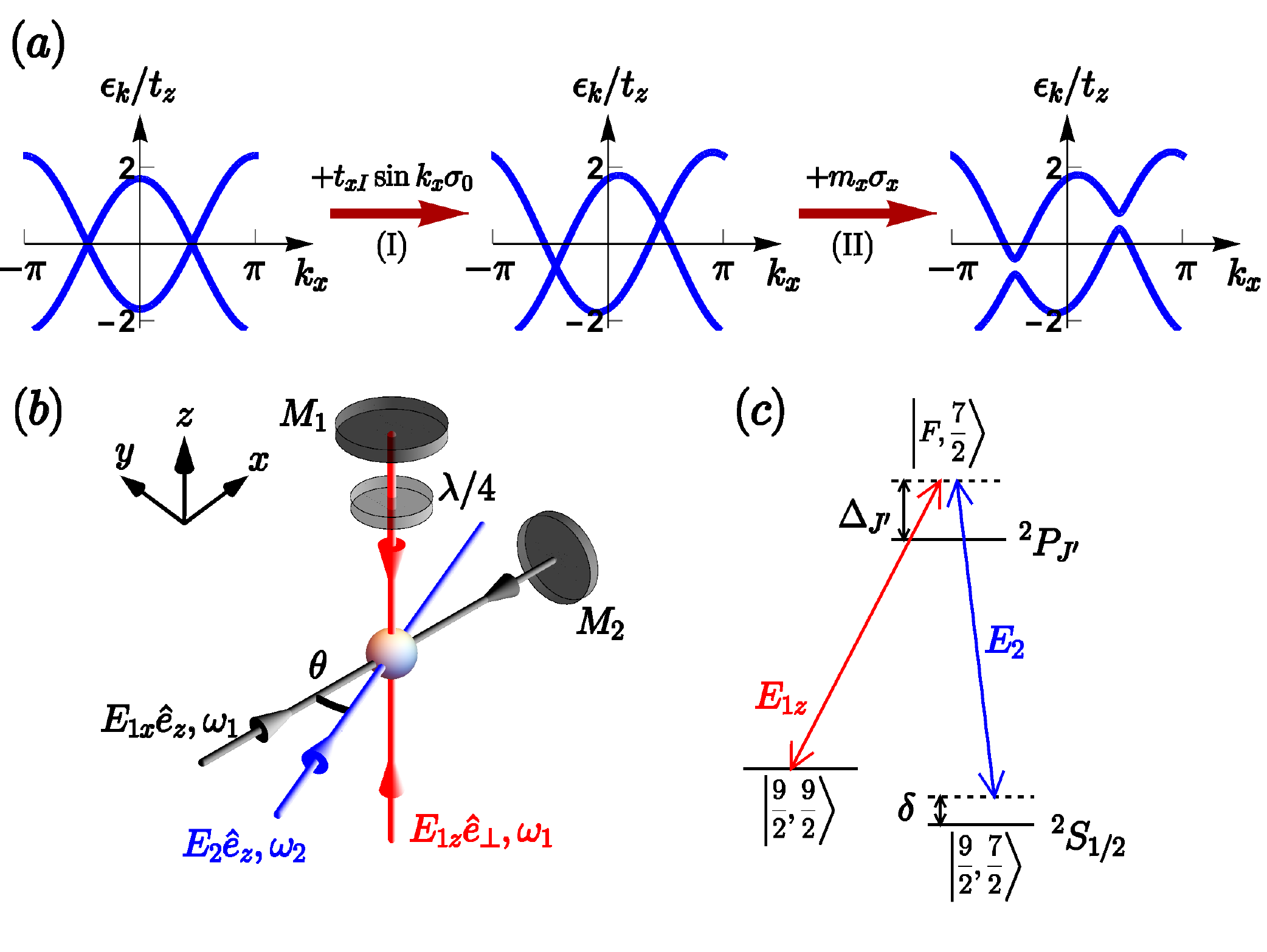}}
\caption{Sketch for band structure and realization of a 2D SO coupled Dirac metal~\cite{Jeffrey2018PRB}. (a) Schematic diagram of the band structure, with the inversion symmetry broken (process I) and gap opening at Dirac points (II) for a 2D SO coupled Dirac semimetal. (b) Proposed experimental setting for realization. The standing wave lights formed by $\bold E_{1x,1z}$ generate a blue-detuned square lattice. The incident polarization of $\bold E_{1z}$ is $\hat{e}_\bot = \alpha \hat e_x + i\beta \hat e_y$, and the $\lambda/4$-wave plate changes the polarization of the reflected field to $\hat{e}'_\bot = \alpha \hat e_x - i\beta \hat e_y$. The Raman coupling, illustrated in (c) for $^{40}$K fermions, is generated by $\bold E_{1z}$ and an additional running light $\bold E_2$ which has tile angle $\theta$ with respect to $x$-$z$ plane.}
\label{Fig:Realization}
\end{figure}

\subsubsection{PDW state for a Dirac metal}

We consider the effective tight-binding Hamiltonian $H_{\text{TB}}$ on a square lattice proposed in Ref.~\cite{Jeffrey2018PRB}
\begin{eqnarray}
H_{\text{TB}} &=& \sum_{\bold k} \left( \begin{matrix}
c_{\bold k\uparrow}^\dagger, c_{\bold k\downarrow}^\dagger
\end{matrix}\right) \mathcal{H}_{\text{TB}}\left( \begin{matrix}
c_{\bold k\uparrow} \\
c_{\bold k\downarrow}
\end{matrix}\right); \nonumber \\
\mathcal{H}_{\text{TB}} &=& \left( m_z - 2t_x \cos k_x - 2t_z \cos k_z\right) \sigma_z \nonumber \\
&+&2t_{so} \sin k_z \sigma_y + t_{xI} \sin k_x \sigma_0 + m_x \sigma_x, \label{HTB}
\end{eqnarray}
where $s=\uparrow,\downarrow$, $t_{x/z}$ is the hoping constant along $x/z$ direction, $t_{so}$ is the strength of spin-flip hopping, and $m_{z,x}$ denote the effective Zeeman couplings. As described in Fig.~\ref{Fig:Realization}(a), the above Hamiltonian describes a Dirac semimetal if $m_x=0$ and $|m_z|<2(t_x+t_z)$, with two Dirac points at $\mathbf{\Lambda_{\pm}} = \left(\pm \cos^{-1} [(m_z-2t_x)/2t_z],0\right)$. The term $t_{xI} \sin k_x \sigma_0$ breaks the inversion symmetry and leads to an energy difference between the two Dirac points. Finally, a nonzero $m_x$ opens a local gap at the two Dirac points. For simplicity we take that $t_z = t_0, t_x = t_0 \cos \theta_0$ and $t_{xI} = 2 t_0 \sin \theta_0$ to facilitate the further discussion. The realization with a new optical Raman lattice is sketched in Fig.~\ref{Fig:Realization}(b,c), where only a single Raman coupling is applied in the scheme. The details of the realization are neglected here, but the readers can refer to Ref.~\cite{Jeffrey2018PRB}. Note that here the 2D Dirac metal is driven by SO interaction, and is distinct from graphene, of which the Dirac points are protected
by symmetry only if SO coupling is absent~\cite{CharlesL2015}.

\begin{figure}[h]
\centerline{\includegraphics[width=\columnwidth]{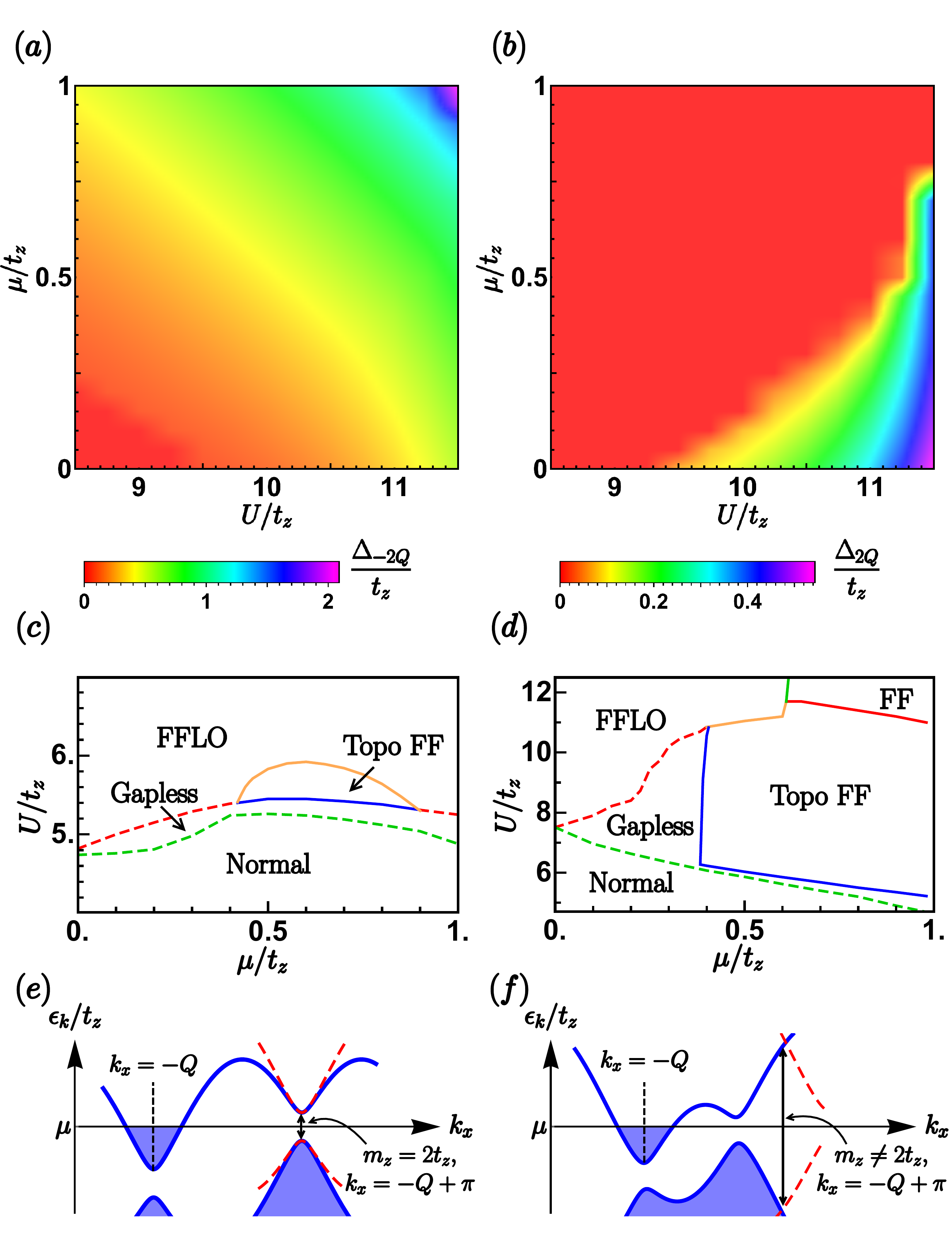}}
\caption{(a)-(b) The superfluid order by self-consistent theory and phase diagram~\cite{Jeffrey2018PRB}. The superfluid order $\Delta_{-2Q}$ dominates over others for $\mu>0$. (c) Phase diagram with normal-size topological region when $t_x=t_{so}=t_z, t_{xI}=0.7t_z, m_z=2t_z, m_x=0.3t_z$. (d) Phase diagram with broad topological region. The values of the parameters in diagrams (a,b,d) are $m_z=2.92t_z, t_x=0.92t_z, t_{so}=t_z, t_{xI}=0.8t_z, m_x=0.3tz$. Schematic diagrams showing the underlying mechanism for the normal-size (c) and broad (d) topological regions.}
\label{DeltaCal}
\end{figure}

The superfluid phase can be induced by considering an attractive Hubbard interaction. The total Hamiltonian
\begin{eqnarray}
H=H_{\rm TB}-U\sum_i n_{i\uparrow} n_{i \downarrow}.
\end{eqnarray}
Due to the existence of multiple FSs corresponding to different Dirac cones, in general we can have intra-cone pairing (FFLO) and inter-cone (BCS) pairing orders, defined by $\Delta_{2q} = \left(U/N\right) \sum_k \langle c^\dagger_{q+k\uparrow} c_{q-k\downarrow}\rangle$, with $q=\pm Q$ or $0$. Note that for the present Dirac system, a BCS pairing cannot fully gap out the bulk but leads to nodal phases, while the FFLO or FF order can~\cite{Jeffrey2018PRB}. On the other hand, since the inversion symmetry is broken, the BCS pairing would be typically suppressed. From the mean field results of a uniform system shown in Fig.~\ref{DeltaCal} (a,b), we find that the BCS pairing nearly vanishes, and $\Delta_{-2Q}$ dominates over $\Delta_{2Q}$, with $\pm Q\approx\bold\Lambda_\pm$ for positive $\mu$. The topology of the superfluid phase can be characterized by the Chern number.

The topology of the superfluid PDW phase can be determined with the theorem presented in formula~\eqref{Chern1}. When there is only one FS, the system with an FF order is topological since $n_L=n_U=n_F^{(i_M)}=0$ and $\int_{\partial\vec S_{(i_M)}} \nabla \theta^{(i_M)}_k\cdot dk=\pm2\pi$, giving the Chern number
\begin{eqnarray}
{\rm Ch}_1=\frac{1}{2\pi}\int_{\partial\vec S_{(i_M)}} \nabla \theta^{(i_M)}_k\cdot dk\mp1
\end{eqnarray}
for the Fermi energy crossing the left and right Dirac cones, respectively. In contrast, when there are two FSs (at both Dirac cones), from the same or different bands, one can readily find that the contributions from both FSs cancels out and the Chern number ${\rm Ch}_1=0$, rendering a trivial phase. This result implies that the system can be topological when it is in an FF phase.

A rich phase diagram is given in Fig.~\ref{DeltaCal} (c,d), where the topological and trivial FF phases with one of $\Delta_{\pm 2Q}$ being nonzero, and FFLO phase with both $\Delta_{\pm 2Q}$ being nonzero, are obtained. The gapless phase with nonzero $\Delta_{\pm 2Q}$ may be obtained due to imperfectly nesting Fermi surfaces. It is particularly interesting that in Fig.~\ref{DeltaCal} (d) a broad topological region is predicted when $m_z$ is away from $m_z=2t_z$. The broad topological region implies that the upper critical value $\Delta_{2q}^{(c)}$, characterizing the transition from topological FF state to other phases, is largely enhanced compared with the case for $m_z=2t_z$. This is a novel effect due to the following mechanism. Note that the pairing $\Delta_{-2Q}$ also couples the particle-hole states at $\tilde{\bold Q}=(\pi-Q,0)$. Increasing $\Delta_{-2Q}$ to $\Delta_{-2Q}^{(c)}$ closes the bulk gap at $\tilde{\bold Q}$ momentum, with the critical value being solved from BdG Hamiltonian as
\begin{equation}
\Delta_{-2Q}^{(c)}=\sqrt{m_x^2+m_p^2-\left(\frac{t_{xI}}{t_x}\sqrt{t_x^2-\frac{m_p^2}{16}}-\mu\right)^2},
\end{equation}
where $m_p = 2(m_z-2t_z)$. For $m_z=2t_z$, we have $Q=\pi/2$ and a small critical value $\Delta_{-2Q}^{(c)}\lesssim m_x$. This is because the gap closes at the right hand Dirac point $\tilde{\bold Q}=(\pi/2,0)$, where the original bulk gap less than $2m_x$ before having superfluid pairing [Fig.~\ref{DeltaCal}(e)]. Importantly, for $m_z=3t_z$, we find that $\Delta_{-2Q}^{(c)} \sim 2t_z$, which is of the order of band width. In this regime $\tilde{\bold Q}$ is away from the right hand Dirac point, and corresponds to a large bulk gap before adding $\Delta_{-2Q}$ [Fig.~\ref{DeltaCal}(f)]. As a result, a large $\Delta_{-2Q}$ is necessary to drive the phase transition, giving a broad topological region, as shown in Fig.~\ref{DeltaCal} (d). Note that an equivalent picture for the phase transition is that increasing the order $\Delta_{-2Q}$ deforms the band structure, which pushes one band (another band) at $\tilde{\bold Q}=(\pi-Q,0)$ toward (away from) the Fermi energy. When $\Delta_{-2Q}$ reaches the critical value, the band is pushed across the Fermi energy, leading to an additional contribution from Fermi surface to the Chern number given in Eq.~\eqref{Chern1}, and the phase transition occurs.

{\it Further discussions.}--The generic theorem given Eq.~\eqref{Chern1} is powerful to determine the topology of the 2D SC/superfluid characterized by Chern numbers. The application can be go much beyond the examples introduced above, which are relatively simple. For example, the topological superfluid phase based on SO coupled QAH model as discussed in Section IV.B exhibit rich phase diagram with different Chern numbers~\cite{Liu2014a}. To determine the full topological phase diagram through the conventional computation of the Chern numbers is some tedious, while using our present generic formula~\eqref{Chern1} can reach all the results with a quick check. We are not going to expand the discussion on the details, but the readers may check by themselves.

\subsection{Non-Abelian Majorana modes in trivial superfluids}

So far we have focused on the topological superconductors which host Majorana modes in the vortex cores or boundary. Note that MZMs in SCs (e.g. bound at vortices) are topological defect modes, which correspond to nonlocal extrinsic deformations in the Hamiltonian of the topological system. For example, MZMs in the chiral $p_x+\mathrm{i}p_y$ SC harbor at vortices which exhibit nonlocal phase windings of the SC order (a global deformation in the original uniform Hamiltonian). This feature tells that the MZMs at vortices are not intrinsic topological excitations, but extrinsic modes of a SC. In this regard, one may conjecture that the existence of MZMs is not uniquely corresponding to the bulk topology of a SC, and there might be much broader range of experimental systems which can host such exotic modes, besides those based on topologically nontrivial SCs. This conjecture was confirmed in a generic theorem shown in the recent work~\cite{Chan2017}.

\subsubsection{Chern-Simons invariant: a generic theorem}

Here we introduce the generic as shown in Ref.~\cite{Chan2017} that the existence of MZMs localized in the
vortex cores does not rely on the bulk topology of
a 2D SC. For a generic 2D normal system with $N$ FSs and gapped out by SC pairings,
one can show that the existence of MZMs at the SC vortices is characterized by an emergent
$\mathbb{Z}_{2}$ Chern-Simons invariant $\nu_{3}$:
\begin{align}
\nu_{3} & =\sum^N_{i}n_{i}w_{i}\;\mathrm{mod}\;2,\label{eq:nw}\\
w_{i} & \equiv\frac{1}{2\pi}\oint_{\mathrm{FS}_{i}}\nabla_{\mathbf{k}}\arg\Delta_{\mathbf{Q}_{i}}(\mathbf{k})\cdot d\bold k,\nonumber
\end{align}
where $\Delta_{\mathbf{Q}_{i}}(\mathbf{k})$ is the SC order projected onto the $i$-th FS and is generically momentum dependent, $w_{i}$ counts the phase winding of $\Delta_{\mathbf{Q}_{i}}(\bold k)$ in the $\bold k$-space around the $i$-th FS loop, and $n_{i}$ denotes the integer vortex winding number (vorticity) attached to $\Delta_{\mathbf{Q}_{i}}\rightarrow\Delta_{\mathbf{Q}_{i}}e^{\mathrm{i} n_i\theta(\bold r)}$, namely the winding in the real space. At least a single MZM is protected when the index $\nu_3=1$ if there is no symmetry protection, even the bulk topology of the SC, characterized by Chern number, can be trivial.

\textit{Proof}.\textendash We introduce the proof of the generic theorem given in Eq.~\eqref{eq:nw} for the Chern-Simons invariant, which governs the existence of the MZM in a 2D SC. The essential idea for the present proof is similar to that done in proving the generic formula of Chern number for chiral SC, as discussed in the previous section. For a system with multiple normal bands and FSs, the superconducting pairings may occur within each FS (intra-FS pairings) and between different FSs (inter-FS pairings). The theorem~\eqref{eq:nw} is not affected by inter-FS pairings. Thus for convenience, we consider first the generic SC Hamiltonian with only intra-FS pairings, given by
\begin{equation}
H=\sum_{\mathbf{k}}C_{\mathbf{k}}^{\dagger}\hat{H}_{0}C_{\mathbf{k}} +\sum_{{i},\alpha,\beta,\mathbf{k}}c_{\mathbf{Q}_{i}/2+\mathbf{k},\alpha}^{\dagger}\hat{\Delta}^{\alpha\beta}_{\mathbf{Q}_{i}} c_{\mathbf{Q}_{i}/2-\mathbf{k},\beta}^{\dagger}+\mathrm{h.c.},\label{eq:H_generic}
\end{equation}
where $C_{\mathbf{k}}=(c_{\mathbf{k},\alpha},c_{\mathbf{k},\beta},\cdots,c_{\mathbf{k},\gamma},\cdots)^{\text{T}}$,
with $\alpha$ incorporating the orbital and spin indices, the normal band Hamiltonian $\hat{H}_{0}(\mathbf{k})$
is considered to have $N$ FSs, and the pairing matrix element $\hat{\Delta}^{\alpha\beta}_{\mathbf{Q}_{i}}\propto\sum_{\mathbf{k}}\langle c_{\mathbf{Q}_{i}/2+\mathbf{k},\alpha}c_{\mathbf{Q}_{i}/2-\mathbf{k},\beta}\rangle$ regarding the $i$-th FS has a central-of-mass momentum $\mathbf{Q}_{i}$. Here for convenience we take that each FS is circular and centered at a
momentum $\mathbf{Q}_{i}/2$. Similar to the proof of Chern number for chiral topological superfluids, as introduced in previous section, one can always continuously deform the FSs to be circular
without changing topology of the system, as long as the bulk gap keeps open during the deformation. In general the SC order exhibits spatial modulation in the real space, rendering the PDW or FFLO state~\cite{FF1964,LO1964}, and bears the form $\hat\Delta(\mathbf{r})=\sum_{i}\hat\Delta_{\mathbf{Q}_{i}}e^{\text{i}\mathbf{Q}_{i}\cdot\mathbf{r}}$.
Note that each PDW component $\hat\Delta_{\mathbf{Q}_{i}}$ possesses a $U(1)$ symmetry, implying that each of them can be attached with a vortex of winding number $n_{i}$ independently, giving $\hat\Delta(\mathbf{r})=\sum_{i}\hat{\Delta}_{\mathbf{Q}_{i}}e^{-\text{i}n_{i}\theta(\mathbf{r})+\text{i}\mathbf{Q}_{i}\cdot\mathbf{r}}$, with  $\theta(\mathbf{r})$ being the vortex phase profile.

The Chern-Simons invariant $\nu_3$ is defined in 3D space, for which one parameterizes the Bogoliubov de Gennes (BdG) Hamiltonian by taking the phase $\phi\in[0,2\pi)$ of the SC order $\hat{\Delta}_{\mathbf{Q}_{i}}e^{-\text{i}n_{i}\phi}$ as a synthetic dimension of ring geometry $S^1$. Together with the 2D physical space, the bulk BdG Hamiltonian can then be written down in a synthetic 3D torus $T^3=T^2\times S^1$ spanned by $(\mathbf{k}, \phi)$. In the synethetic 3D space, the $\mathbb{Z}_{2}$ Chern-Simons invariant~\cite{Pontryagin,JeffreyTeo2010,WenXiao-Gang2008} is computed by
\begin{align}
\nu_{3} & =-\frac{1}{4\pi^{2}}\int_{T^{2}\times S^{1}}\mathcal{Q}_{3}\;\mathrm{mod}\;2\label{eq:nu_3}\\
\mathcal{Q}_{3} & =\mathrm{Tr}\left[\mathcal{A}d\mathcal{A}-\frac{2\text{i}}{3}\mathcal{A}^{3}\right],\nonumber
\end{align}
where the elements of one-form Berry connection are given by $\mathcal{A}_{\lambda\lambda'}(\bold k,\phi)=\text{i}\langle \psi_{\lambda}|\bold d\psi_{\lambda'}\rangle$, with $|\psi_{\lambda}\rangle$ denoting the corresponding eigenvector of the BdG Hamiltonian, and the trace is performed on the filled bands.

A direct computation of the index $\nu_3$ for the generic case is not realistic. To simplify the study we again take the advantage that the topology of the system is unchanged under any kind of continuous deformation without closing bulk gap. For this we further adiabatically deform the Hamiltonian $H$ to a new form
\begin{eqnarray}
H'\equiv H[\hat{\Delta}_{\mathbf{Q}_{i}}\rightarrow\hat{\Delta}_{\mathbf{Q}_{i}}\Omega_{\mathbf{Q}_{i}}(\mathbf{k})],
\end{eqnarray}
where $\Omega_{\mathbf{Q}_{i}}(\mathbf{k})$ is a positive real smooth
truncation function with $\Omega_{\mathbf{Q}_{i}}(\vec{S}_{i})=1$ inside
the orientable vector area $\vec{S}_{i}$ enclosed by the $i$-th FS loop centered at $\mathbf{Q}_{i}/2$, and decays to zero at a short
distance beyond this area. Since the system remains fully
gapped for the continuous deformation, the invariant $\nu_{3}$
can be evaluated over $H'$. Denoting by $\mathcal{\vec{F}}_{i}$ the vector area with $\Omega_{\mathbf{Q}_{i}}(\mathbf{k})\neq0$, The invariant given in Eq.~\eqref{eq:nu_3} can be reduced to the integral over the disjoint union $\bigsqcup_{i}\vec{\mathcal{F}}_{i}\times S^{1}$, as shown in Ref.~\cite{Chan2017}, which facilitates the further study.

While in general the Hamiltonian $\hat H_0$ incorporates multiple normal bands, one can consider the weak SC pairing regime, in which case only the states around each FS will be effectively paired up. The coupling between states from different FSs and that from different bands can be ignored due to the energy detuning. In this way, the BdG $H'$ further reduces to an effective one-band form in the eigen-basis $u_{\mathbf{k}}$ of $\hat{H}_{0}$. In particular, for the momentum $\mathbf{k}\in\vec{\mathcal{F}}_{i}$ around a specific FS centered at momentum $\mathbf{Q}_{i}/2$, the effective BdG Hamiltonian takes the form
\begin{align}
h_{{i}}(\mathbf{k},\phi) =\left[\begin{array}{cc}
\epsilon_{\mathbf{Q}_{i}/2+\mathbf{k}} & \Delta_{\mathbf{Q}_{i}}(\mathbf{k})\Omega_{\mathbf{Q}_{i}}e^{-\text{i}n_{i}\phi}\\
\Delta_{\mathbf{Q}_{i}}^{*}(\mathbf{k})\Omega_{\mathbf{Q}_{i}}e^{\text{i}n_{i}\phi} & -\epsilon_{\mathbf{Q}_{i}/2-\mathbf{k}}
\end{array}\right]\label{eq:h_Q}
\end{align}
where $\Delta_{\mathbf{Q}_{i}}(\mathbf{k})\equiv\langle u_{\mathbf{k}}|\hat{\Delta}_{\mathbf{Q}_{i}}|u^{*}_{-\mathbf{k}}\rangle$ is the pairing term projected onto the $i$-th FS. Note that $\Delta_{\mathbf{Q}_{i}}(\mathbf{k})$ has captured the original band topology.  The eigenstates of  $h_{\mathbf{Q}_{i}}$ take the form $|\psi_{\mathbf{k}\pm}\rangle=(\alpha_{\mathbf{k}\pm}u_{\mathbf{k}},\beta_{\mathbf{k}\pm}u_{-\mathbf{k}}^{*})^{\text{T}}$ (refer to also the proof of Chern number for chiral topological superfluid in the previous section). Then $\nu_{3}$ can be decomposed into $\nu_{3}=\sum_i\nu_3^{(i)}$
(`mod 2' temporarily omitted), and
\begin{eqnarray}\label{eq:nu3}
\nu_{3}^{(i)} =-\frac{1}{4\pi^{2}}\int_{\mathcal{\vec{F}}_{i}\times S^{1}}[\mathcal{A}_{\phi}\nabla_{\mathbf{k}}\times\mathcal{A}_{\mathbf{k}}+ \mathcal{A}_{\mathbf{k}}\times\nabla_{\mathbf{k}}\mathcal{A}_{\phi}]d\phi d^{2}\mathbf{k}\nonumber
\end{eqnarray}
for each $\vec{\mathcal{F}_{i}}$, where $\mathcal{A}_{\phi}=\text{i}\langle\psi_{\mathbf{k}-}|\partial_{\phi}|\psi_{\mathbf{k}-}\rangle$,
and $\mathcal{A}_{\mathbf{k}}\equiv(\mathcal{A}_{k_{x}},\mathcal{A}_{k_{y}})=\text{i}\langle\psi_{\mathbf{k}-}|\nabla_{\mathbf{k}}|\psi_{\mathbf{k}-}\rangle$,
with $\nabla_{\mathbf{k}}\equiv(\partial_{k_{x}},\partial_{k_{y}})$.
The above result can be further simplified by taking the weak pairing order limit $\Delta_{\mathbf{Q}_i}\rightarrow0^{+}$, in which case the gap becomes infinitesimal at the FSs, and the contribution to $\nu_3$ will completely come from the FS states. It can be derived directly on $\vec{\mathcal{F}_{i}}$ that
$\mathcal{A}_{\phi} =- n_i\Theta_{\vec{S}_i}$ and
$\mathcal{A}_{\mathbf{k}} =(1-2\Theta_{\vec{S}_i})\mathcal{A}_{0,\mathbf{k}}+\Theta_{\vec{S}_i}(\nabla_{\mathbf{k}}\arg\Delta_{\mathbf{Q}_i} +\mathbf{A}^i_{d})$,
where $\Theta_{\vec{S}_i}$ is a step
function equal to $1$ within $\vec{S}_i$ and $0$ otherwise, $\mathcal{A}_{0,\mathbf{k}}\equiv\text{i}u_{\mathbf{k}}^{\dagger}\nabla_{\mathbf{k}}u_{\mathbf{k}}$ represents the Berry connection for the normal band, and $\mathbf{A}^i_{d}$ is the \textit{defect gauge field} as a consequence of the multivalueness of $\arg\Delta_{\mathbf{Q}_i}$ \cite{Kleinert2008}. Substituting these results into the formula of $\nu_3$ yields
\begin{align}
\nu_3=\sum_i\frac{n_i}{2\pi}\int_{\vec{\mathcal{F}_i}}\Theta_{\vec{S}_i}\nabla_{\mathbf{k}}\times(\nabla_{\mathbf{k}}\arg\Delta_{\mathbf{Q}_i} +\mathbf{A}^i_{d})\cdot d^{2}\mathbf{k}.
\end{align}
The above result is exactly the one given in Eq.~\eqref{eq:nw} by observing that the curl of gradient of SC phase vanishes, while the contribution from the defect gauge field $\bold A^i_d$ renders the phase winding of SC order in the momentum space around FS loop~\cite{Chan2017}. This completes the proof. The theorem is still valid if the system has inter-FS pairings and connected FSs~\cite{Chan2017}. Particularly, in the one band case, the $\mathbb{Z}_{2}$ Chern Simons invariant readily reduces to the Hopf invariant.

The above result shows that the existence of MZMs at vortex cores is essentially protected not by the bulk topology of the 2D SC, but by an emerging Chern-Simons invariants $\nu_3$, implying that a non-Abelian MZM can exist in a trivial SC. A famous example can be obtained from a Rashba SO coupled semiconductor with Zeeman splitting and in the presence of an $s$-wave superconductivity~\cite{Sato2009,Sau2010Majorana}, as introduced in the previous section. To obtain a chiral topological SC the chemical potential has to lie within the Zeeman gap and cross the bulk band for once. According to the above theorem, even the chemical potential is above the Zeeman gap and crosses two FSs, MZMs can in principle be generated if the SC orders in the two FSs are independent and only one of them is attached with vortex.

\subsubsection{Majorana zero modes in 2D trivial superfluids}

\begin{figure}
\includegraphics[width=1.0\columnwidth]{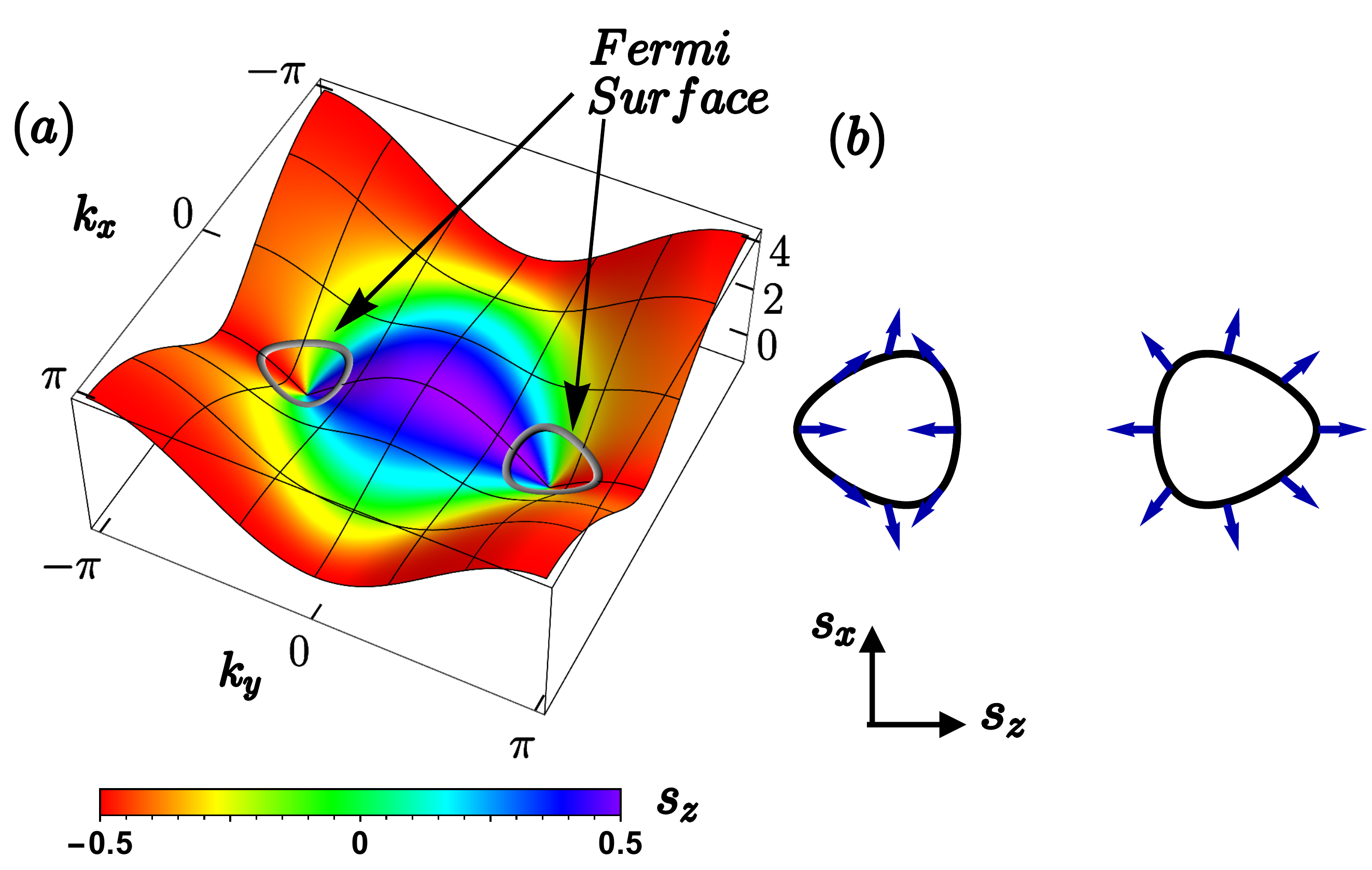}
\caption{\label{fig:band} An example of 2D SO coupled semimetal~\cite{Chan2017}. (a) The band structure of 2D topological Dirac metal
with two Dirac points located at $\mathbf{Q}_{\pm}$, the gray thick
loops around two Dirac points represent the Fermi surfaces, and the
color represents the average value of the spin component $\left\langle s_{z}\right\rangle $.
(b) Schematic of the spin orientations, shown by blue arrows, at
the Fermi surfaces around the Dirac points. Parameters: $t_{x,y}=t_{\mathrm{so}}=m_{z}=1$, $\mu=0.8$,
and the corresponding Dirac node momenta $\mathbf{Q}_{+}=-\mathbf{Q}_{-}=(0,2\pi/3)$. }
\end{figure}

\textit{2D Dirac metal}.\textendash
The theorem in~\eqref{eq:nw} suggests that MZMs can exist in broader range of physical systems. Now we introduce a minimal scheme, which can be readily achieved based on the optical Raman lattice scheme~\cite{Wu2016,Jeffrey2018PRB}, for the realization of MZMs. The total Hamiltonian takes the form
\begin{eqnarray}
H&=&H_{0}+H_U,\\
H_{0} & = & \sum_{\mathbf{k}}(c_{\mathbf{k},\uparrow}^{\dagger},c_{\mathbf{k},\downarrow}^{\dagger})\mathcal{H}_{0}\left(\begin{array}{c}
c_{\mathbf{k},\uparrow}\\
c_{\mathbf{k},\downarrow}
\end{array}\right)\label{eq:H_0}\nonumber\\
\mathcal{H}_{0} & = & (m_{z}-2t_{x}\cos k_{x}-2t_{y}\cos k_{y})\sigma_{z}\nonumber\\
&&+2t_{\mathrm{so}}\sin k_{x}\sigma_{x}-\mu,\nonumber\\
H_U&=&-U\sum_{i}n_{i\uparrow}n_{i\downarrow},\nonumber
\end{eqnarray}
where the Hubbard interaction is attractive $U>0$.
As is known that the Hamiltonian $H_0$ describes
a topological Dirac semimetal for $|m_{z}|<2(t_{x}+t_{y})$, with two
Dirac points at $\mathbf{Q}_{\pm}=(0,\pm\cos^{-1}((m_{z}-2t_{y})/2t_{x}))$
and possesses non-trivial spin texture on the FSs (Fig.~\ref{fig:band}). The difference of the present case from the one in Eq.~\eqref{HTB} is that here no inversion symmetry is broken and no gap opening at the Dirac points.

\textit{Self-consistent phase diagram}.\textendash The superfluid (SC) states can be studied with the
above model. Having multiple FSs around various Dirac cones, generically
one shall consider both the inter-cone (BCS) and the intra-cone (PDW)
pairing orders, described by $\Delta_{\mathbf{q}}=(U/N)\sum_{\mathbf{k}}\langle c_{\mathbf{q}/2+\mathbf{k},\uparrow}c_{\mathbf{q}/2-\mathbf{k},\downarrow}\rangle$,
with $\mathbf{q}=2\mathbf{Q}_{\pm}$ or 0
and $N$ is total number of lattice sites. Generally, the order parameter
in real space takes the form
\begin{eqnarray}
\Delta(\mathbf{r})=\Delta_{0}+\Delta_{2\mathbf{Q}_{+}}e^{2\text{i}\mathbf{Q}_{+}\cdot\mathbf{r}}+\Delta_{2\mathbf{Q}_{-}}e^{2\text{i}\mathbf{Q}_{-}\cdot\mathbf{r}}
\end{eqnarray}
and the BCS and PDW orders may compete with each other. The intra-cone
PDW order can fully gap the bulk while reducing the translation symmetry.
On the other hand, owing to the different spin-momentum lock at the
FSs of the two Dirac cones {[}Fig.~\ref{fig:band} (b){]}, the inter-cone
BCS pairing cannot fully gap out the bulk spectrum, and leaves four
nodal points. These nodal points can be further gapped by charge density
wave (CDW) orders
\begin{eqnarray}
\rho_{\mathbf{q},\sigma}=(U/N)\sum_{\mathbf{k}}\langle c_{\mathbf{k}-\mathbf{q}/2,\sigma}^{\dagger}c_{\mathbf{k}+\mathbf{q}/2,\sigma}\rangle.
\end{eqnarray}
The BCS, PDW, and CDW orders may compete to dominate in different parameter regimes, and can be solved self-consistently.

\begin{figure}
\includegraphics[width=0.99\columnwidth]{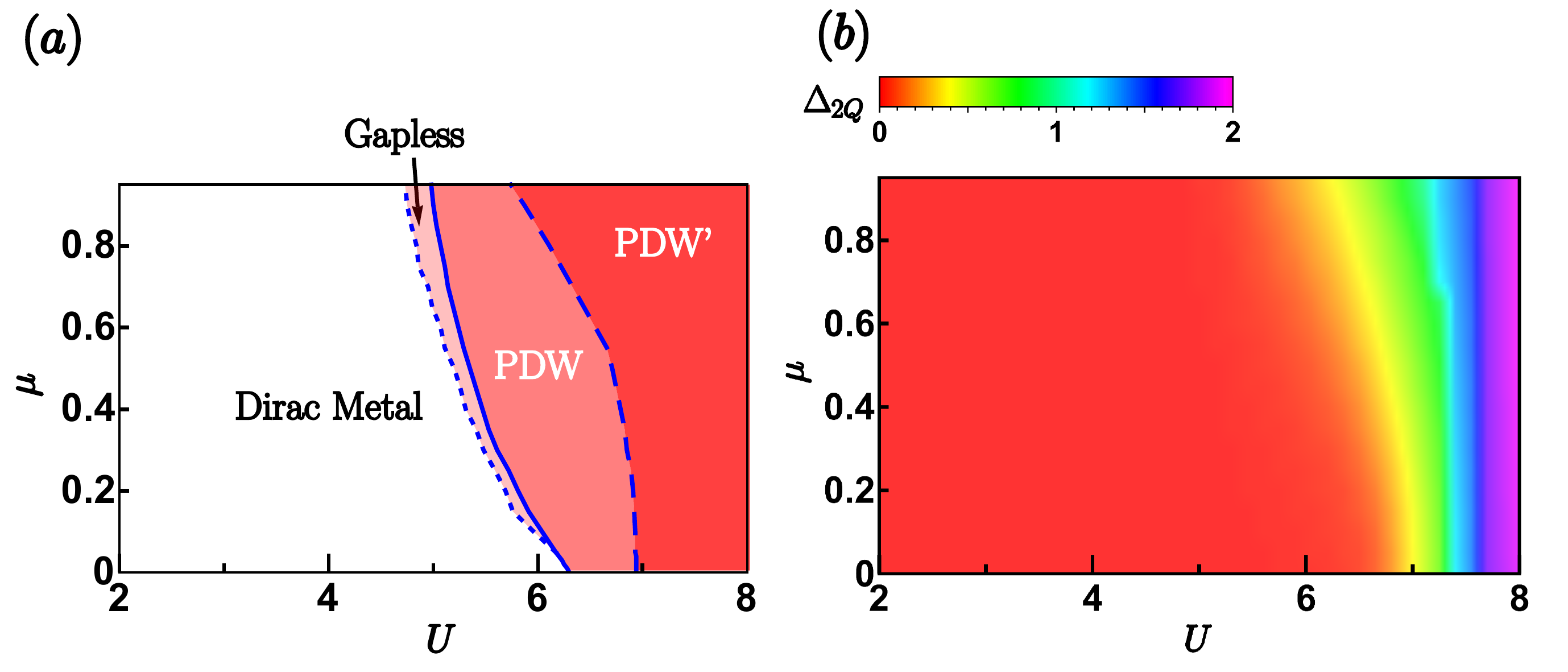}
\caption{\label{fig:phsDiag} Self-consistent result for the order parameters~\cite{Chan2017}. (a) Mean field phase diagram of the Dirac-Hubbard
Hamiltonian versus attractive Hubbard interaction $U$ and chemical
potential $\mu$. In the ``Dirac metal'' phase, all $\Delta_{\mathbf{q}}=0$;
in the narrow ``Gapless'' region, $\Delta_{2\mathbf{Q}_{\pm}}$
are finite but not strong enough to fully gap the system. In the ``PDW''
phase, the system is fully gapped by finite PDW.  In the ``$\mathrm{PDW}^{\prime}$'' phase at large $U$, BCS and
CDW orders of small magnitude also appear and accompany the PDW. (b) Magnitude of
PDW order $\Delta_{2\mathbf{Q}_{\pm}}$.
The parameters are the same as those in Fig.~\ref{fig:band}.}
\end{figure}
The phase diagram are obtained by self-consistent calculation with
proper parameters so that the Dirac points are located at $\mathbf{Q}_{\pm}=(0,\pm2\pi/3)$, as shown
in Fig.~\ref{fig:phsDiag}. The phase diagram is dominated by PDW
order with $|\Delta_{2\mathbf{Q}_{\pm}}|\neq0$, which appears only
for finite $U$. With increasing chemical potential, the Dirac cone
becomes less isotropic (Fig.~\ref{fig:band}) and the FSs are less
well-nested. As a consequence, a narrow gapless region with nonzero
PDW orders $|\Delta_{2\mathbf{Q}_{\pm}}|\neq0$ is obtained for $\mu>0.05$
{[}Fig.~\ref{fig:phsDiag}(a){]}, while the spectrum becomes fully
gapped when $|\Delta_{2\mathbf{Q}_{\pm}}|$ increases exceeding some
finite value. In the fully gapped region at larger $U$, one can readily
check that the Chern number vanishes ${\rm Ch}_1=0$ from the generic result shown in formula~\eqref{Chern1}, and the bulk is trivial for the
present (class D) superfluid~\cite{Jeffrey2018PRB}. In the
$\mathrm{PDW}^{\prime}$ phase at large $U$, BCS and CDW orders of
small magnitude also appear and accompany the PDW. They can be regarded
as small perturbations to the PDW phase and hence the $\mathrm{PDW}^{\prime}$
phase is topologically equivalent to the PDW phase.

\begin{figure}
\includegraphics[width=1.0\columnwidth]{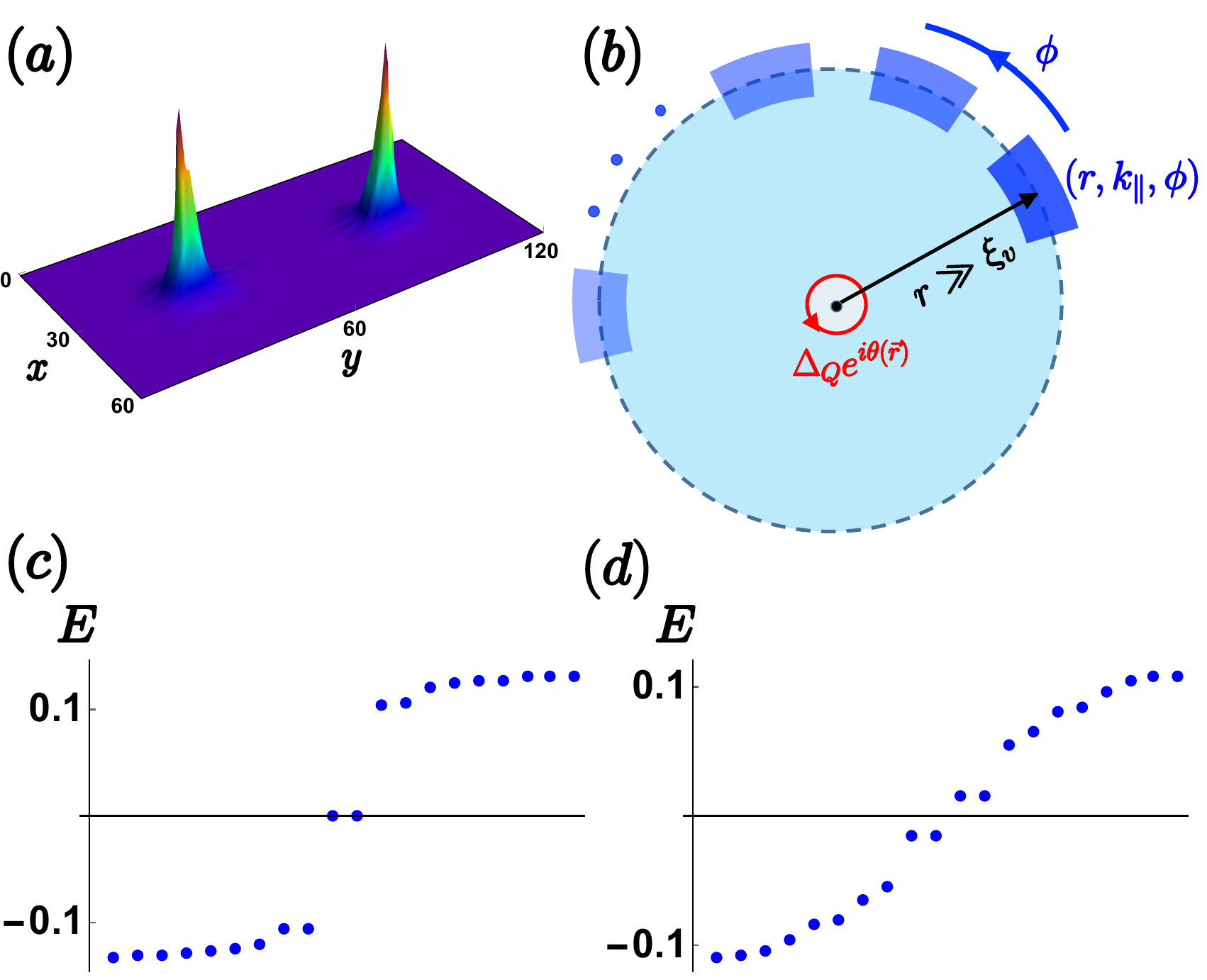}
\caption{\label{fig:MZM} MZMs in the 2D trivial superfluid~\cite{Chan2017}. (a) MZM wave function density $\sum_{s=\uparrow,\downarrow}|\psi_{s}(\mathbf{r})|^{2}$
computed in the Dirac-Hubbard model, with $\mu=0.8$, $U=5.5$,
which gives the pairing order $\Delta_{2\mathbf{Q}_{\pm}}=0.23$. System size: $N_{x}=N_{y}/2=60$. The vortices
with opposite unit vorticities are located at $(30,30)$ and $(30,90)$
and the vortex field $e^{\text{i}\theta(\mathbf{r})}$ is attached only to $\Delta_{2\mathbf{Q}_{+}}$. (b) Schematic of the physical origin of MZMs at vortex cores. The vortex core can be viewed approximately as the open boundary of $r$-dimension in the 3D space spanned by $(r, k_{\parallel},\phi)$. Energy spectrum for the half vortex (c) and full vortex (d) regime. The MZMs are obtained in the former case. Other parameters are the same as those in Fig.\ \ref{fig:band}.}
\end{figure}

{\it MZMs for the trivial superfluid}.--Despite the topologically trivial superconducting state here, the system
can host non-trivial MZM bound to vortices and protected by the Chern-Simons invariant shown in the present work. In general, the vortices proliferated to the PDW order read
\begin{eqnarray}
\Delta(\mathbf{r})&=&\Delta_{2\mathbf{Q}_{+}}e^{2\text{i}\mathbf{Q}_{+}\cdot \mathbf{r}+\mathrm{i}n_{+}\theta(\bold r)}+\Delta_{2\mathbf{Q}_{-}}e^{2\text{i}\mathbf{Q}_{-}\cdot \mathbf{r}+\mathrm{i}n_{-}\theta(\bold r)}\nonumber\\
&=&2\Delta_{2\mathbf{Q}_{\pm}}e^{\mathrm{i}(n_{+}+n_{-})\theta(\bold r)/2}\cos\bigr[2\mathbf{Q}_{+}\cdot \mathbf{r}+\nonumber\\
&&+(n_+-n_-)\theta(\bold r)/2\bigr].
\end{eqnarray}
The minimal allowed defect corresponds to a half-vortex, given by $n_++n_-=\pm1$, while a full vortex is given by $n_++n_-=\pm2$~\cite{Agterberg2008,Ji2008}. In particular, in Fig.~\ref{fig:MZM} (a,c) we consider the half vortex regime with two unit vortices of opposite vorticities $\pm2\pi$ (i.e. $n_+=\pm1$) attached only to $\Delta_{2\mathbf{Q}_{+}}$
and located with a finite distance between each other in the real space. The real space BdG Hamiltonian with vortices is then numerically
solved and the two lowest energy modes with finite-size energies $E=\pm1.039\times10^{-4}$
are obtained {[}Fig.\ \ref{fig:MZM}(c){]}. Spatial wave function
density $\sum_{s=\uparrow,\downarrow}|\psi_{s}(\mathbf{r})|^{2}$ for one of
the solutions (the other is the same) is plotted in Fig.~\ref{fig:MZM}(a), showing that it is in the zero angular-momentum channel
and well-localized at vortex cores, thus being a MZM. The robustness of MZMs against impurities can be shown straightforwardly~\cite{Chan2017}. The physical origin of the exsistence of MZMs can be viewed as a direct consequence of \textit{bulk-boundary correspondence}, as illustrated in Fig.~\ref{fig:MZM}(b). Consider the region far away enough form the vortex core so that at each azimuthal angle $\phi$ we can find a microscopically large region with approximately constant SC phase $\theta$. This region can be thought of as a 2D system in $(r, k_{\parallel};\phi)$ with fixed $\phi=\theta$, periodic boundary along $k_{\parallel}$ direction and open boundary along $r$ direction. Combining all such 2D systems with $\phi \in [0,2\pi)$ yields an effective 3D space with periodic boundary with respect to $k_{\parallel}$ and $\phi$, while open boundary along $r$ axis due to the existence of vortex. With this picture when the parameterized 3D system has a nontrivial Chern-Simons invariant $\nu_3$, which is the case for half-vortex regime based on a direct numerical check, MZM is obtained as a boundary zero mode at the vortex core.
In comparison, we have performed a similar calculation by attaching a full vortex with $n_++n_-=2$ to $\Delta_{2\mathbf{Q}_{\pm}}$, which gives a null $\nu_{3}$. In Fig.~\ref{fig:MZM}(d), the corresponding low energy spectrum
reveals that no zero mode but finite energy Andreev bound states are present in the system, consistent
with the $\nu_{3}$ result.

\subsubsection{Chiral Majorana modes in 3D trivial superfluids}

The existence of non-Abelian Majorana modes in the trivial phase of a superconducting 2D Dirac semimetal can be generalized to the 3D case, as studied in~\cite{Chan2017}. In this case a 3D Weyl semimetal together with an attractive Hubbard interaction was considered, with the Hamiltonian given by
\begin{eqnarray}
H & = & \sum_{\mathbf{p}}\psi_{\mathbf{p}}^{\dagger}h_{\mathbf{p}}\psi_{\mathbf{p}}-U\sum_{\mathbf{i}}n_{\mathbf{i}\uparrow}n_{\mathbf{i}\downarrow},\label{eq:ham_0}\\
h_{\mathbf{p}} & = & [m_{z}-2t_{0}(\cos p_{x}+\cos p_{y})-2t_{z}^{c}\cos p_{z}]\sigma_{z}\nonumber \\
 &  & +2t_{{\rm SO}}(\sin p_{x}\sigma_{x}+\sin p_{y}\sigma_{y})-\mu-2t_{z}^{s}\sin p_{z}.\nonumber
\end{eqnarray}
Here the hopping terms along $z$ direction with
$(t_{z}^{c},t_{z}^{s})=t_{z}(\cos\varphi_{0},\sin\varphi_{0})$,
which break inversion symmetry of the Weyl semimetal unless $\varphi_{0}=n\pi/2$
with integer $n$. For $m_{z}=4t_{0}+2t_{z}^{c}\cos Q$ ($0<Q<\pi$),
the Weyl semimetal has two nodal points of chiralities $\chi=\pm$
located at $\mathbf{Q}_{\chi}=(0,0,\chi Q)$ and with energies $E_{\pm}=-\mu\mp2t_{z}^{s}\sin Q$,
respectively. It is convenient to
choose $Q=\frac{2\pi}{3}$ and $t_{0,z}=t_{{\rm SO}}=1.0$ to facilitate
further discussion. We note that the Weyl semimetal can be realized by generalizing the optical Raman lattice scheme to 3D regime, which is currently considered in both theory and experiment for ultracold atoms.

\begin{figure*}
\includegraphics[width=0.8\textwidth]{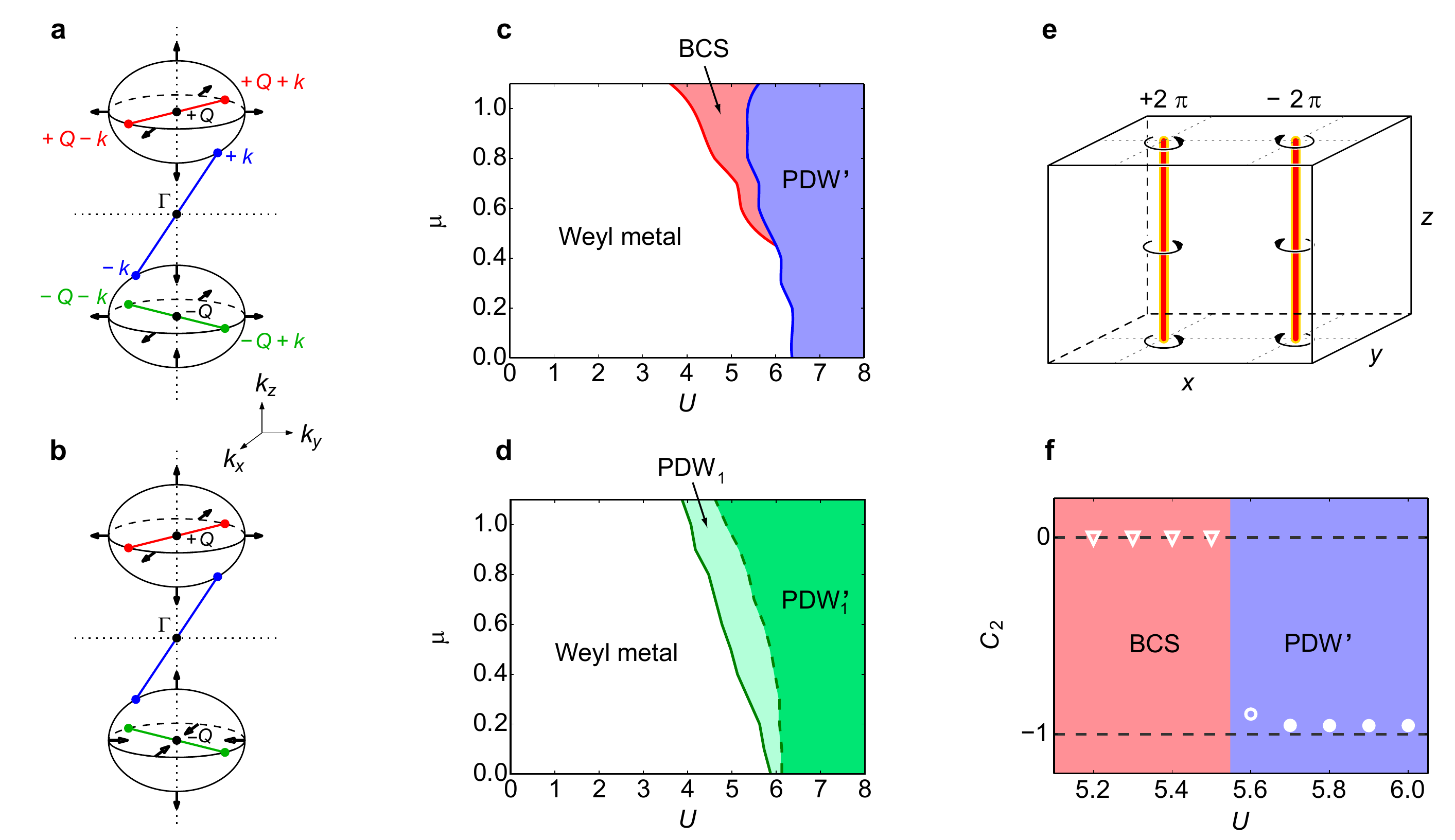}
\caption{\label{fig:Weyl} Weyl metals, mean field phase diagrams, vortex line
configuration, and 2nd Chern numbers~\cite{Chan2016}. (a,b) Schematic for FSs
and pairing order of Weyl semimetals with (a) or without (b) inversion symmetry. The two ellipsoids centered at
$\mathbf{Q}_{\pm}$ are the FSs. The spin orientations (black arrows) at the FS around $\mathbf{Q}_{-}$
is flipped when $0<\mu<2t_{z}^{s}\sin|Q_{\pm}|$. The colored lines represent the pairing order parameters $\Delta_{\mathbf{q}}$ with
$\mathbf{q}=0,2\mathbf{Q}_{\pm}$. (c,d) Phase diagram for cases with (c) or without (d) inversion
symmetry versus attractive Hubbard interaction $U$ and chemical potential
$\mu$. (e) Configuration of two vortex lines with opposite vorticities.
(f) The emergent 2nd Chern number $C_{2}$ for a 4D synthetic space
generalized from the 3D physical system in the inversion symmetric
case and with $\mu=0.7$. Upon increasing $U$, the system undergoes
a transition from BCS to $\mathrm{PDW}^{\prime}$ phase. The circles
(triangles) gives the 2nd Chern number in BCS ($\mathrm{PDW}^{\prime}$)
phase.}
\end{figure*}
The SC phases can be induced with an attractive interaction $U>0$. Similar to the results introduced in the previous sections for the 2D Dirac semimetal, the possible
pairing orders are of two distinct types, namely the $s$-wave BCS
and PDW phases, as sketched in Fig.~\ref{fig:Weyl}(a). The former
describes a uniform order $\Delta_{0}$ occurring between
two different Weyl cones and with zero center-of-mass momentum of Cooper
pairs, while the latter are spatially modulated
orders $\Delta_{2\mathbf{Q}_{\pm}}$ occurring within each Weyl cone
and the Cooper pairs have nonzero center-of-mass momentum.
These pairing orders read $\Delta_{\mathbf{q}}=\frac{U}{2N}\sum_{\mathbf{k}}\langle c_{\mathbf{q}/2+\mathbf{k}\uparrow}c_{\mathbf{q}/2-\mathbf{k}\downarrow}\rangle$,
with $\mathbf{q}=0,2\mathbf{Q}_{\pm}$, and $N$ being number
of sites, and the interaction is decoupled into $\mathcal{H}_{\mathrm{MF}}=\sum_{\mathbf{k},\mathbf{q}}\Delta_{\mathbf{q}}\thinspace c_{\mathbf{q}/2+\mathbf{k}\uparrow}^{\dagger}c_{\mathbf{q}/2-\mathbf{k}\downarrow}^{\dagger}+\mathrm{h.c}$, with $\mathbf{k}$ being summed over the entire Brillouin
zone. Again the order parameter takes
the following generic form in real space $\Delta(\mathbf{r})=\Delta_{0}+\Delta_{+2Q}e^{i2\mathbf{Q}_{+}\cdot\mathbf{r}}+\Delta_{-2Q}e^{i2\mathbf{Q}_{-}\cdot\mathbf{r}}$.
The numerical simulation reveals different phase diagrams for the inversion-symmetric {[}Fig.~\ref{fig:Weyl}(b){]}
and inversion symmetry broken {[}Fig.~\ref{fig:Weyl}(d){]} Weyl
metals. For the inversion-symmetric Weyl metal with $\varphi_{0}=0$,
a direct transition from Weyl metal phase to the $\mathrm{PDW}'$
phase, which has $|\Delta_{+2Q}|=|\Delta_{-2Q}|\gg|\Delta_{0}|\neq0$,
is obtained by increasing $U$ in the low chemical potential regime
with $|\mu|<\mu_{c}$ and $\mu_{c}\sim0.4$. The equality $|\Delta_{+2Q}|=|\Delta_{-2Q}|$
in the $\mathrm{PDW}'$ phase is a consequence of the inversion symmetry.
However, when $\mu$ is tuned beyond the
critical value $|\mu|>\mu_{c}$, the BCS phase with $\Delta_{0}\neq0$
and $\Delta_{\pm2Q}=0$ appears between the Weyl metal and $\mathrm{PDW}'$
phase. This result reflects that the pairings within each Weyl Fermi
surface dominates in relatively small $\mu$ regime.
Moreover, if the inversion symmetry is broken, the BCS
phase is suppressed, and the system enters from Weyl metal into
$\mathrm{PDW}_{1}$ state first and then $\mathrm{PDW}_{1}^{\prime}$
phase by increasing $U$, as shown in Fig.~\ref{fig:Weyl}(d) with
$\varphi_{0}=\frac{3}{32}\pi$. The $\mathrm{PDW}_{1}$ phase is characterized
by $|\Delta_{+2Q}|\geq|\Delta_{-2Q}|$ for $\mu\geq0$ and $\Delta_{0}=0$,
while in the $\mathrm{PDW}_{1}^{\prime}$ phase a small
$|\Delta_{0}|$ ($<|\Delta_{\pm2Q}|$) also emerges.

\begin{figure*}
\includegraphics[width=0.95\textwidth]{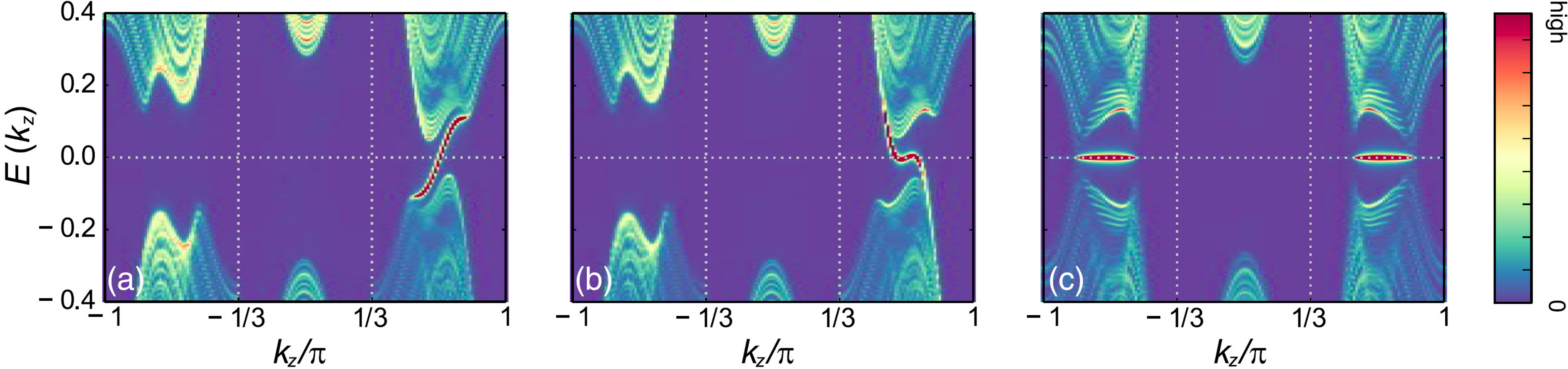}
\caption{\label{fig:lineModes}Chiral Majorana modes as in-gap bound states
to a vortex line~\cite{Chan2016}. (a-b) Chiral vortex line modes shown from spectral functions $A(k_{z},E)$ of the inversion
symmetric Weyl metal with PDW orders based on self consistent calculation for $\mu=0.7$ and $U=5.8$. The vortex line is attached
to $\Delta_{+2Q}$, with winding number $n=+1$ for (a) and $n=-1$ for (b). (c) Spectral function $A(k_{z},E)$ for the BCS dominated
phase in the inversion symmetric Weyl metal with $\Delta_{0}=0.25$. Two segments of Majorana flat bands are obtained.
Periodic boundary conditions are considered and the system size is $N_x/2=N_y=55$ and $N_z=162$.}
\end{figure*}
\textit{Vortex Line Modes for trivial superfluid phase.\textendash }Both BCS and PDW states
have only the particle-hole symmetry, so they belong to
class D according to the Altland-Zirnbauer symmetry classification~\cite{AZ1997}.
Both BCS and PDW phases are $p_{x}+ip_{y}$ SCs stacked
along $z$ direction and are topological in the 2D subspace, but are trivial in the 3D space. Namely, any invariant defined in the 3D space is zero for the present BCS and PDW phases.

While the phases are trivial for the 3D space, the nontrivial vortex line modes can be obtained. We consider first the PDW dominated
phase, and take that $|\Delta_{+2Q}|=|\Delta_{-2Q}|$
and $\Delta_{0}=0$, since a small perturbative BCS order does not
affect the results. Let a vortex line of winding $n$ be
along $z$-axis and attached to $\Delta_{+2Q}$, so that
$\Delta_{+2Q}=|\Delta_{+2Q}|\exp[in\phi(\mathbf{r})]$, with $\tan(\phi)=y/x$.
In the low-energy limit the effective Hamiltonian can be obtained by linearizing the 3D Weyl cone
Hamiltonian around $\mathbf{Q}_{+}$ point
\begin{align}
H_{\mathrm{eff}}= & -(\sum_{j=x,y,z}iv_{j}\sigma_{j}{\partial}_{j}+\mu_{\mathrm{eff}})\tau_{z}\label{eq:H_eff}\\
 & +\bar{\Delta}(\rho)\cos(n\phi)\tau_{x}+\bar{\Delta}(\rho)\sin(n\phi)\tau_{y},\nonumber
\end{align}
where $v_{j}$ is the Fermi velocity along $j$-th direction, $\mu_{\mathrm{eff}}$
is the chemical potential measured from the Weyl node, and $\tau$'s
are Pauli matrices for the Nambu space spanned by $\hat{f}(\mathbf{r})=[c_{\uparrow}(\mathbf{r}),c_{\downarrow}(\mathbf{r}),c_{\downarrow}^{\dagger}(\mathbf{r}),-c_{\uparrow}^{\dagger}(\mathbf{r})]^{T}$.
For the vortex line located at $x=y=0$ along $z$-axis, one can choose cylindrical
coordinate $(\rho,\phi,z)$, with $\bar{\Delta}(\rho\rightarrow0)=0$, $\bar{\Delta}(\rho\rightarrow\infty)=\mathrm{constant}$,
and $n\phi$ the phase winding of the PDW component. For $n=\pm1$ the Majorana in-gap
modes can be obtained analytically and satisfy~\cite{Chan2017}
\begin{eqnarray}
H_{\mathrm{eff}}\gamma_{z}(\rho,\phi,k_{z})&=&{\cal E}_{k_{z}}\gamma_{z}(\rho,\phi,k_{z}),\\
{\cal E}_{k_{z}}&=&{\rm sgn}(nv_{x}v_{y})v_{z}k_{z},\nonumber
\end{eqnarray}
which implies that the vortex Majorana modes are chiral, with chirality $\chi_{\mathrm{M}\ell}={\rm sgn}(nv_{x}v_{y}v_{z})$
being related to $n$ and
Weyl node chirality $\chi$. The Majorana
operator reads $\hat{\gamma}_{z}(k_{z})=\int\!d^{2}\mathbf{r}\thinspace\gamma_{z}(\rho,\phi,k_{z})\hat{f}(\rho,\phi,k_{z})$,
with $\hat{\gamma}_{z}(k_{z})=\hat{\gamma}_{z}^{\dagger}(-k_{z})$
for real Majorana states. This solution can be readily generalized to the case of a generic
winding $n={\cal N}$. Actually, such a vortex line is topologically
equivalent to $|{\cal N}|$ vortex lines with unity winding $n={\rm sgn}({\cal N})$.
In the later case each line hosts a branch of chiral Majorana
modes. Due to the chiral property putting the $|{\cal N}|$ branches of vortex
modes together with couplings can at most deform their dispersions, but cannot annihilate them, yielding $|{\cal N}|$ chiral Majorana
modes.

The chirality of Majorana vortex modes imply that these modes are
gapless and traverse the bulk gap of the PDW phase. This property can be further confirmed
by performing a full real-space numerical calculation based on the lattice model without linearity assumption. In particular,
the inversion symmetric Weyl metal with $\mu=0.7$ and $U=5.8$ is considered. The
self-consistent calculation for this regime reveals a $\mathrm{PDW}^{\prime}$ phase
with $\Delta_{\pm2Q}\approx0.201$ and $\Delta_{0}\approx-1.85\times10^{-2}$,
and the system has a bulk gap $E_{{\rm gap}}\approx0.21$. A vortex line ($n=1$) and anti-vortex line ($n=-1$) are considered along $z$-axis,
separating from each other in $x$-$y$ plane, and are attached to one
of $\Delta_{0,\pm2Q}$, as sketched in Fig.~\ref{fig:Weyl}(e). With
this configuration appropriate periodic boundary condition can be
applied in the numerical calculation. The local spectral function
$A(x,y,k_{z},E)$ can be obtained from the retarded Green's function
$G^{R}(E)$ of the system
\begin{equation}
A=-\frac{1}{\pi}\sum_{s=\uparrow,\downarrow}{\Im}\langle x,y,k_{z},s|G^{R}(E)|x,y,k_{z},s\rangle,
\end{equation}
where $|x,y,k_{z},s\rangle$ is the Bloch basis with momentum $k_{z}$. Computing $A(k_{z},E)$ near
the vortex core gives the energy spectra of the bulk and vortex line modes.

Fig.~\ref{fig:lineModes}(a) and (b) show the spectra measured from
the vortex line ($n=1$) and anti-vortex line ($n=-1$), respectively.
In both cases the chiral Majorana modes traverse
the bulk gap connecting the lower and upper bands. The chirality
of these modes depends on the vortex line winding number,
consistent with the previous analytic solution. This result is fundamentally
different from that for a BCS phase, as shown in Fig.~\ref{fig:lineModes}(c),
where we compute the Majorana modes by attaching vortex and anti-vortex
lines to $\Delta_{0}$ with $\Delta_{0}=0.25$ and $\Delta_{\pm2Q}=0$.
Majorana zero-energy flat bands for the vortex lines are
obtained. These Majorana zero modes are simply the vortex modes of
$p_{x}+ip_{y}$ SCs with different momenta $k_{z}$.

Chiral gapless modes have to be protected by chiral topological invariants,
e.g. the Chern numbers. However, it can be verified that for any
2D sub-plane incorporating $z$-axis the 1st Chern number is zero. Similar to the case in 2D superconducting Dirac semimetal system, an emerging invariant defined in a higher-dimensional space shall provide the protection of the present chiral Majorana modes. Similarly, the SC order can be parameterized
by its phase factor $\Delta_{\mathbf{q}}e^{in\theta}$, where the constant $\theta\in[0,2\pi)$
forms a 1D periodic parameter space $S^{1}$. Together with the 3D
lattice, we construct a 4D synthetic space $T^{4}=T^{3}\times S^{1}$
spanned by $\mathsf{p}=(\mathbf{p},p_{\theta})$ with $\mathbf{p}=(p_{x},p_{y},p_{z})$
and $p_{\theta}=\theta$. In this synthetic 4D space we define the
2nd Chern number by $C_{2}=\frac{1}{32\pi^{2}}\int_{T^{4}}d^{4}\mathsf{p}\thinspace\epsilon_{ijk\ell}\mathrm{Tr}[\mathtt{F}_{ij}\mathtt{F}_{k\ell}]\in\mathbb{Z}$,
where $\epsilon$ is the antisymmetric tensor and $\mathtt{F}_{ij}$
are the gauge field strengths calculated {by diagonalizing Hamiltonian $H_{\rm MF}(\mathbf{p},p_{\theta})$ for every $\bold p$ and $p_\theta$.
The second Chern number $C_{2}$ can be numerically computed for BCS and PDW
phases, as shown numerically in Fig.~\ref{fig:Weyl}(f). It is seen
that $C_{2}=0$ for the BCS dominated phase, while $C_{2}\approx-0.956$
for the PDW phase obtained with the same parameters except
for $U$ as in Fig.~\ref{fig:lineModes}(a). The deviation of $C_{2}$
from an integer is due to finite size effect. It was further shown and proposed that when considering the ring configuration of vortex lines, a 3D non-Abelian loop-loop braiding statistics can be obtained~\cite{Chan2017}. These results reveal a real physical system to explore the exotic 3D non-Abelian braiding statistics, and shall attract further studies based on realistic ultracold atom platforms.

\section{Discussion and Outlook}

In conclusion, in the present review we have pedagogically introduced the realization of spin-orbit (SO) coupling and topological quantum phases for ultracold atoms, and have focused on the latest progresses in theory and experiment, particularly for the SO coupling beyond one-dimension (1D) in optical lattices, the topological insulating states, and topological superfluid phases. We systematically discussed the optical Raman lattice schemes, with which the 1D and 2D Dirac types, 2D Rashba type, and 3D Weyl type SO couplings can be synthesized. Being of the high feasibility, the proposed SO couplings and topological quantum physics have motivated several experimental studies, with some of the proposals having been successfully realized. The topological superfluids and Majorana modes have also been discussed. After a brief introduction to the background, we showed a generic theorem for the chiral topological superfluid/superconductor phases, which provides a simple but generic approach to determine the topology of 2D chiral superfluids/superconductors. Moreover, the existence of non-Abelian Majorana modes at vortices (or vortex lines for 3D systems) is revisited and is found to
be irrespective of the bulk topology of the superfluid/superconductor phases, and is protected by emergent topological numbers defined in the synthetic spaces with dimension higher than the physics system. As a direct consequence, the non-Abelian Majorana modes can exist in trivial superfluids/superconductors, with the minimal experimental schemes for the realization have been proposed and studied.

The realization of high-dimensional SO couplings has opened intriguing opportunities to explore novel quantum physics with ultracold atoms, ranging from traditional spintronic effects to the exotic topological physics. In particular, the recently proposed new optical Raman lattice scheme~\cite{Wang2018} exhibits high controllability in engineering various types of high-dimensional SO couplings, with the realized SO coupled quantum gases having long lifetime (up to several seconds depending on atom candidates). Especially, with such scheme a long-lived topological Bose gas has been achieved in experiment, confirming the high feasibility of the new proposal. With these progresses the research of the high-dimensional SO coupled quantum gases is developing into a mature topic, and the realization of high-dimensional SO couplings shall become routine studies in experiment in the near future. The next important issues in this direction include to investigate the quantum-far-from equilibrium dynamics with novel topology and the interacting topological quantum states for the SO coupled systems. For ultracold atoms, the nonequilibrium dynamics could be particularly useful to explore topological quantum physics, since the equilibrium ground state is not easy to reach due to the heating and loss in the system. Nonequilibrium dynamics may bring up new physics beyond the achievability of equilibrium studies. The interacting topological quantum physics, e.g. the topological superfluids and fractional topological insulating states, are highly-sought-after but very challenging due to heating/loss in the real experiments. Note that the optical Raman lattice schemes, introduced in the present review, can be generically applied to any type of atom candidates, including the alkali earth and lanthanide atoms which have been confirmed to have much less heating in observing SO effects. It is therefore of high interests to investigate the novel interacting topological physics with high-dimensional SO coupled alkali earth and lanthanide quantum gases.

\section{Acknowledgment}

This work was supported by the National Key R\&D Program of China (2016YFA0301604), National Nature Science Foundation of China (under grants N0. 11574008 and No. 11761161003), and the Thousand-Young-Talent Program of China.


\noindent

\end{document}